\documentclass[aps,pre,longbibliography,notitlepage,floatfix,nofootinbib]{revtex4-1}

\usepackage{bm,amsmath,amssymb,graphicx,stmaryrd,float,subcaption,upgreek,sidecap,MnSymbol,mathtools}
\usepackage{epstopdf}
\usepackage[toc]{appendix}
\usepackage[justification=raggedright]{caption}
\usepackage[abs]{overpic}
\usepackage{comment,footnote}

\newcommand{\bX}{{\bf X}}

\usepackage{fullpage}

\begin{document}
	\title{Bifurcations of buckled, clamped anisotropic rods and thin bands under lateral end translations}
	\author{T. Yu}
	\email{jhyutian@vt.edu}
	\affiliation{Department of Biomedical Engineering and Mechanics, Virginia Polytechnic Institute and State University, Blacksburg, VA 24061, U.S.A.}
	\author{J. A. Hanna}
	\email{hannaj@vt.edu}
	\affiliation{Department of Biomedical Engineering and Mechanics, Department of Physics, Center for Soft Matter and Biological Physics, Virginia Polytechnic Institute and State University, Blacksburg, VA 24061, U.S.A.}
	\date{\today}
	
\begin{abstract}
Motivated by observations of snap-through phenomena in buckled elastic strips subject to clamping and lateral end translations, we experimentally explore the multi-stability and bifurcations of thin bands of various widths and compare these results with numerical continuation of a perfectly anisotropic Kirchhoff rod.
Our choice of boundary conditions is not easily satisfied by the anisotropic structures, forcing a cooperation between bending and twisting deformations.
We find that, despite clear physical differences between rods and strips, a naive Kirchhoff model works surprisingly well as an organizing framework for the experimental observations.
In the context of this model, we observe that anisotropy creates new states and alters the connectivity between existing states.
Our results are a preliminary look at relatively unstudied boundary conditions for rods and strips that may arise in a variety of engineering applications, and may guide the avoidance of jump phenomena in such settings.  We also briefly comment on the limitations of current strip models.
\end{abstract}
\maketitle

\section{Introduction}

Thin rods and bands, the latter known also as strips or ribbons, display complex geometric response under simple end loadings and clampings. 
While much work has been done to explore these phenomena, most of the literature pertains either to periodic boundary conditions or highly symmetric end loadings such as a wrench, in which the end-to-end vector, loading vector, and twist are coaxial.  
Most of this is also limited to the analytically tractable case of isotropic rods; here and elsewhere in this paper, the term isotropic refers to the structure rather than the material, such that the cross section has no preferred bending direction.
However, the space of possible boundary conditions is much wider, and includes conditions that can interact strongly with the anisotropy of a strip or any other elastic structure with a distinguished material frame.
In many practical situations, the two ends of the structure may be clamped such that their material frames take any orientation with respect to each other and the end-to-end vector.  As we will show, certain conditions conspire with the anisotropy to frustrate the system and couple its twist and writhe response.  Clamped boundary conditions can not only create energy barriers through frustration, but may even introduce topological barriers between an undeformed ground state and excited states \cite{baez1991topological}.
The present study is a preliminary exploration covering a small piece of this wider parameter space, as applied to anisotropic rods and bands.
We begin with symmetrically clamped buckled strips of varying width, and subject their ends to a lateral displacement parallel to the width direction.
The introduction of this ``shearing'' motion reveals a rich set of stable configurations and jump phenomena, including several snap-through instabilities, that to our knowledge do not appear in the literature (we encourage the reader to take a quick glance now at the supplementary video {\texttt{widtheffect45.mp4}} \cite{videos} to see examples of such stable states and snap-throughs).
We compare experimentally determined stability ranges of various configurations to results from numerical continuation of the Kirchhoff rod equations, 
and find that a perfectly anisotropic rod model captures the complicated choreography of bifurcations of narrow bands, and provides much of the backbone of the behavior of wider bands.  
We reveal connections between various states, including higher-order unstable \emph{elastica} modes and stable twisted states created by the rod's anisotropy.

While the Kirchhoff equations show themselves to be a surprisingly useful tool in the analysis of strip behavior, we wish to emphasize that there is no reason to assume that such a model, which assumes that cross sections remain perpendicular to the centerline, would be appropriate for strips.
On the other hand, the common assumption that transverse bending of strips is governed by the constraint of developability can lead to difficulties of its own, particularly for narrow strips, issues that we will briefly touch upon in an appendix.  Until such issues are resolved, it is advantageous to employ an easily implemented rod model from which the strips inherit most, or even all, of their bifurcations.
However, use of such a model should not be taken to imply that a narrow strip is equivalent to a rod.

 Boundary conditions like those we explore here are potentially of interest in helping to avoid violent snap-throughs of connectors, hinges, and umbilicals in flexible and deployable systems.  Geometries similar to ours appear 
 as slipping folds \cite{arya2015wrapping} in deployable space membranes, buckled elements in flexible electronics and robotics, and decorative streamers in childrens' toys \cite{princesswand}.
Multi-stable structures find use in compliant mechanisms \cite{Howellbook} at all length scales.
The behavior of strips under our loading conditions is likely related to the phenomenon of lateral-torsional buckling, known to structural engineers \cite{mandal2002lateral}.

There is much prior work on the configurations of naturally straight rods.  Work on the general behavior and classification of solutions includes that of Antman \cite{antman1981large,antman1975}, Maddocks \cite{kehrbaum1997elastic}, Nizette and Goriely \cite{nizette1999towards}, and Cognet and co-workers \cite{ameline2017classifications}.
Neukirch and Henderson made a detailed investigation into the connectivity of solutions for rods subject to end thrusts and coaxial twists \cite{neukirch2002classification,henderson2004classification}. Theory, numerics, and experiment show that circular cross section rods, an integrable system, when subject to such boundary conditions will buckle, hockle into a loop, or snarl into a self-contacting twisted structure \cite{coyne1990analysis,Yabuta1982cable,miyazaki1997analytical,goss2005experiments,van2003instability,thompson1996helix,van2000helical,goyal2005nonlinear,coleman1995theory}. Anisotropic rods, those with preferred bending directions, display even more complicated and potentially non-integrable behavior due to non-conserved twist \cite{mielke1988spatially,champneys1996multiplicity,van1998lock,beda1992postbuckling,buzano1986secondary, goriely2001dynamics}.  van der Heijden and Thompson \cite{van1998lock} distinguish between weakly anisotropic and strongly anisotropic ``tape-like'' behavior such as that we will discuss in this paper. Integrability can also be destroyed by the addition of gravity \cite{antman1981large,lu1995complex}, but can be preserved under addition of extensibility and shearability \cite{stump99hockling,shi1995elastic}.  
Early experiments by Green \cite{green1936equilibrium, green1937elastic} showed that twist under tension makes strips unstable to the formation of multiple loops, with only a single loop forming in the absence of tension.  Recent work by Chopin and Kudrolli \cite{chopin2013helicoids} extends these findings and reveals a rich set of possible deformations and patterns under tension and twist, many of them involving stretching.
Aside from the present investigation, the only work we know of featuring lateral displacements is that of Morigaki and co-workers \cite{morigaki2016stretching}, who begin with a slightly laterally displaced loop configuration of a strip, pull the ends, and find behavior similar to the hockling and pop-out regimes of isotropic rods.
Other interesting boundary conditions include freely hinged conditions \cite{perkins1990planar,lu1995complex}, and asymmetric rotation in the plane of buckling leading to snap-throughs \cite{plaut2009vibration}.  
Another much-studied corner of parameter space, due to its supposed relevance to DNA, is that of pre-twisted rings composed of isotropic or anisotropic rods \cite{coleman2004theory,manning1999symmetry, hoffman2003link,tanaka1985elastic,wadati1986elastic,tobias1994dependence,starostin1996three}.  The latter includes, as one particular case, the configurations of a narrow M{\"{o}}bius band \cite{mahadevan1993shape,moore2015computation}.
Dichmann, Li and Maddocks \cite{dichmann1996hamiltonian}, Li and Maddocks 1996 \cite{li1996computation}, and Domokos and Healey \cite{domokos2001hidden} provide insight into the connectivity of various solutions of this type.

The current paper is organized as follows. 
We introduce the geometry of our problem and describe our experiments in Section \ref{description}, and present the anisotropic Kirchhoff rod model in Section \ref{rodmodel}.
Results from experiments on narrow bands and numerical continuation of the Kirchhoff equations are compared in Section \ref{results}, through a series of slices through parameter space.  In Section \ref{loci}, a more global view is presented through the loci of bifurcation points in parameter space, which also delineate regions of stability for different states.
Experiments on bands of varying width are presented in Section \ref{widtheffect} and compared with the results for narrow bands and rods.
 Many smooth and discontinuous bifurcations for both narrow and wide bands are shown in several supplementary videos \cite{videos}, which complement the diagrams in Sections \ref{results}-\ref{widtheffect}.  We discuss a few additional points of interest in Section \ref{discussion}.
  In the Appendices, we show two types of configuration that exist as the bands approach the limits of developable deformation, where elastic energy focuses in conical defects at the clamped ends, give details on solving the boundary value problem, briefly discuss the minor effects of Poisson's ratio on the loci of bifurcation points, briefly contrast our results with those for isotropic (square) rods, and discuss problems that arise when employing strip models to describe the behavior of bands such as those in our experiments.

\section{Geometry of the experiments, methods, and errors} \label{description}

\begin{figure}[h!]
	\centering
	\begin{subfigure}[t]{0.48\textwidth}
		\centering
		\includegraphics[height=1.75in]{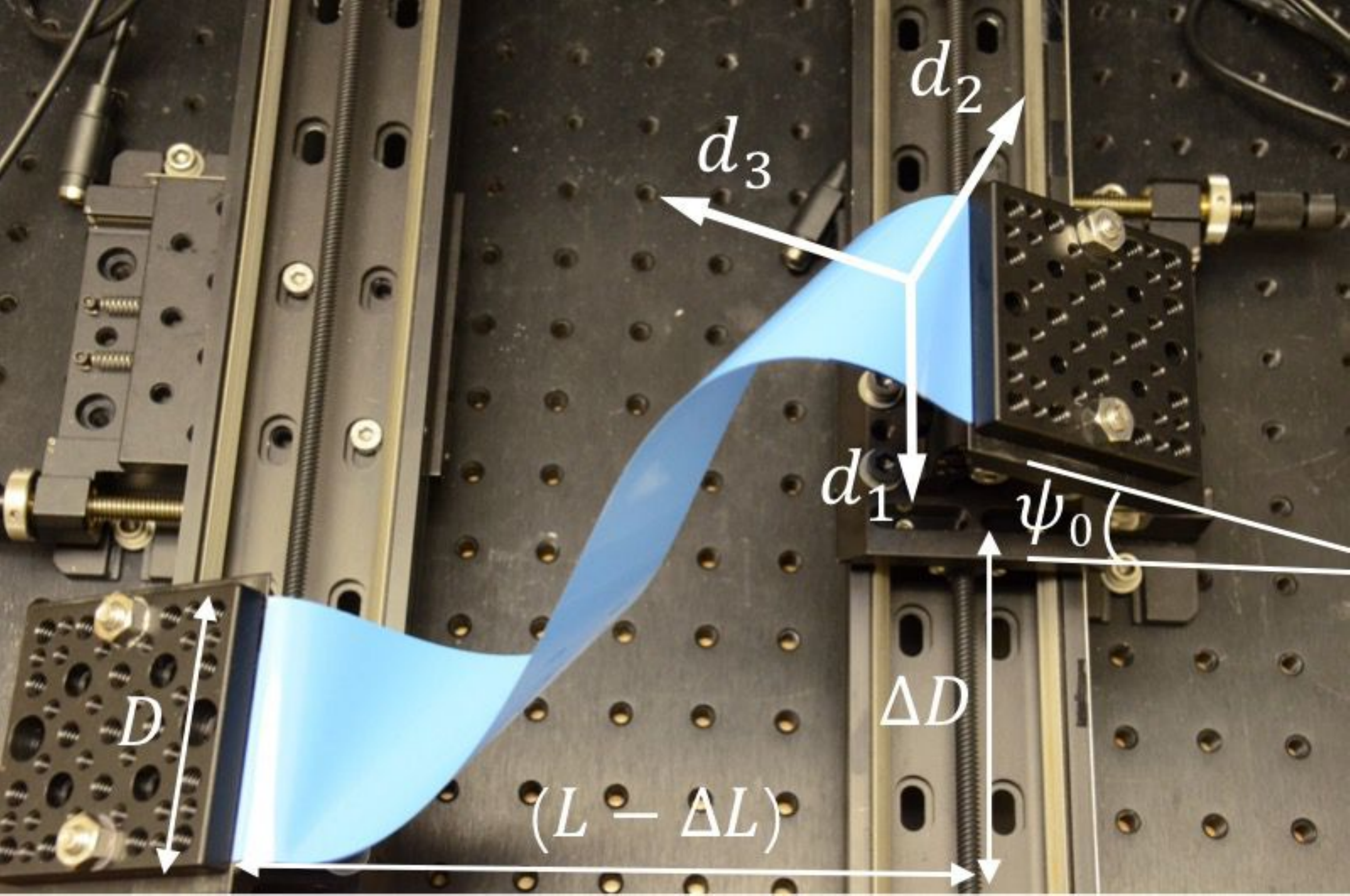}
		\caption{}\label{fig:SlikeRl}
	\end{subfigure}  \hspace{-30pt}
	\begin{subfigure}[t]{0.48\textwidth}
		\centering
		\includegraphics[height=1.75in]{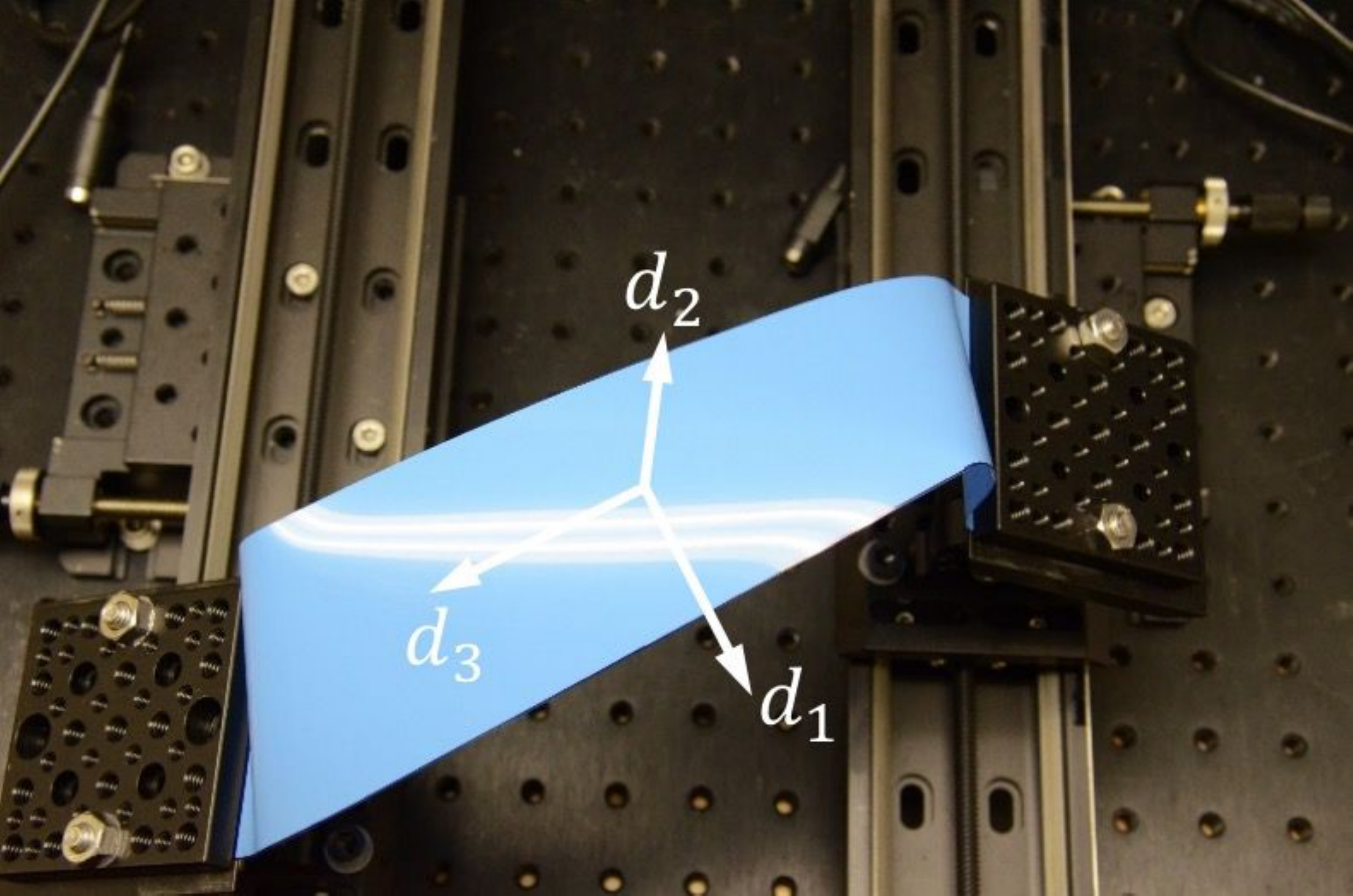}
		\caption{}\label{fig:UlikeRl}
	\end{subfigure}                \\
	\vspace{-0.1in}
	\captionsetup[subfigure]{labelfont=normalfont,textfont=normalfont}
	\caption{A thin, rectangular band of width $D$ and length $L$ is symmetrically clamped with an angle $\psi_0$, ``compression'' $\Delta L$, and ``shear'' $\Delta D$.  The centerline of the band carries an orthonormal director frame $(\bm{d_1}, \bm{d_2}, \bm{d_3})$ corresponding to the width direction, the surface normal, and the tangent, respectively.  Shown here are (\subref{fig:SlikeRl}) $S$-like and (\subref{fig:UlikeRl}) $U$-like configurations.
	}\label{fig:Intros}
\end{figure}

The boundary conditions we impose are shown in Figure \ref{fig:Intros}.  We use thin (0.005 $\pm$0.0005 inch / 0.127 $\pm$ 0.013 mm) bands cut from polyester shim stock (Artus Corp., Englewood, NJ) with free length $L = 240 \pm 0.5$ mm ($\approx 5$ mm clamped length on either side) and various widths, the most common being $D = 3 \pm 0.05$ mm, $30 \pm 0.5 $ mm, and $60 \pm 0.5$ mm, corresponding to aspect ratios $D/L$ of $1/80$, $1/8$, and $1/4$, respectively.  We use a Silhouette Cameo 3 cutting machine (Silhouette America, Lindon, UT) for the narrow bands, and a paper trimmer for the wider bands.  The additional accuracy was necessary for narrow bands because bifurcations of highly twisted configurations were found to be sensitive to non-uniformities in width.
  With respect to an initially flat configuration, the boundary conditions consist of a symmetric tilt angle $\psi_0$ ($\pm 1^{\circ}$), a ``compression'' $\Delta L$ ($\pm 0.5$ mm) in the  length direction, and a lateral ``shear'' $\Delta D$ ($\pm 0.5$ mm) parallel to the width direction at the ends. 
The clamping is parallel to the width direction with an accuracy of $\pm 2^{\circ}$ for $D/L =1/80$, $\pm0.2^{\circ}$ for $D/L = 1/8$, and $\pm0.1^{\circ}$ for $D/L=1/4$.  Typically, we fix the compression and clamping angle, and use the normalized lateral shear displacement $\Delta D/L$ as the primary bifurcation parameter.  We align the shear direction with gravity, which mitigates its influence.  A bias of $\approx 0.5^{\circ}$ in this alignment was introduced by the slope of the laboratory floor. We measure the range of stability 
 for all observed configurations of the bands, many of which require manual manipulation to obtain.  
The bands are never kept in any particular deformed state for longer than a few minutes, to avoid possible viscous response that could affect results.  With these precautions, results are reproducible with a typical variation of about $\pm 2$ mm (corresponding to $\pm1/120$ $\Delta D/L$) between trials, although a very few narrow band states such as the $US\pm$ and $W$ states at $\psi_0=60^{\circ}$ show deviations of as much as $\pm 5$ mm.  Near bifurcations that change connectivity, multiple trials may have qualitative differences; these are discussed when they arise in Section \ref{results}.   All narrow band data come from averaging three trials; wide band data are single trials, other than some additional trials performed to estimate variation.  

We describe the bands in terms of the geometry of a rod, a description that is most suitable for narrow bands, though quite distinct from the description of a band as a developable surface.  The mechanics of an anisotropic rod will be discussed in the following section. The description involves an orthonormal material frame $(\bm{d_1}, \bm{d_2}, \bm{d_3})$ attached to the centerline of the band, with the three directors corresponding to the width direction, the surface normal, and the tangent, respectively.   Figure \ref{fig:Intros} shows such a frame superimposed on two example configurations of a wide band, which we refer to as $S$-like and $U$-like, and which we will describe in more detail in Section \ref{results}.  

Bands can, in theory, be approximated as developables until the shear displacement $\Delta D/L$ approaches a limiting value where such an isometric description is no longer possible.  We discuss this further in Appendix \ref{limitstates}.  
Note that in describing the bands as developable surfaces, the straight line generators of the surface would not coincide with the material directors of the rod description; the geometry of a developable strip differs from that of a rod with undeformed cross section.  Further comment can be found in Appendices \ref{bvp} and \ref{stripmodel}.

\section{Anisotropic rod model}\label{rodmodel}

We compare experimental results with a simple model of a perfectly anisotropic rod.  This model assumes that the only way a band can deform is by bending around its width direction and twisting around its tangent.  Bending around the surface normal is forbidden.  The rod is inextensible and unshearable; its centerline is given by $\bm{X}(s)$, where $s$ is the arc length, and the tangent can be identified with one of the directors, $\bm{X}' = \bm{d_3}$ (throughout this paper a prime will denote an $s$-derivative).  The kinematics of the frame $(\bm{d_1}, \bm{d_2}, \bm{d_3})$ are given by 
\begin{align}\label{rotationframe}
\bm{d_i}'&=\bm{\omega} \times \bm{d_i} \, , \\
\bm{\omega}&=\kappa_1 \bm{d_1} + \tau \bm{d_3} \, ,
\end{align}
where the Darboux vector $\bm{\omega}$ has no component normal to the strip.  The generalized strains $\kappa_1$ and $\tau$ are the curvature in the easy (only) direction, and the twist about the tangent.  For a perfectly anisotropic strip, the frame  $(\bm{d_1}, \bm{d_2}, \bm{d_3})$ can be identified with the Frenet-Serret frame as $(\bm{b}, \bm{-n}, \bm{t})$, and the curvature $\kappa_1$ and twist $\tau$ with the curvature and torsion.  This type of model has been used previously as an approximate model for the shape of elastic strips \cite{mahadevan1993shape}; our present interest is primarily in bifurcations rather than shapes.

Linear and angular momentum balances are provided by the Kirchhoff equations for the contact force and moment $\bm{N}$ and $\bm{M}$ in the absence of gravity or other distributed loads or couples,
\begin{equation}\label{F&Mequilibrium}
\begin{aligned}
\bm{N} ' &=\bf{0} \, ,  \\
\bm{M} '+\bm{d}_3  \times \bm{N}  &=\bf{0} \, .  \\
\end{aligned}
\end{equation}
 Three quantities are conserved along the centerline \cite{van2000helical},

\begin{equation}\label{FirstIntegral} 
\begin{aligned}
C_1 &=\tfrac{1}{2}\bm{M} \cdot \bm{\omega} +\bm{N} \cdot \bm{d_3} \, ,\\
C_2 &=\bm{N} \cdot \bm{N} \, ,\\
C_3 &=\bm{N} \cdot \bm{M} \, .\\
\end{aligned}
\end{equation}
Isotropic rods conserve the twist as a fourth quantity; our system does not.  The general anisotropic rod is known to be non-integrable, but we are unaware of any published results on the presence or lack of integrability for the perfectly anisotropic case.
We resolve $\bm{N}$ and $\bm{M}$ on the moving frame as $\bm{N} =N_i\bm{d}_i$ and $\bm{M} =M_i\bm{d}_i$, and assume linear constitutive relations $M_1=EI_1 \kappa_1$ and $M_3=GJ \tau$, where  $E$ is the Young's modulus and $G$ is the shear modulus, $I_1$ is the principal moment of inertia of the cross-section in the easy direction, and $GJ$ is the torsional rigidity. The other moment $M_2$ is a Lagrange multiplier enforcing the vanishing of curvature in the hard $\bm{d_2}$ direction.  The ratio of $E$ to $G$ involves the elastically isotropic Poisson's ratio $\nu$, which we set to $0.25$ for the present study; this choice makes little difference to the results, as shown in Appendix \ref{poisson}.
 There are thus six scalar balance equations.  Reconstruction of the rod centerline and frame orientation is achieved through a quaternion representation leading to a set of thirteen equations.
We solve these using the continuation package AUTO 07P \cite{doedel2007auto}.  Details, along with the specification of boundary conditions, are discussed in 
Appendix \ref{bvp}.

\section{Numerical and experimental results for narrow bands}\label{results}

In this section, we present experimental results on narrow bands ($D/L = 1/80$) and compare them with numerical results from the anisotropic Kirchhoff rod model.  
We restrict our experimental parameter space to a single compression $\Delta L/L = 1/2$ and clamping angles $0 \le \psi_0 \le 60^{\circ}$.
Using the shear $\Delta D/L$ as a bifurcation parameter, we deform the bands to near the isometric limit, and  find a rich and complicated landscape of stable configurations and both smooth and violent transitions.
These observations, which depend strongly on clamping angle, are described surprisingly well by the naive model.  We keep gravity out of the model, as its effects are easily accounted for, and in practical terms it would simply break some of the symmetry of the solutions we wish to explore and create more complicated and potentially confusing figures.

The boundary conditions we impose might be easily accommodated by an isotropic rod bending in what is a forbidden direction for the perfectly anisotropic strip.  Instead, our frustrated system is induced to find some combination of allowed bending and twist in order to satisfy the constraints.  In short, the shear indirectly causes a non-uniform twist, and as there are multiple ways for bending and twisting energy to compete, creates a highly multi-stable system.  As the shear increases, the system shifts from being compressed to being in tension.   To our knowledge, these boundary conditions have not been explored before in the literature.  However, Morigaki's \cite{morigaki2016stretching} recent experiments on tensioned loops can be interpreted in terms of our boundary conditions as a clamping angle $\psi_0$ of $180^{\circ}$, large compression ($\Delta L / L > 1$), and small shear $\Delta D/L$. 

In general, the clamping angle will bias the strip towards two general types of configurations, with some seeming similarity to primary buckling modes of planar \emph{elastica}.  Small clamping angles favor $S$-like shapes that live both above and below the plane of clamping, and come in chiral pairs.  High clamping angles favor $U$-like shapes that live mostly on one side of the plane of clamping, and are symmetric about their midpoints.  Examples are shown in Figure \ref{fig:Intros}.
This is consistent with what is known about the rod equations, 
namely that all solutions are either reversibly symmetric about their midpoint or are reversibly symmetric pairs \cite{van2003instability,neukirch2002classification,domokos2001hidden}.
As we increase the clamping angle, we gradually lose many of the states that exist at low angles.

\begin{figure}[h!]
	\captionsetup[subfigure]{labelformat=empty}
	\centering
	\begin{subfigure}[t]{0.13\textwidth}
		\centering
		\includegraphics[width=0.8in]{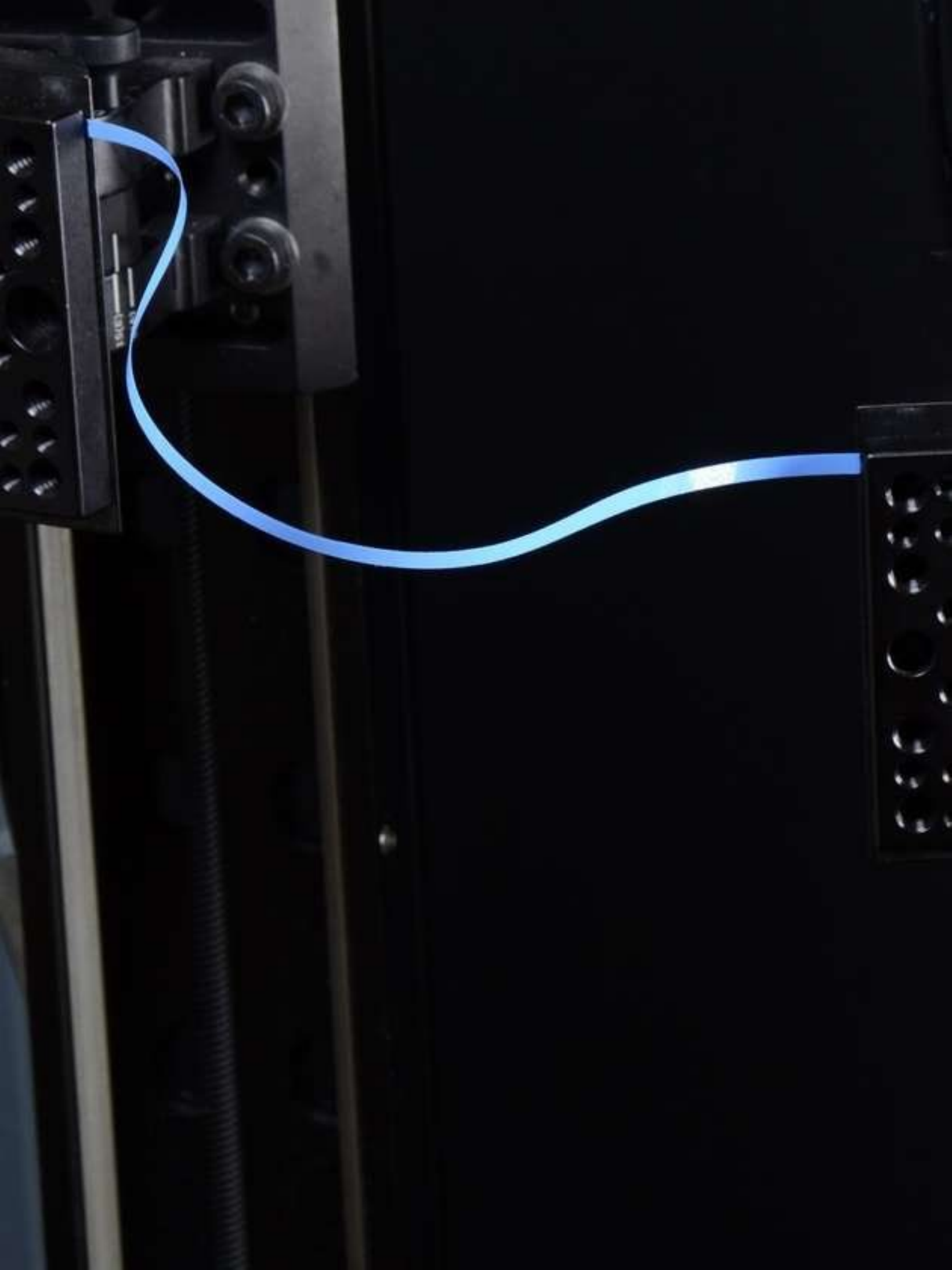}
		\caption{$U$}
	\end{subfigure}%
	\hspace{0.6pt}
	\begin{subfigure}[t]{0.13\textwidth}
		\centering
		\includegraphics[width=0.8in]{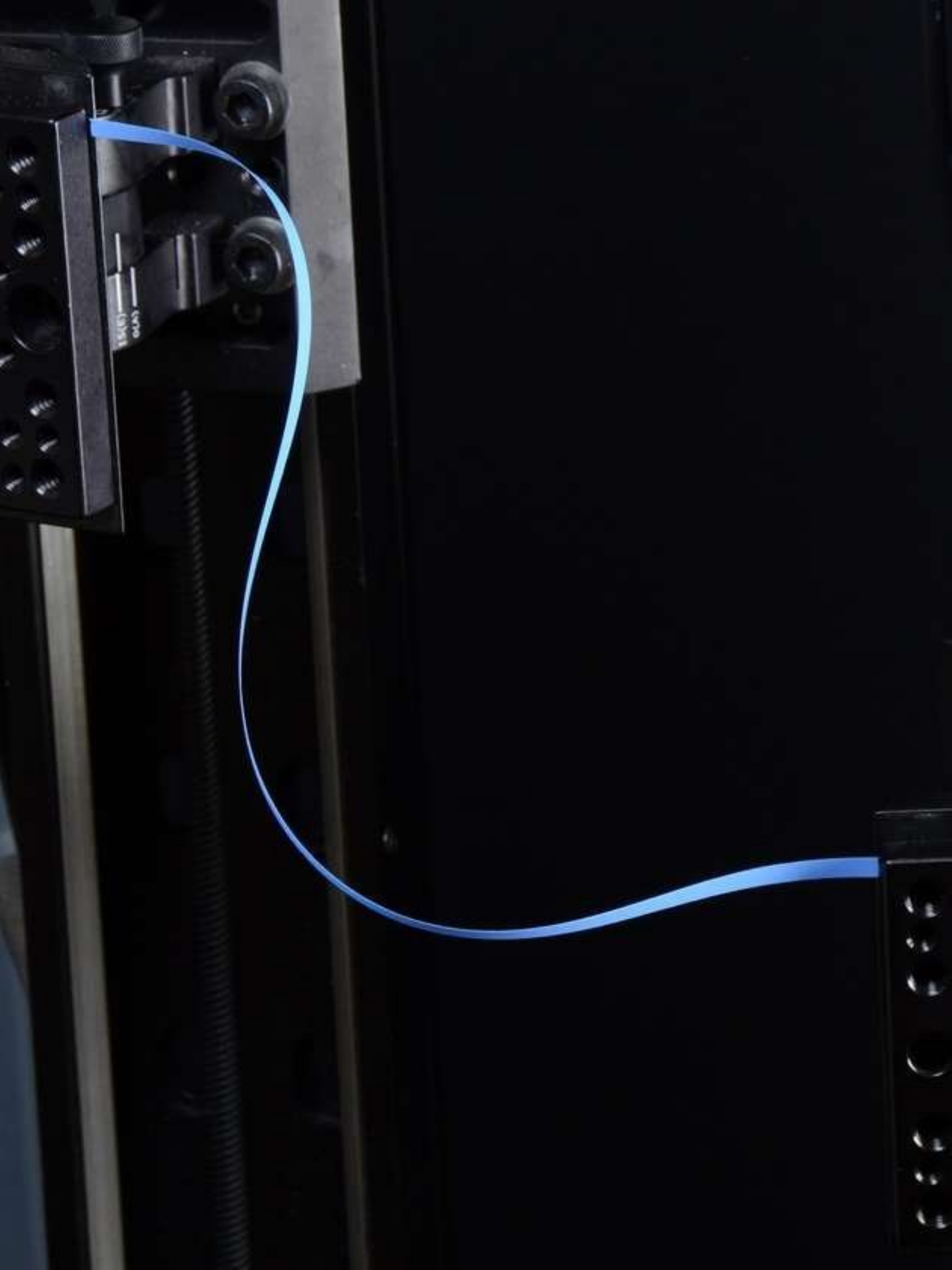}
		\caption{$US+$}
	\end{subfigure} 
	\begin{subfigure}[t]{0.13\textwidth}
		\centering
		\includegraphics[width=0.8in]{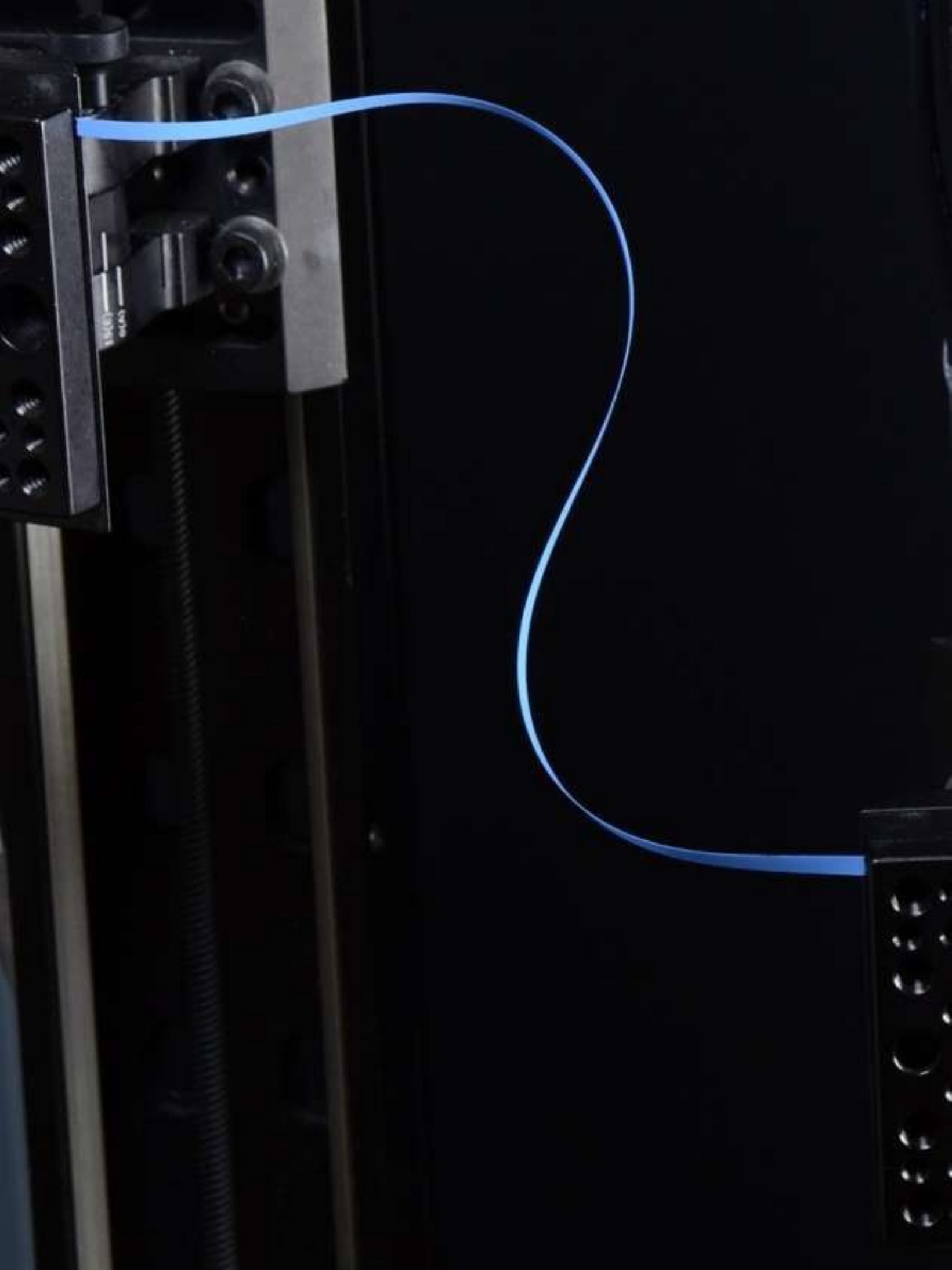}
		\caption{$WS+$}
	\end{subfigure}
	\begin{subfigure}[t]{0.13\textwidth}
		\centering
		\includegraphics[width=0.8in]{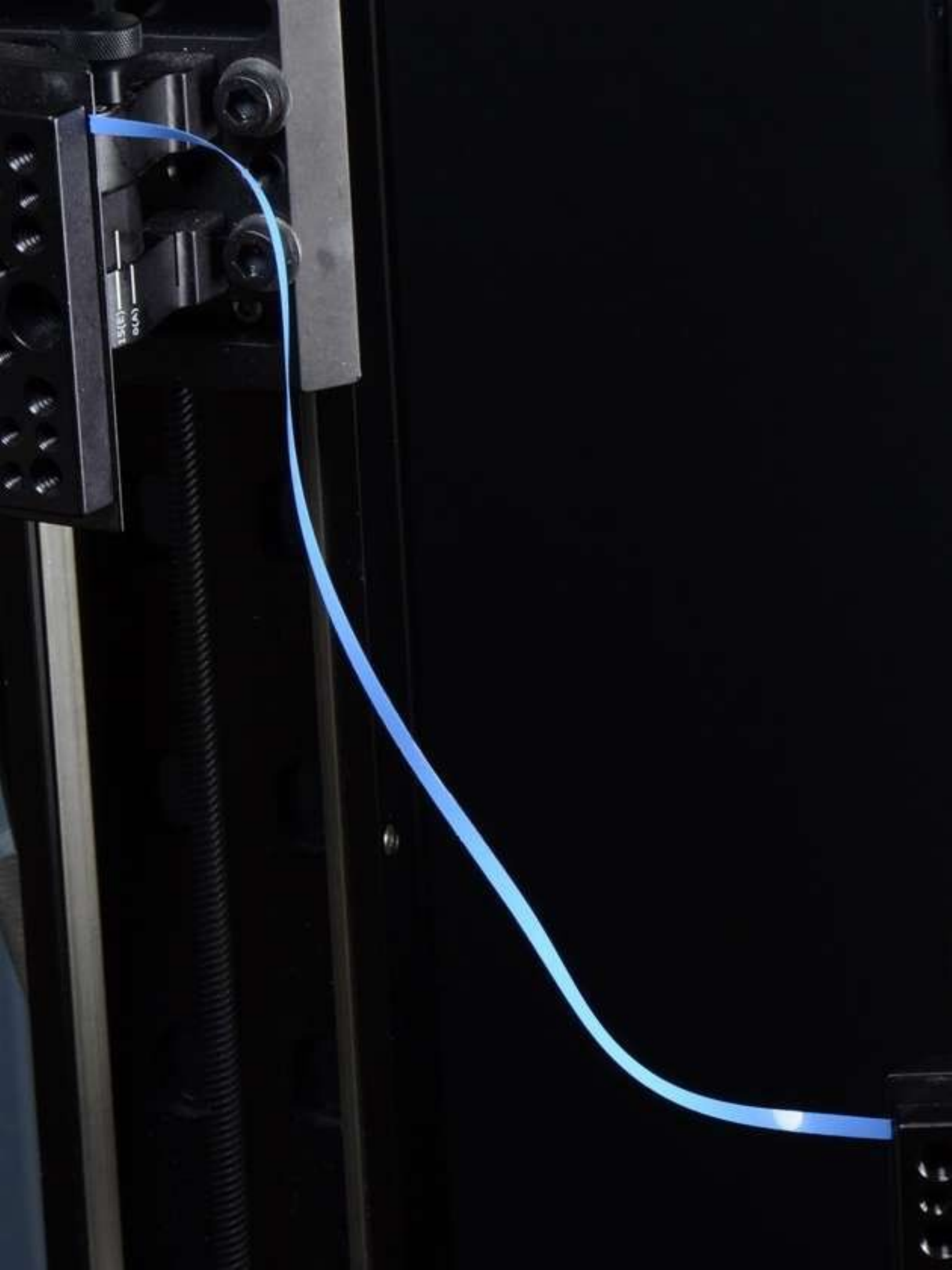}
		\caption{$uUu$}
	\end{subfigure}
	\begin{subfigure}[t]{0.13\textwidth}
		\centering
		\includegraphics[width=0.8in]{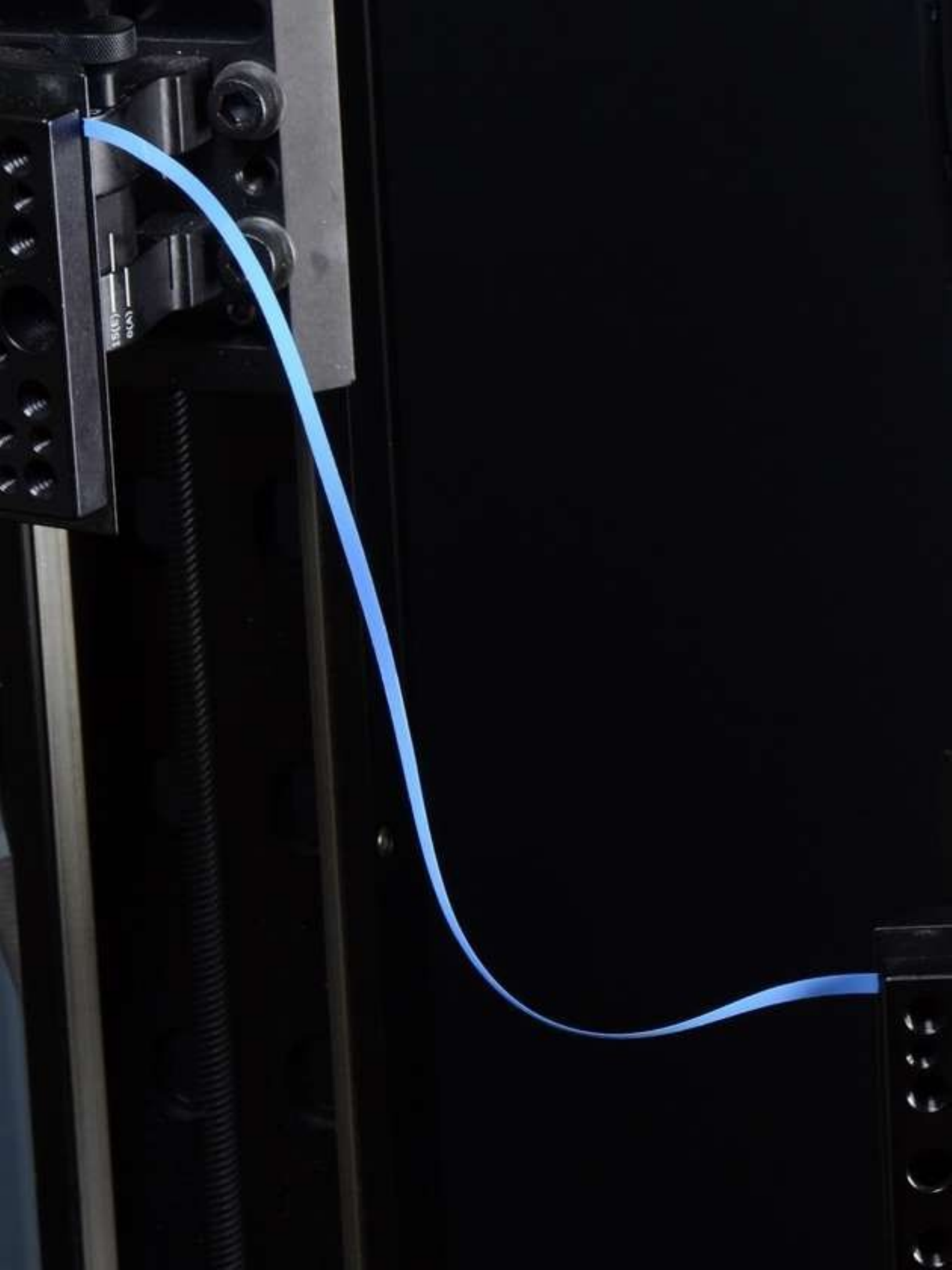}
		\caption{$w$}
	\end{subfigure}
	\begin{subfigure}[t]{0.13\textwidth}
		\centering
		\includegraphics[width=0.8in]{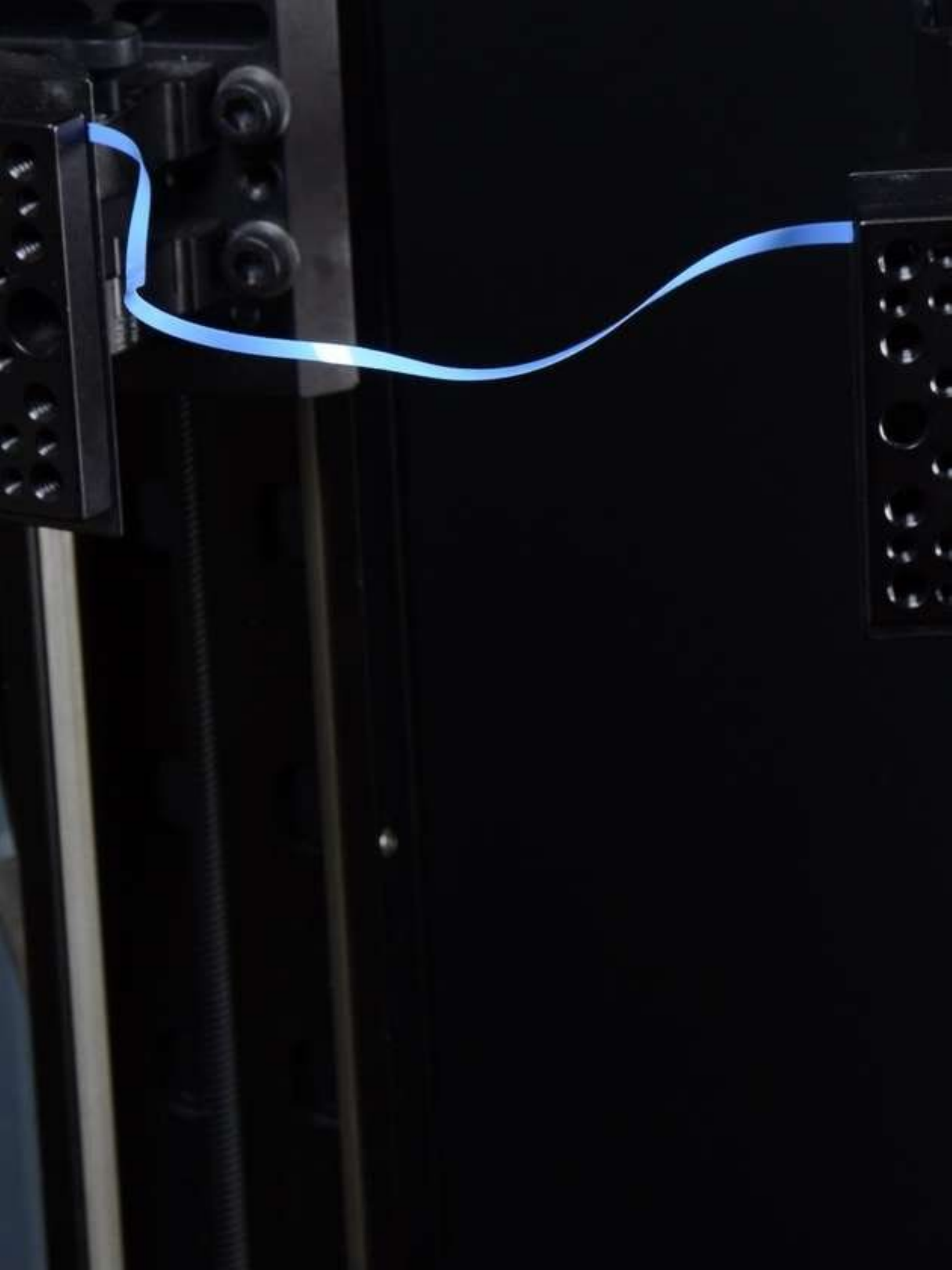}
		\caption{$TU+$}
	\end{subfigure}    
	\begin{subfigure}[t]{0.13\textwidth}
		\centering
		\includegraphics[width=0.8in]{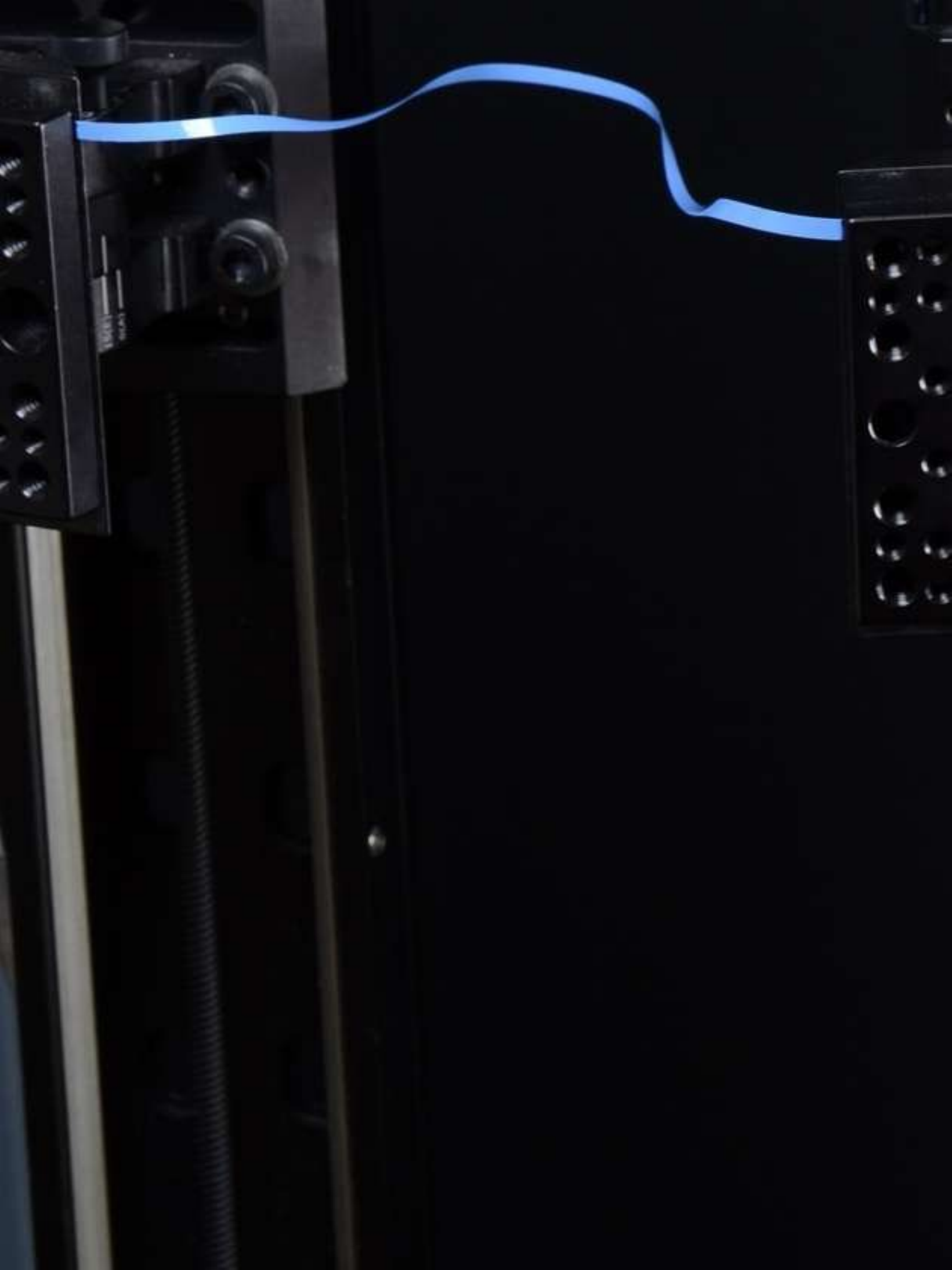}
		\caption{$TW+$}
	\end{subfigure}                \\
	\vspace{-3pt}
	\begin{subfigure}[t]{0.13\textwidth}
		\centering
		\includegraphics[width=0.8in]{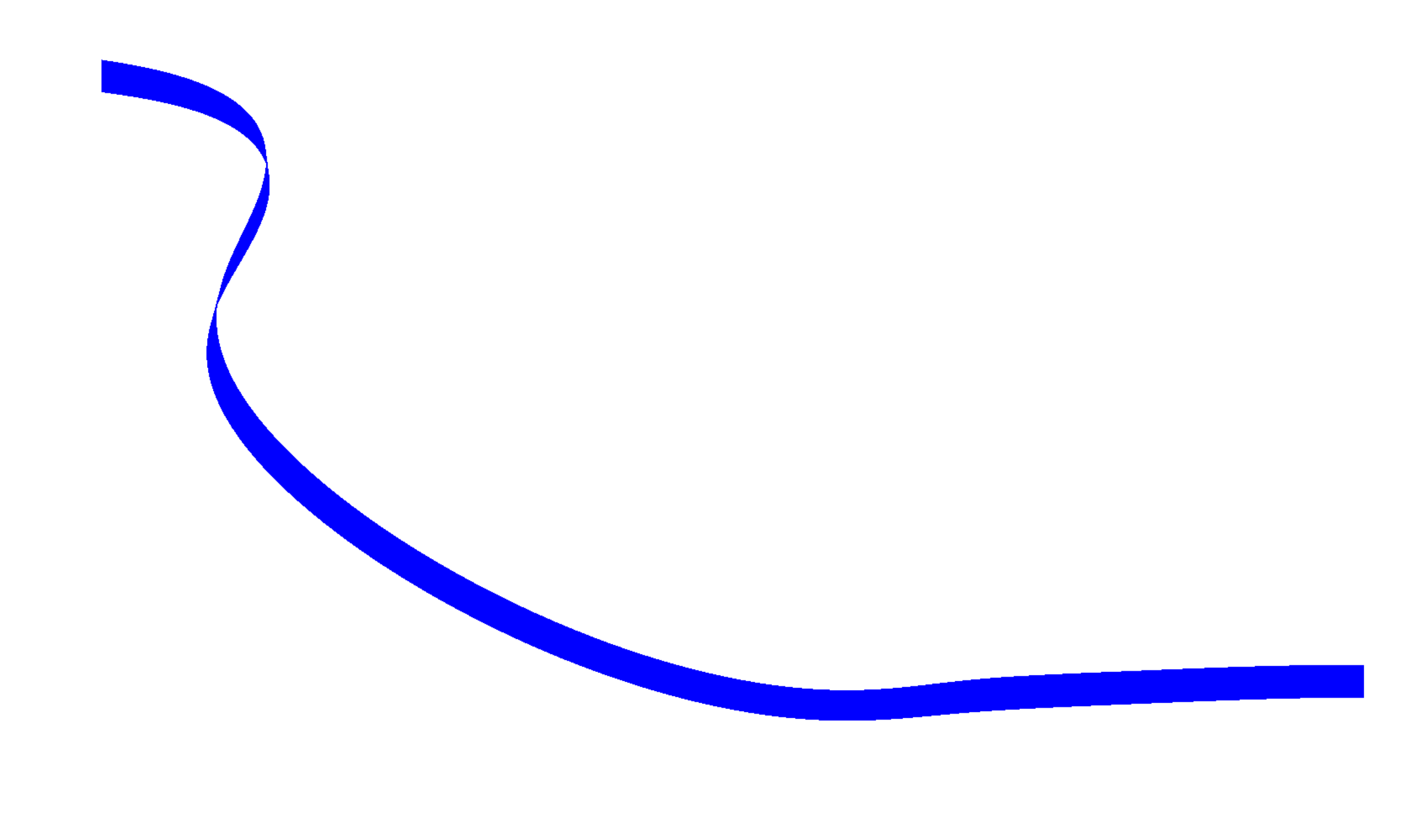}
	\end{subfigure}%
	\hspace{0.6pt}
	\begin{subfigure}[t]{0.13\textwidth}
		\centering
		\includegraphics[width=0.8in]{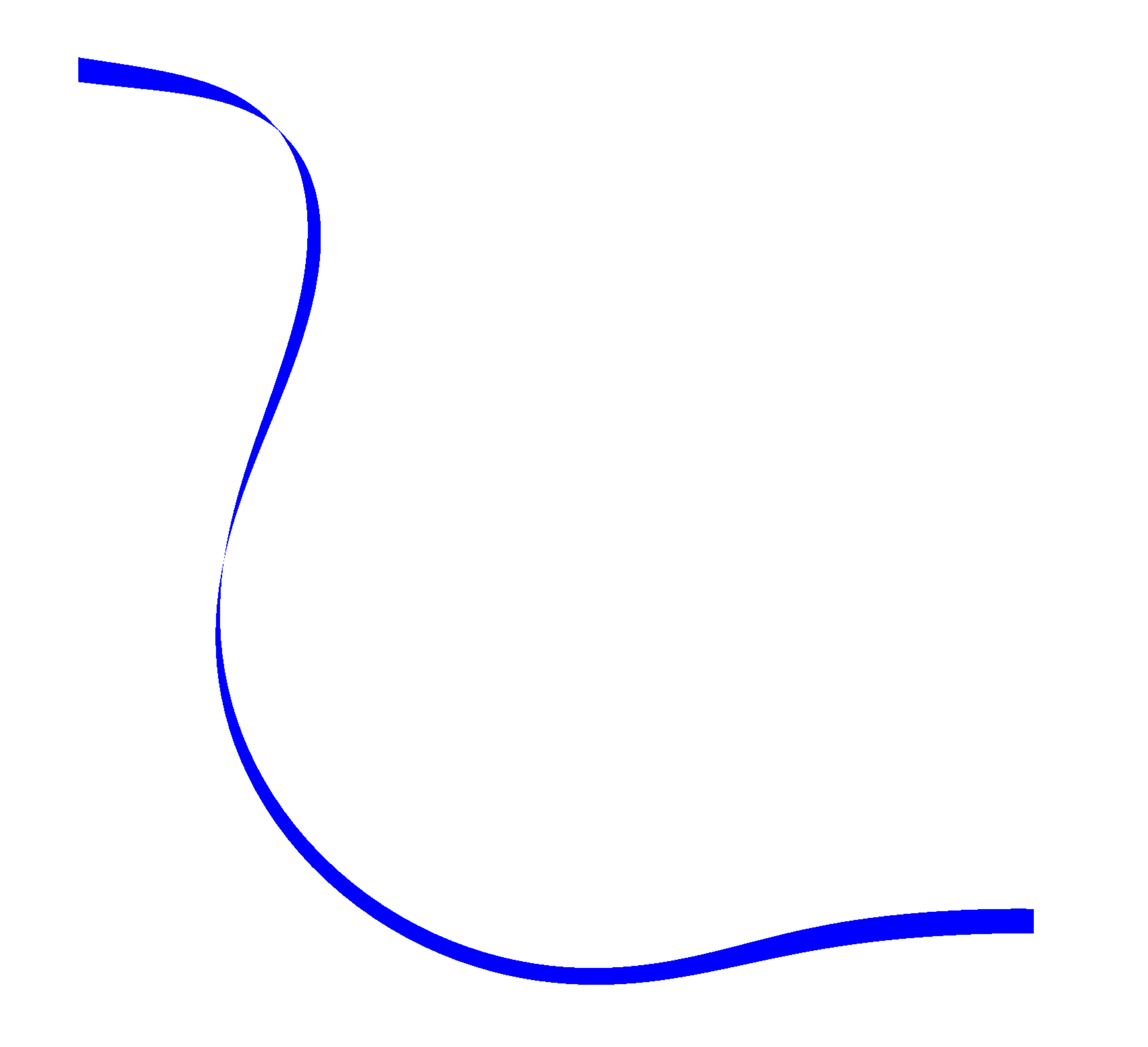}
	\end{subfigure} 
	\begin{subfigure}[t]{0.13\textwidth}
		\centering
		\includegraphics[width=0.8in]{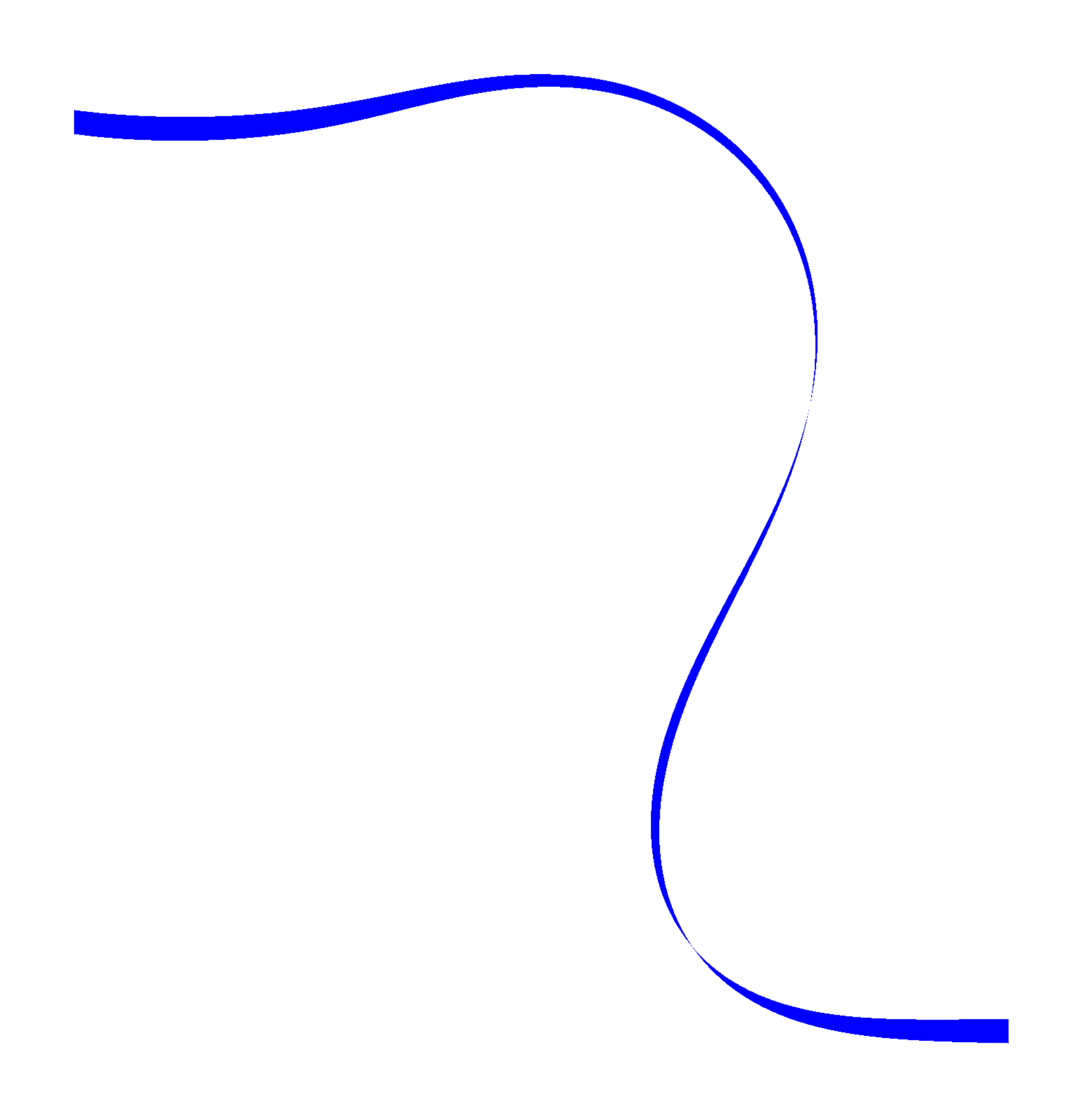}
	\end{subfigure}
	\begin{subfigure}[t]{0.13\textwidth}
		\centering
		\includegraphics[width=0.8in]{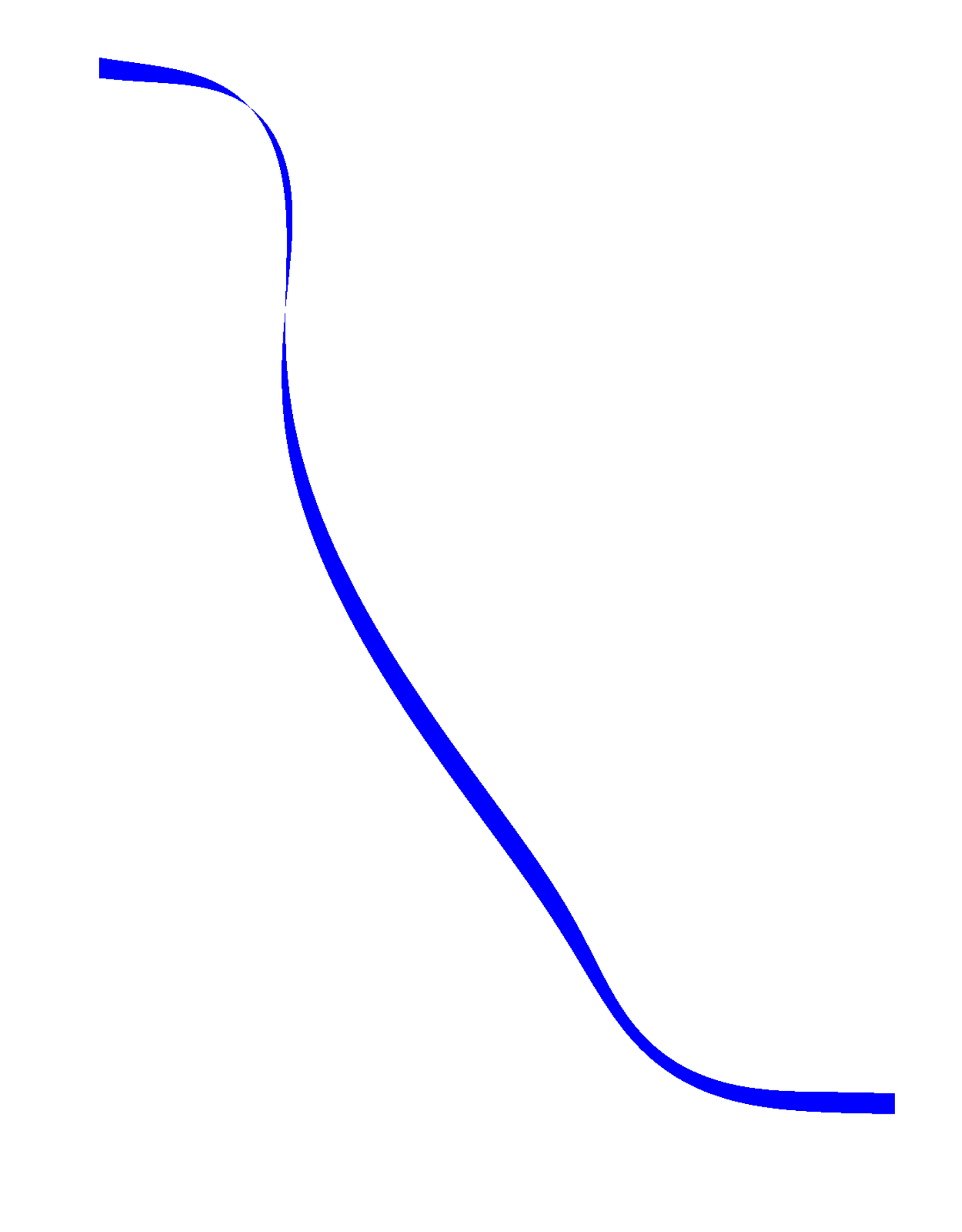}
	\end{subfigure}
	\begin{subfigure}[t]{0.13\textwidth}
		\centering
		\includegraphics[width=0.8in]{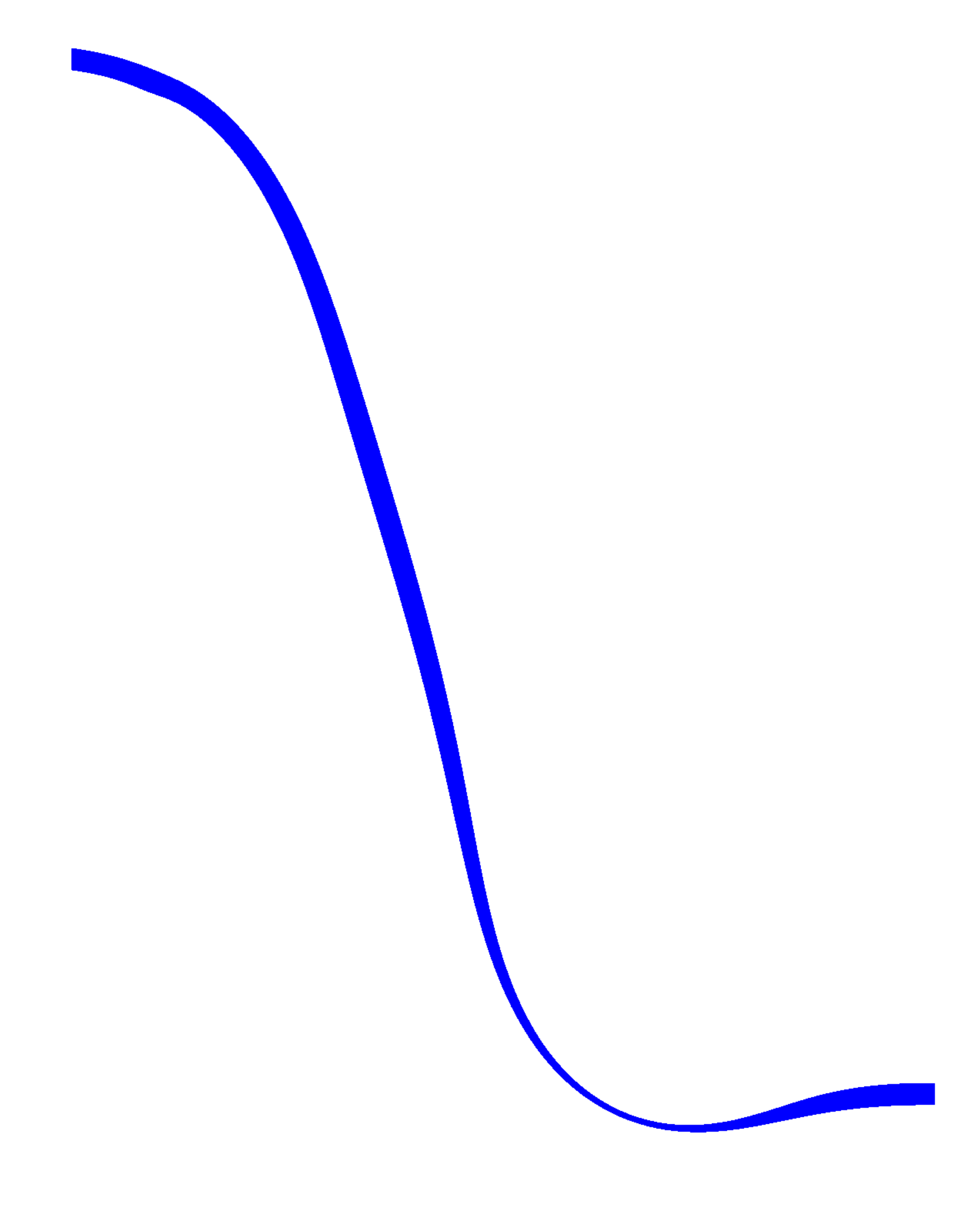}
	\end{subfigure}
	\begin{subfigure}[t]{0.13\textwidth}
		\centering
		\includegraphics[width=0.8in]{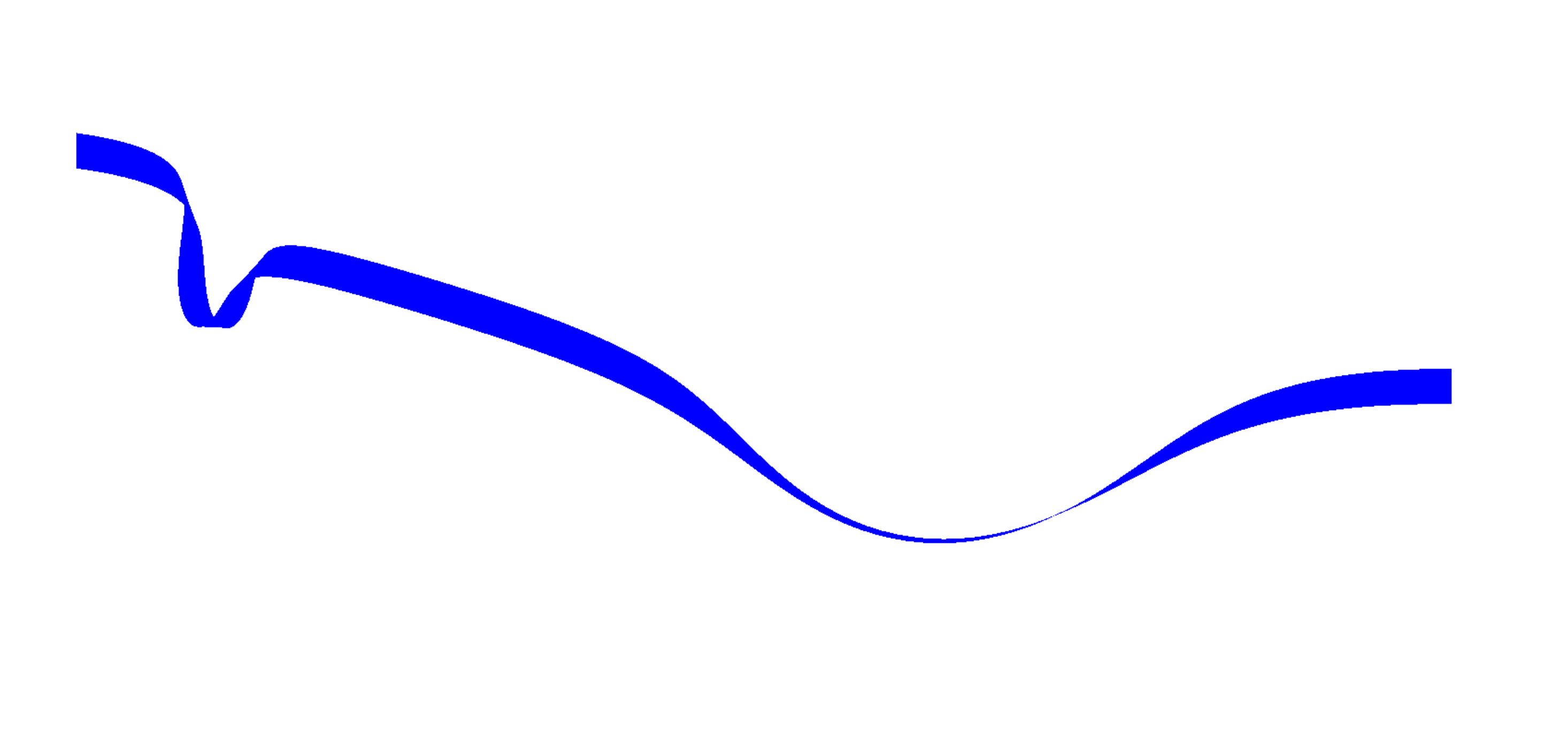}
	\end{subfigure}    
	\begin{subfigure}[t]{0.13\textwidth}
		\centering
		\includegraphics[width=0.8in]{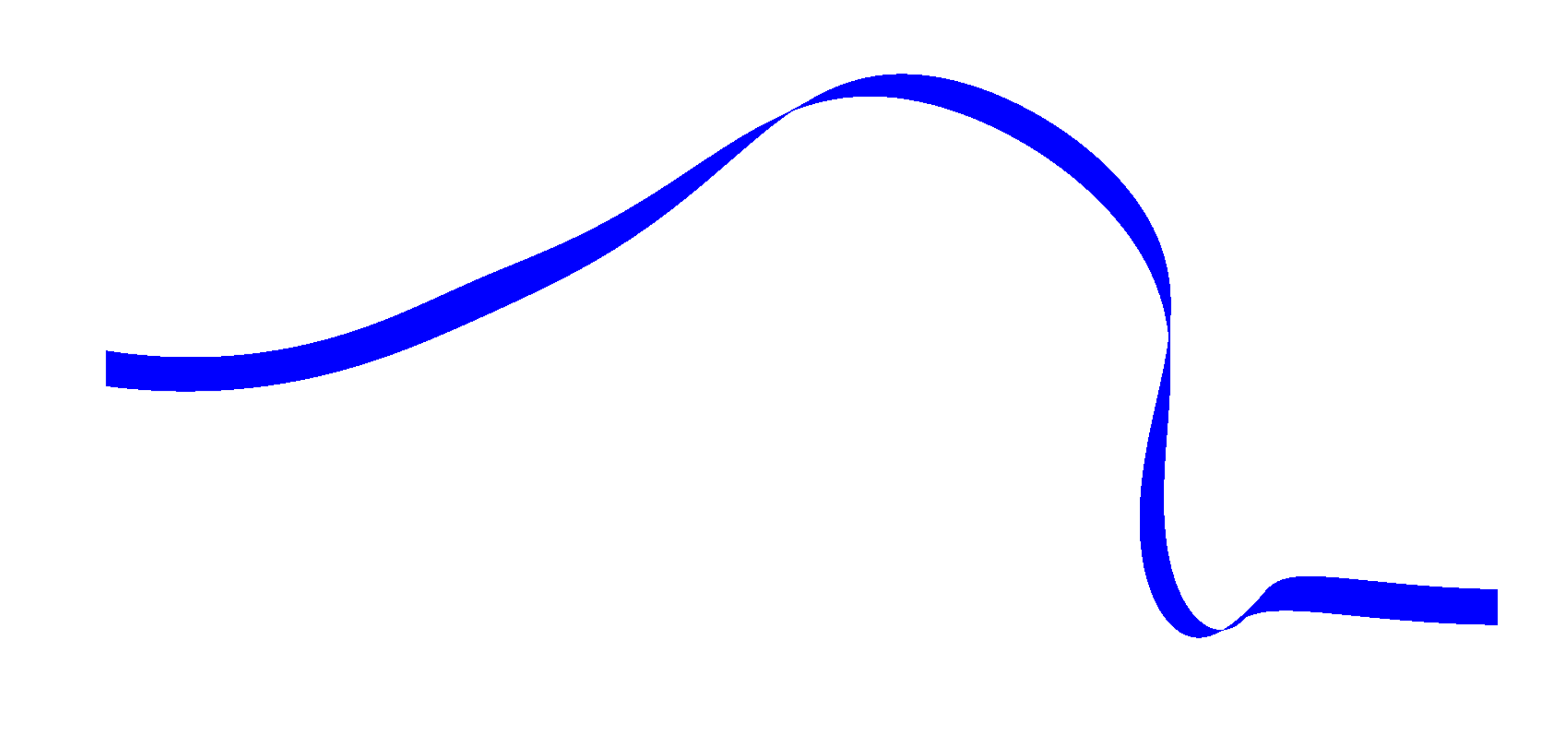}
	\end{subfigure}                  \\
\vspace{15pt}
	\begin{subfigure}[t]{0.13\textwidth}
		\centering
		\includegraphics[width=0.8in]{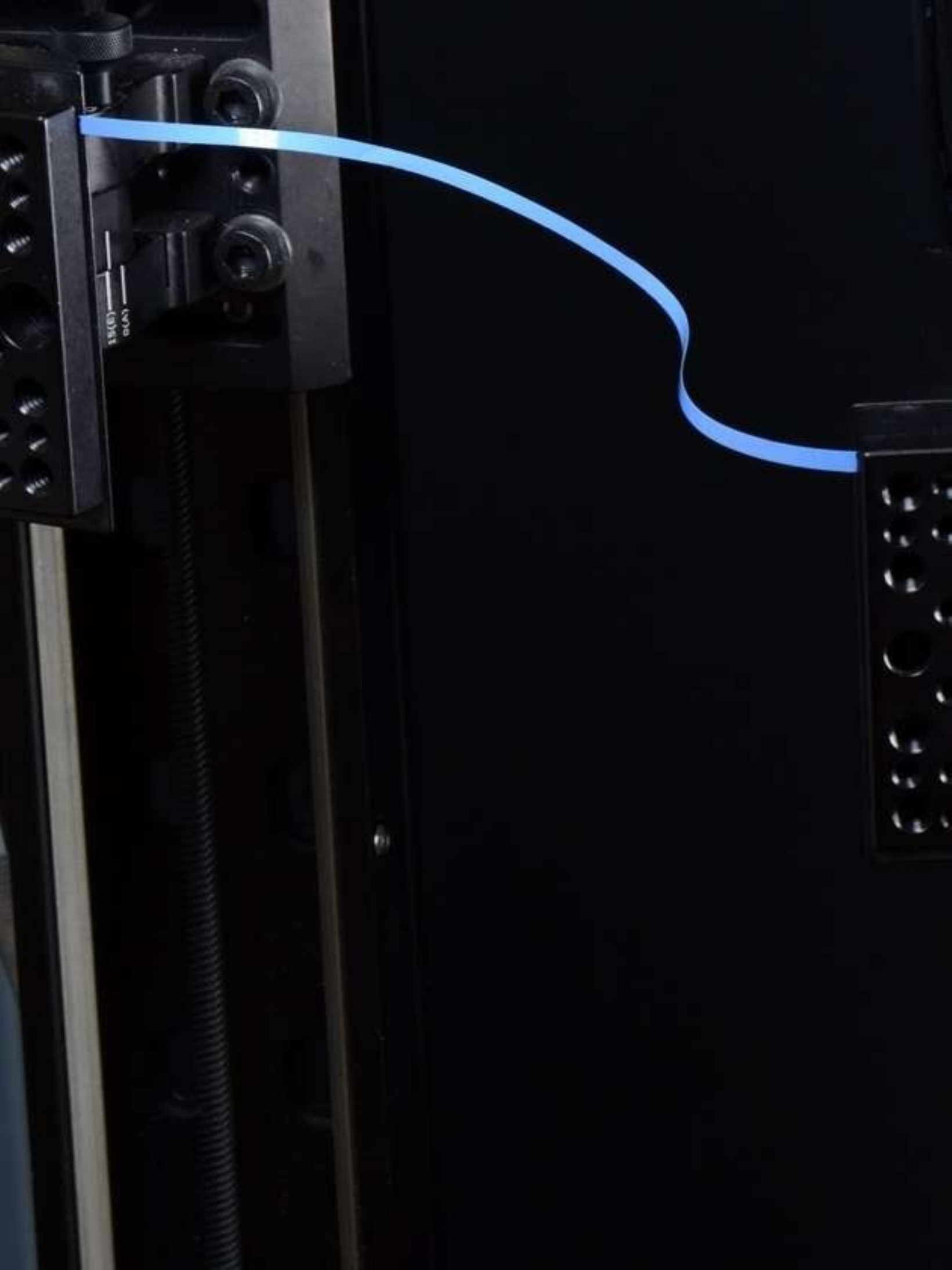}
		\caption{$W$}
	\end{subfigure}%
	\hspace{0.6pt}
	\begin{subfigure}[t]{0.13\textwidth}
		\centering
		\includegraphics[width=0.8in]{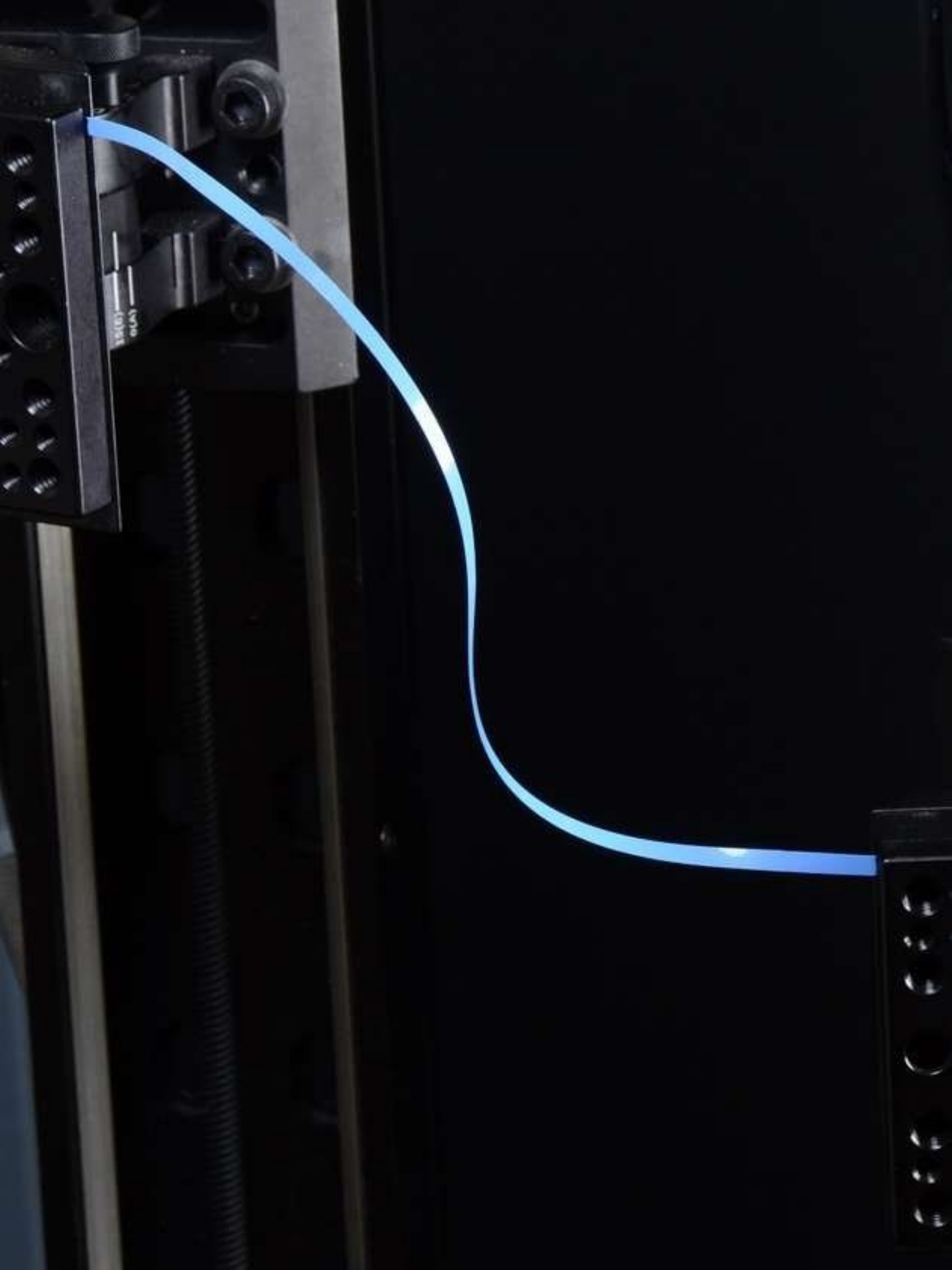}
		\caption{$US-$}
	\end{subfigure} 
	\begin{subfigure}[t]{0.13\textwidth}
		\centering
		\includegraphics[width=0.8in]{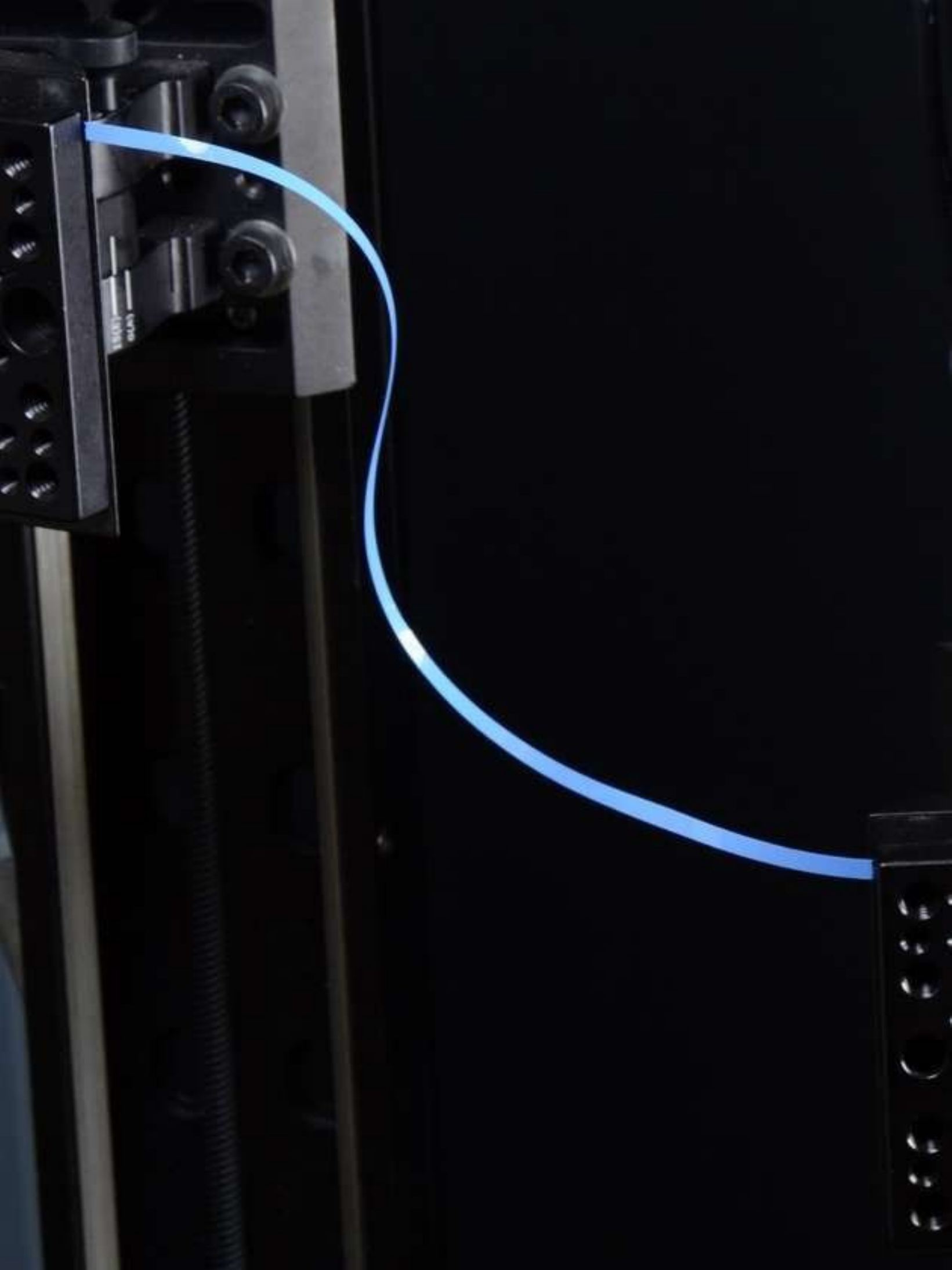}
		\caption{$WS-$}
	\end{subfigure}
	\begin{subfigure}[t]{0.13\textwidth}
		\centering
		\includegraphics[width=0.8in]{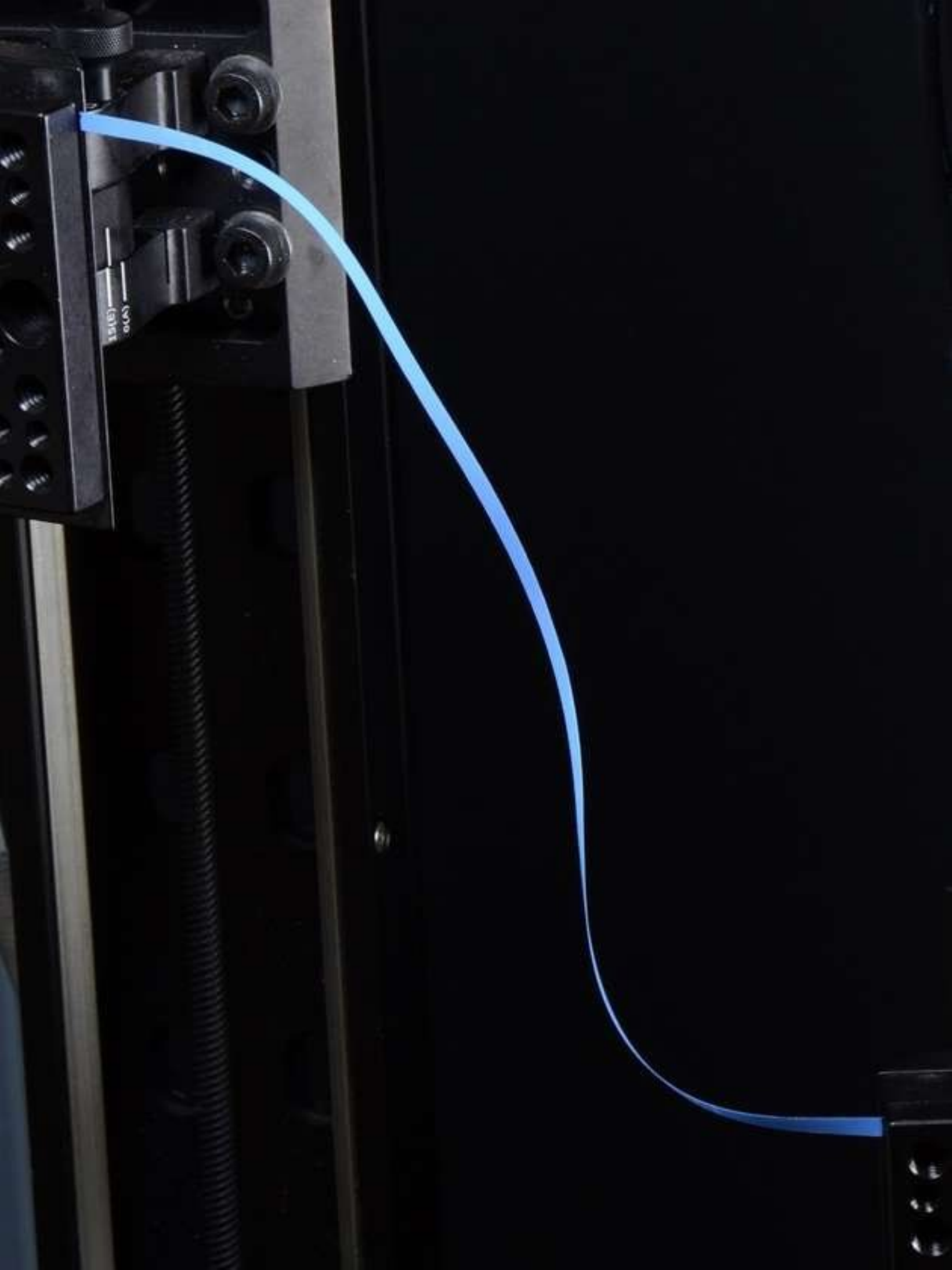}
		\caption{$uUui$}
	\end{subfigure}
	\hspace{0.13\textwidth}
	\begin{subfigure}[t]{0.13\textwidth}
		\centering
		\includegraphics[width=0.8in]{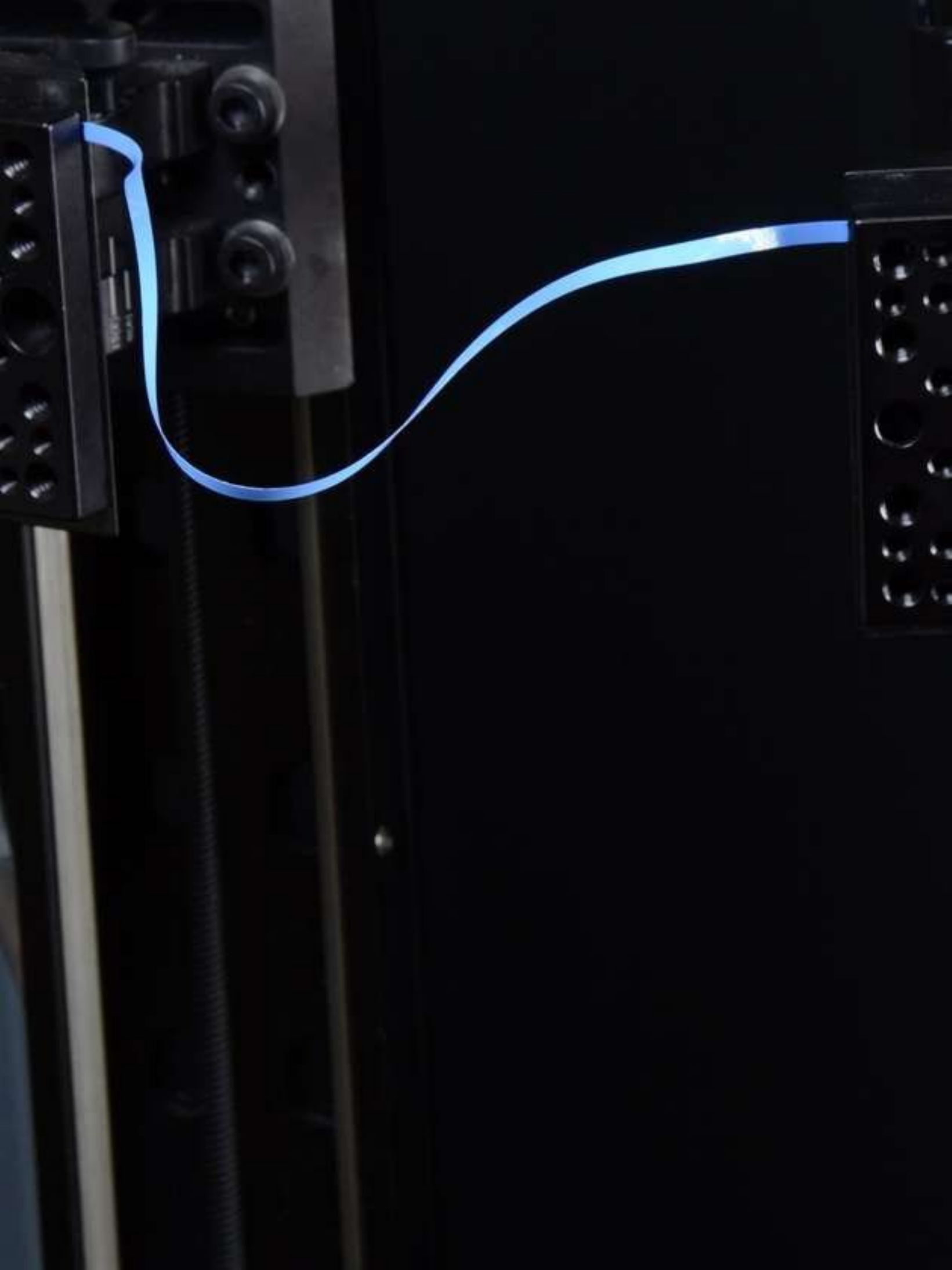}
		\caption{$TU-$}
	\end{subfigure}
	\begin{subfigure}[t]{0.13\textwidth}
		\centering
		\includegraphics[width=0.8in]{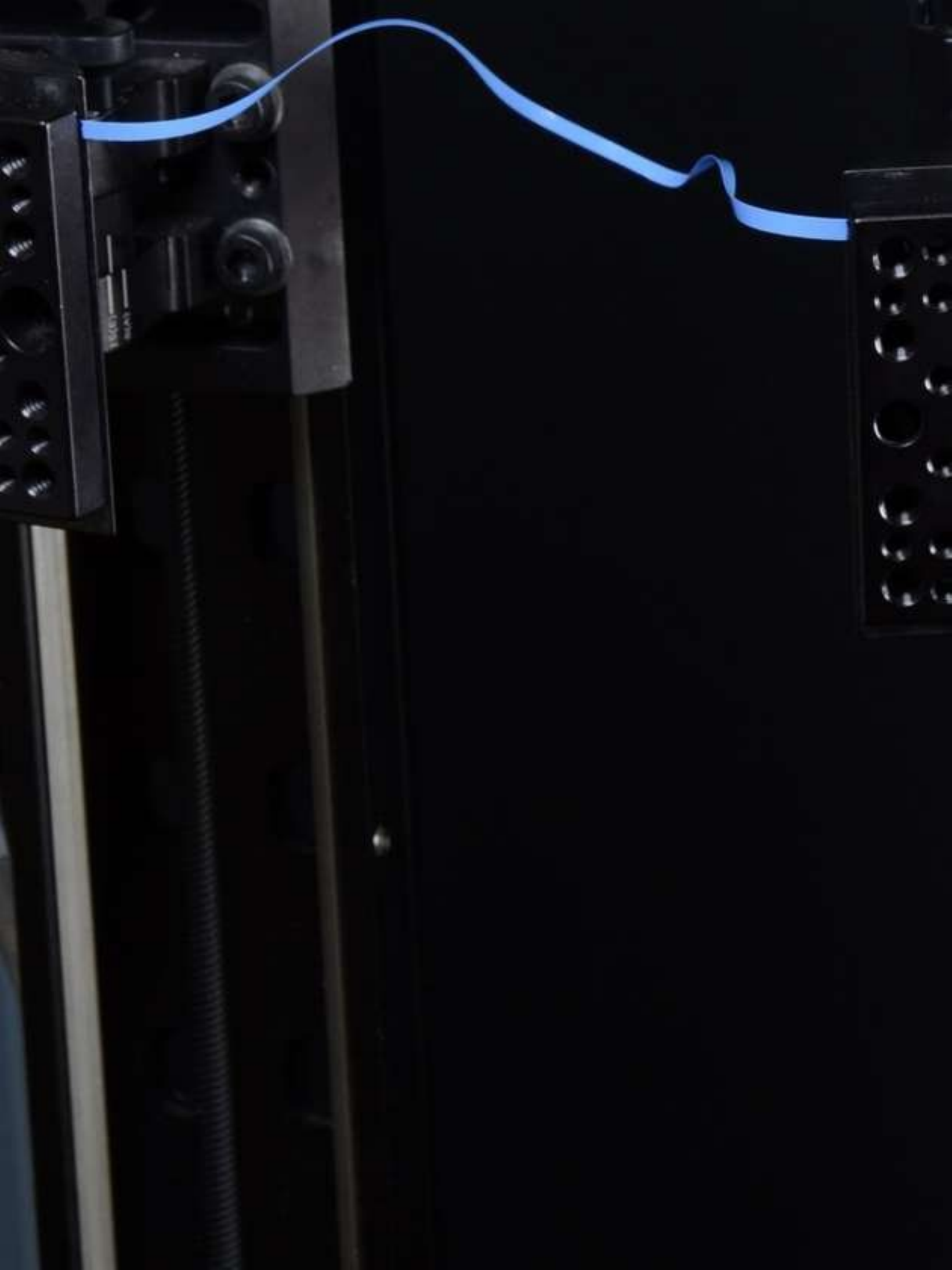}
		\caption{$TW-$}
	\end{subfigure}              \\
	\vspace{-3pt}
	\begin{subfigure}[t]{0.13\textwidth}
		\centering
		\includegraphics[width=0.8in]{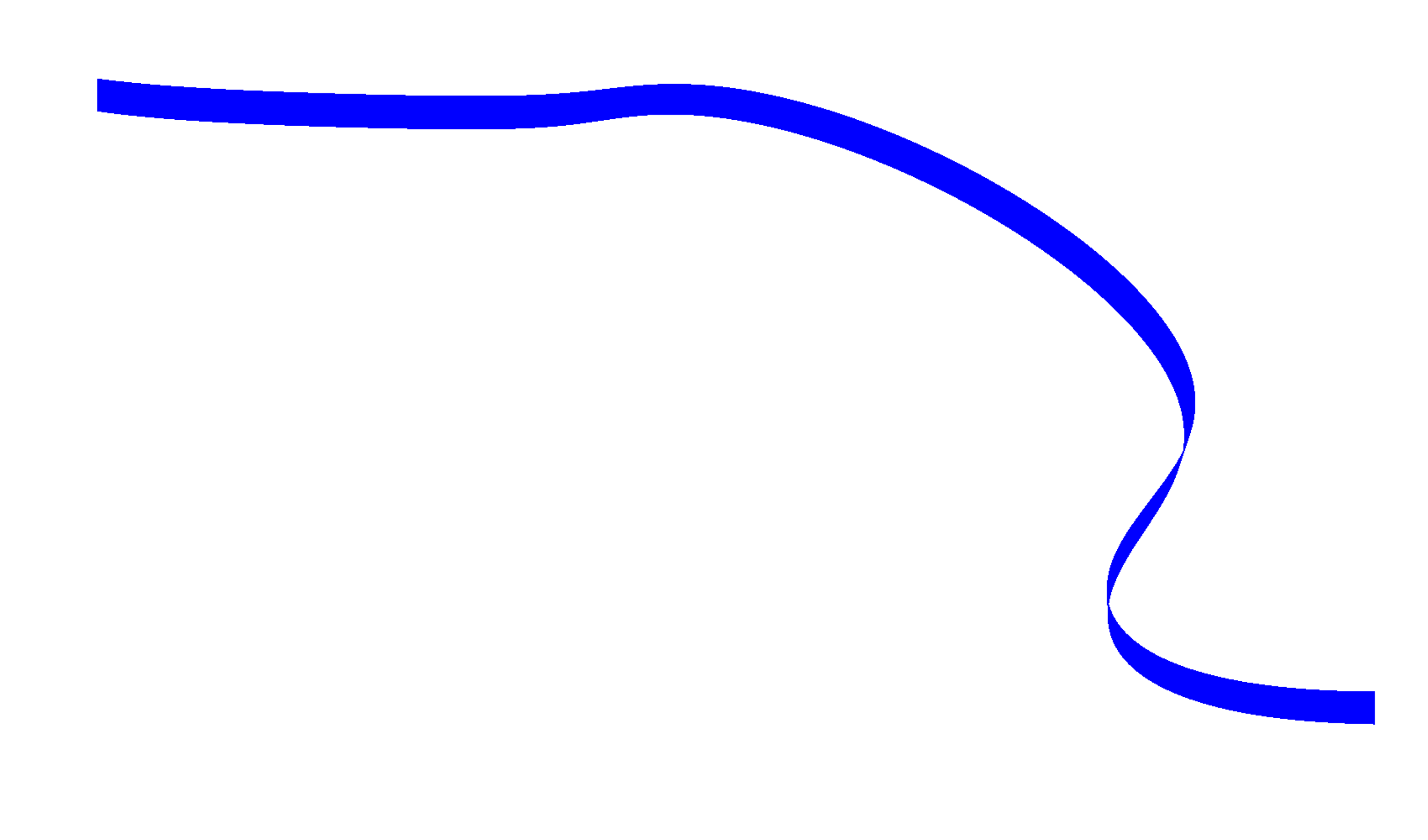}
	\end{subfigure}%
	\hspace{0.6pt}
	\begin{subfigure}[t]{0.13\textwidth}
		\centering
		\includegraphics[width=0.8in]{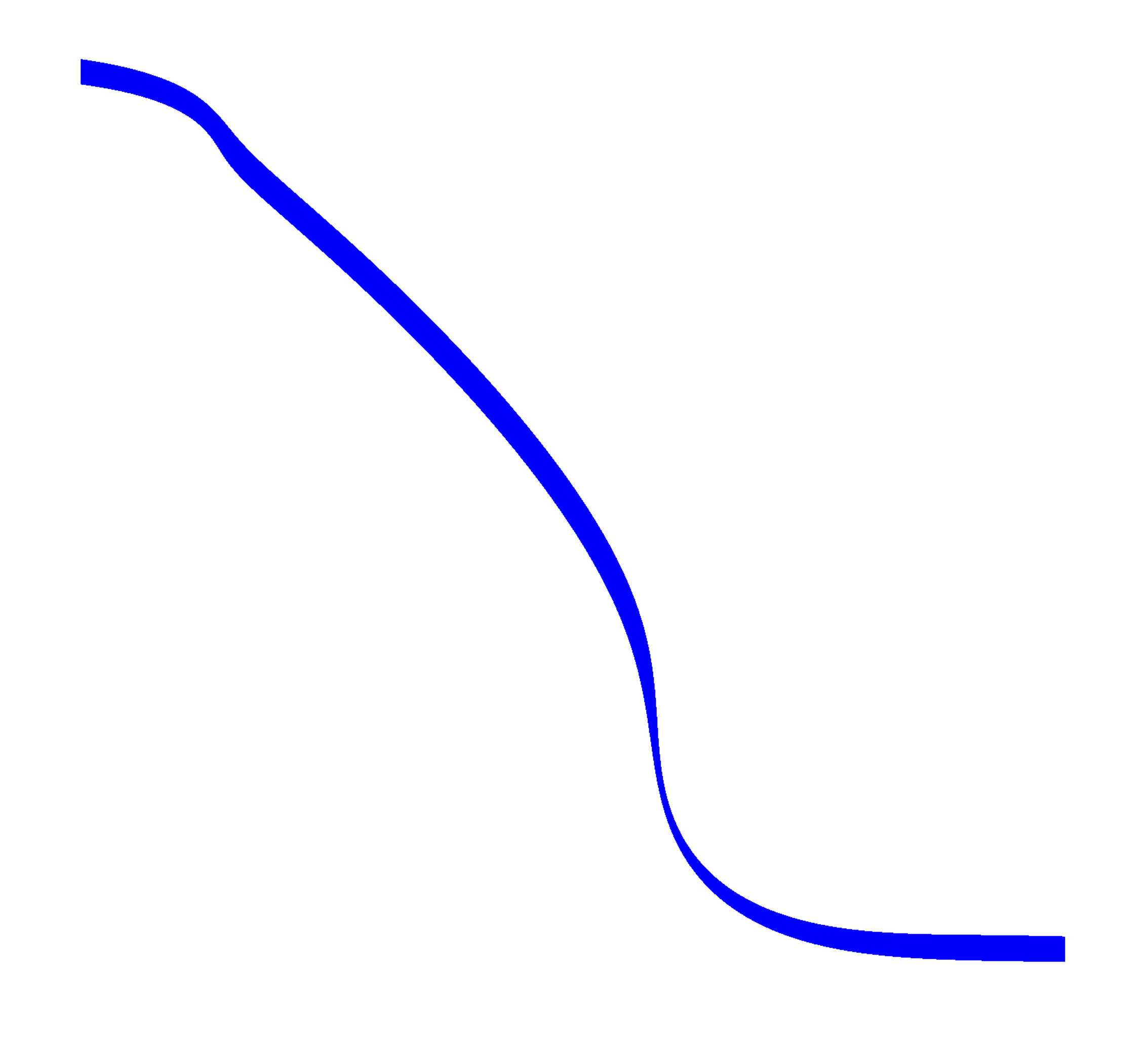}
	\end{subfigure} 
	\begin{subfigure}[t]{0.13\textwidth}
		\centering
		\includegraphics[width=0.8in]{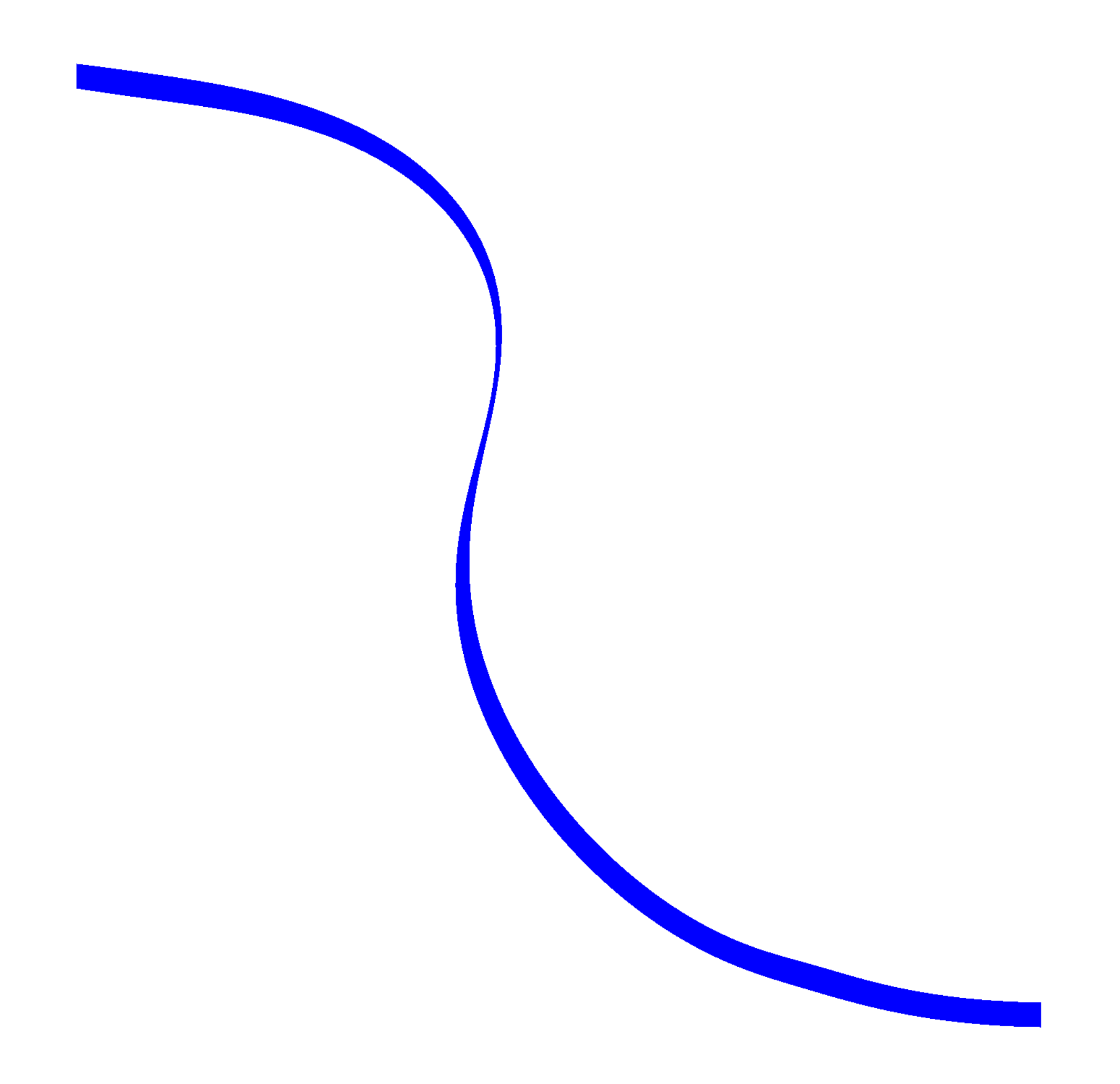}
	\end{subfigure}
	\begin{subfigure}[t]{0.13\textwidth}
		\centering
		\includegraphics[width=0.8in]{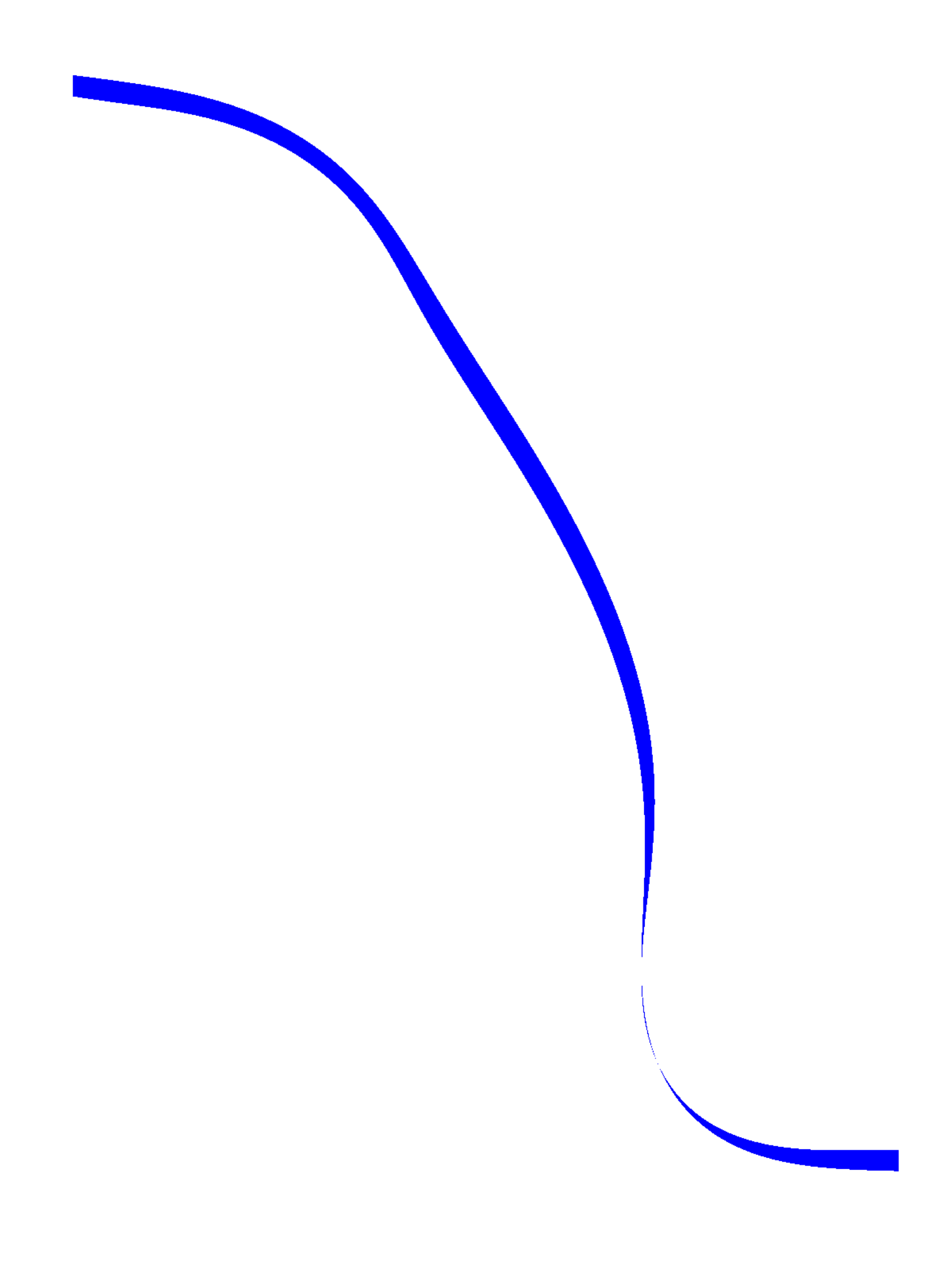}
	\end{subfigure}
	\hspace{0.13\textwidth}
	\begin{subfigure}[t]{0.13\textwidth}
		\centering
		\includegraphics[width=0.8in]{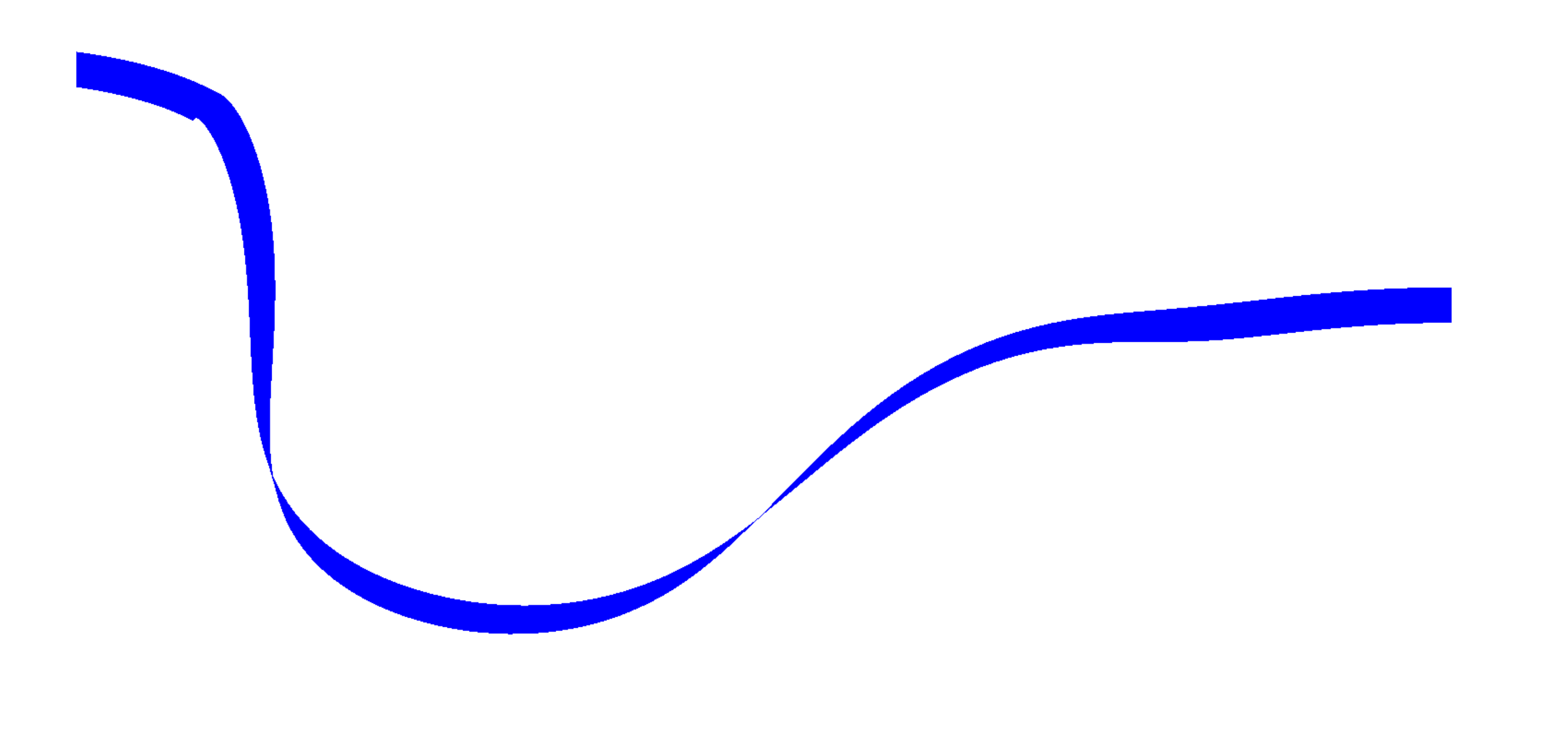}
	\end{subfigure}
	\begin{subfigure}[t]{0.13\textwidth}
		\centering
		\includegraphics[width=0.8in]{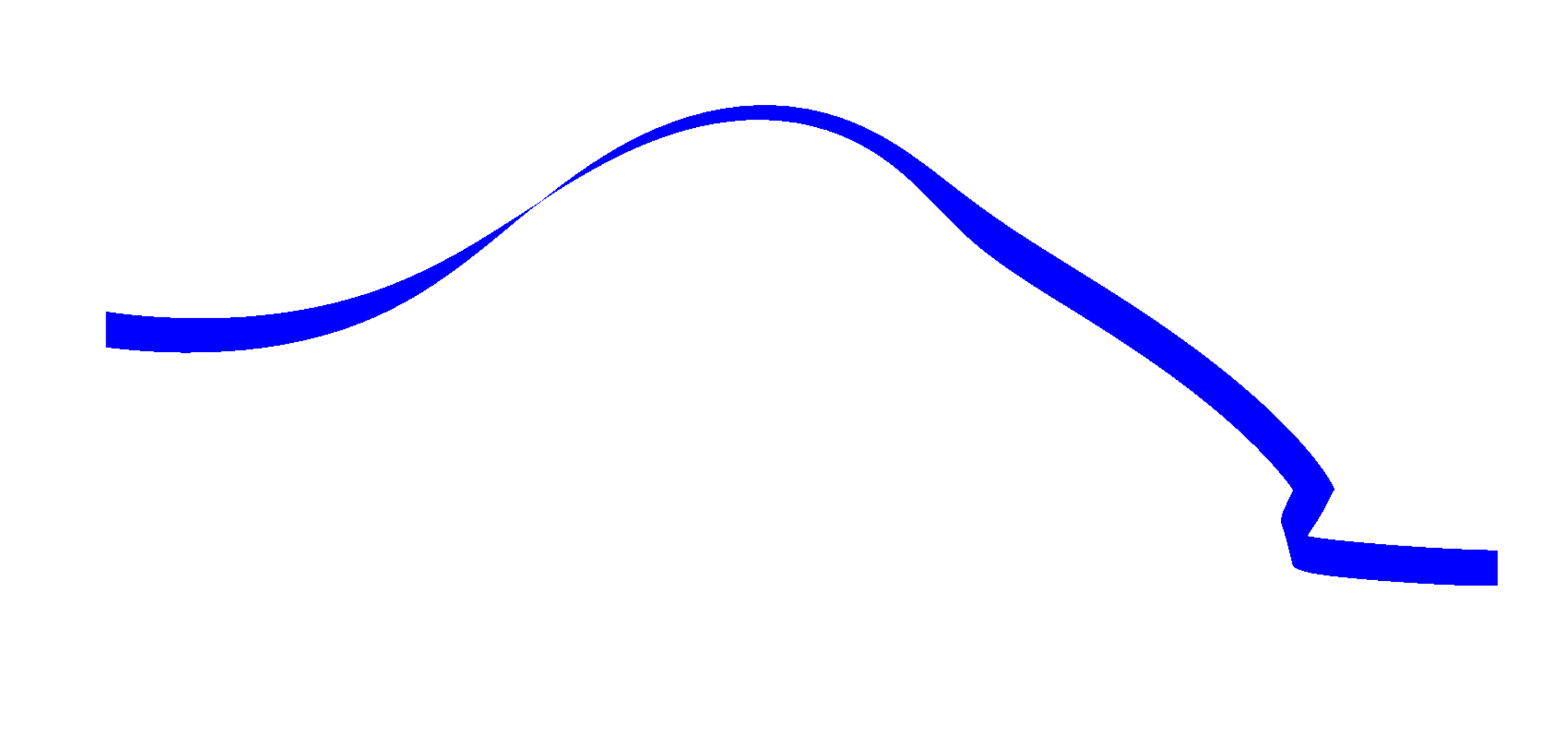}
	\end{subfigure}
	\caption{Comparison between experimental configurations of a narrow band ($D/L=1/80$) and renderings of the rod frame based on numerical solutions of the perfectly anisotropic rod equations, with compression $\Delta L/L=1/2$, clamping angle $\psi_0 = 15^{\circ}$, and various values of shear $\Delta D / L$.
	Note that the bands deform into a surface different than the rod frame rendering.
	  There are no fitting parameters; boundary conditions and viewing angle are the same between experiments and numerics.  Gravity is roughly vertical in the experimental images, and is absent in the numerical solutions; its effects are greatest on the twisted solutions $TU\pm$ and $TW\pm$. Thirteen states are shown, including four symmetric $\pm$ pairs.
}
	\label{fig:ExpConfigurationwithNu}
\end{figure}

Figure \ref{fig:ExpConfigurationwithNu} shows all the types of states we observe in narrow bands, alongside renderings based on numerical solutions of the perfectly anisotropic rod equations, for a shallow clamping angle $\psi_0 = 15^{\circ}$ and various values of shear $\Delta D / L$.  We name the states in a manner that roughly describes their shapes.  There are no fitting parameters; boundary conditions and viewing angle are the same between experiments and numerics.  
The numerical solutions are rendered as strips representing the rod frame (Appendix \ref{bvp}), with the same width as the actual bands, but we note that \emph{the actual bands will deform into a surface that is different than the surface representing the rod frame}, so comparisons must be made carefully.  Throughout this paper, we are able to identify experimental and numerical states using a combination of factors including symmetry of the shapes and the number of inflections in centerline curvature and twist, rather than from details of the shape adopted by material off of the centerline.
For later comparison with Figures \ref{fig:30states} and \ref{fig:jumpenergy} one needs to know that $s/L \in [0,1]$ increases from right to left in the renderings.

 Gravity is roughly vertical in the experimental images, and is absent in the numerical solutions.  Its effects are generally weak, although stronger on some solutions such as the overhanging twisted solutions $TU\pm$ and $TW\pm$.  Overall, the Kirchhoff equations reproduce the shapes of stable states quite well.  Thirteen states are shown, but this includes four symmetric $\pm$ pairs, so only nine distinct states exist.
We may classify them into three families.  First we define a coordinate $y$ perpendicular to the clamping plane, sharing the same sign as $\psi_0$.  The $U$ family ($U$, $US\pm$, $uUu$ and $w$) and $W$ family ($W$, $WS\pm$ and $uUui$) tend to sit on the side of positive and negative $y$, respectively, and are  mirror images when $\psi_0 = 0^{\circ}$.  A family of twisted states ($TU \pm$ and $TW\pm$, mirror images when $\psi_0 = 0^{\circ}$) exists at low values of shear.  These states, which clearly display the non-conservation of twist in anisotropic rods, can be achieved by applying a twist near the center of a $U$ or $W$ state.   Alternatively, we may separate the states into reversibly symmetric $\pm$ pairs and reversibly symmetric single solutions.

We now present a sequence of slices through parameter space for increasing values of clamping angle $0 \le \psi_0 \le 60^{\circ}$, showing the evolving solution manifolds and corresponding rod shapes for the Kirchhoff equations alongside experimentally determined ranges of stability of narrow band states.  It is not difficult to link observed states with numerical solutions through qualitative comparison of the shapes and inferences about stability information from the types of bifurcations encountered. 
 Changes in connectivity of the solution manifolds are also reflected in experimental data and verified in supplementary videos \cite{videos}. 
The slices will display the connectivity of the solution curves before and after certain transitions.  These transitions are pinpointed more accurately using two-parameter continuation of bifurcation points in clamping angle-shear space, as will be shown in more detail in Figure \ref{fig:phasediagram} in Section \ref{loci}.
The shear $\Delta D / L$ is the bifurcation parameter.  The system is mirror-symmetric around zero shear, but we plot a small portion of the numerical negative shear results to show the loop structure of various states near the origin.  For the vertical (response) axis, we choose the integrated height above the plane of clamping $\int_{0}^{1}\! y \, ds$, a quantity that converges to zero for all states as the limiting shear deformation is approached, and which is identical for each $\pm$ pair.  A strip of finite width has a shear limit, discussed in Appendix \ref{limitstates}, beyond which stretching must occur.  
For a compression $\Delta L /L = 1/2$, this limit is $\Delta D/L \approx 0.854$ for our narrow bands, and $\Delta D/L = \sqrt{3}/2 \approx 0.866$ for an ideal rod with zero width.
 We perform narrow-band experiments and continue solutions only up to $0.82$, which avoids damaging the band as well as numerical stiffness issues.
Solution manifolds are obtained by continuation of angle and/or displacement boundary conditions from known solutions, typically from a circle deformed through the first buckled mode of planar \emph{elastica}. 
Some branches, such as twisted state branches, are isolated on the cross sections we present but can be reached by continuation in the full parameter space.
Curves we wish to emphasize are plotted in black, while other closely related or connected curves are shown in grey; often, particularly at higher angles, black and grey will be used for different portions of a single continuous curve.     We show more of these grey curves at lower angles on some of the plots, and replace them with dashed lines at higher angles, and often remove them entirely to overlay additional numerical or experimental results.  Stability information is not shown anywhere on these plots, although it can often be inferred.  An infinite number of other states exist, and are of course not shown. Branch points, and occasionally some fold bifurcations, are marked with symbols; unmarked intersections of the manifolds do not correspond to any bifurcation.  Because of the $\pm$ symmetry of the integrated height response parameter $\int_{0}^{1}\! y \, ds$, most pitchfork bifurcations look like half of a pitchfork, as two lines overlap.  
  Numbers on the figures identify particular bifurcations whose loci will be shown later in Figure \ref{fig:phasediagram} in Section \ref{loci}.

\begin{figure}[h!]
	\centering
	\includegraphics[width=0.85\textwidth]{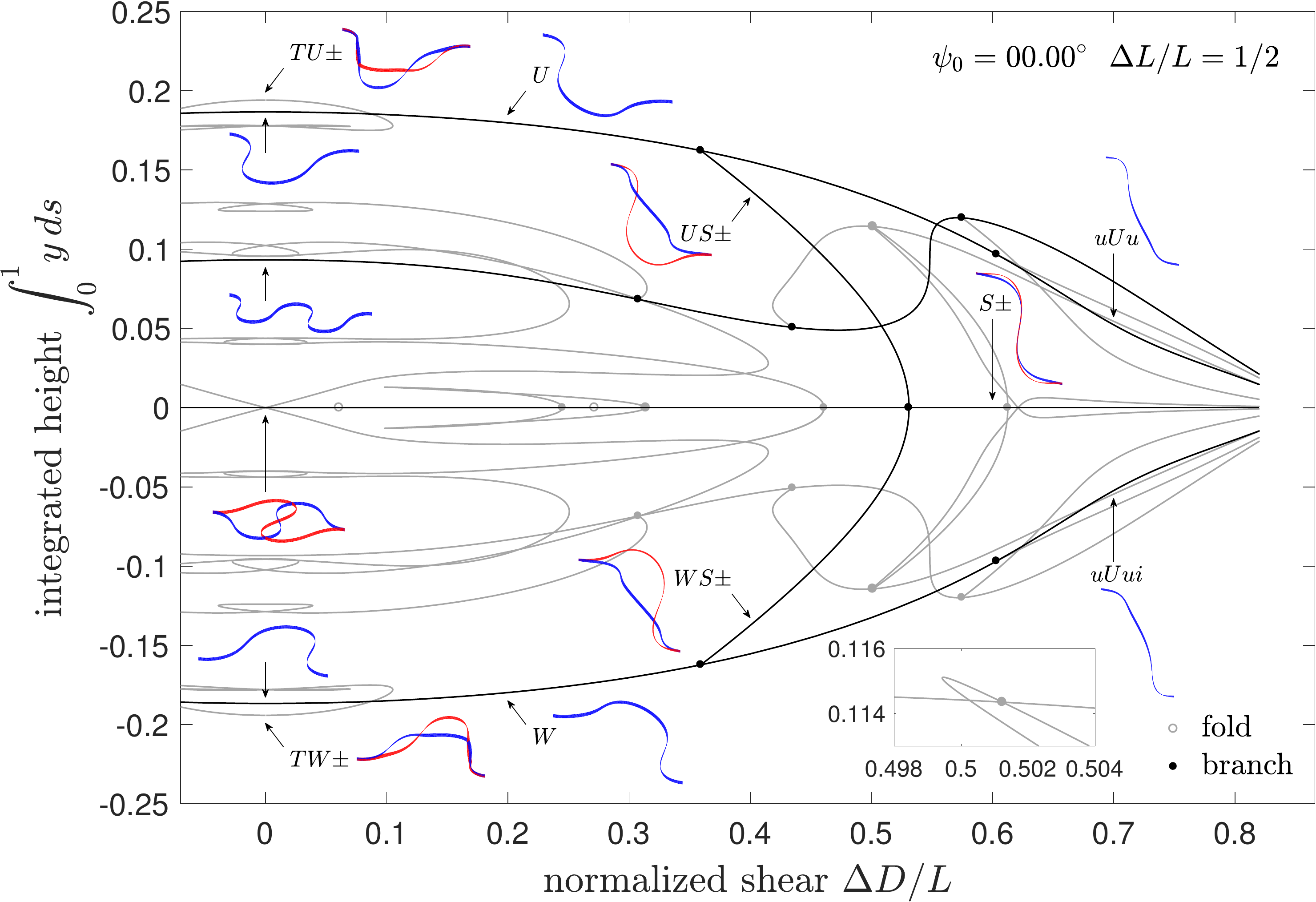}
	\caption{Some solutions (curves and renderings from the arrowed locations) and bifurcations (disks) of the perfectly anisotropic Kirchhoff rod for clamping angle $\psi_0=0^{\circ}$.  Stability information is not shown; black and grey are used for emphasis only. The solution curves are symmetric about the zero-height axis and the zero-shear axis.  Red and blue shapes are $\pm$ pairs that share a single curve on the diagram.}
	\label{fig:Config0}
\end{figure}

Figure \ref{fig:Config0} shows some of the solution manifolds for the symmetric case of zero clamping angle $\psi_0=0^{\circ}$.  Also shown are rod frame renderings of numerical solutions at several points along the curves, many of which we identify with the configurations named in Figure \ref{fig:ExpConfigurationwithNu}.  The $\pm$ pairs are drawn as red and blue.  Several turning points (fold bifurcations) and branch points are observed, some of which are overlapping pairs.  
The states shown include the first few stable and unstable planar modes of Euler \emph{elastica}, which are unlabeled.
The loop-like curves near zero shear are highly twisted states, many of which are unstable, some of which connect with the \emph{elastica} modes.  All even-numbered modes of planar elastica and states continued from these will sit on top of one another on the horizontal axis of symmetry (zero integrated height).  The connectivity along this axis is very complicated, including many (possibly an infinite number) of branch and fold bifurcations, as will be revealed when we proceed to a nonzero value of $\psi_0$.  We show only a few branch points here, and our choice of response parameter hides the presence of folds when $\psi_0=0$.
This raises interesting questions. Can we assume that the entire infinite family of planar buckled modes connect through bifurcations to one or more twisted states?  And how are the pitchforks distributed along the axis?

We are able to identify these numerical states with the stable states observed experimentally, and infer information about stability and bifurcation types.  We now recognize that the $U$ and $uUu$ states lie on a single branch connected to the first-mode planar \emph{elastica}, but are separated by two bifurcations and an unstable stretch.  We will refer to this entire branch as the $U$ branch, except when it may cause confusion.  There is a supercritical pitchfork at  $\Delta D/L \approx 0.36$ that connects the $U$ and $US\pm$ states and causes loss of stability of the $U$ branch; stability is regained through a subcritical pitchfork at $\Delta D/L\approx0.60$, with the second set of stable configurations referred to as $uUu$.
There are two supercritical pitchforks on the zero-height axis of symmetry at $\Delta D/L\approx0.53$, one linking the $US+$ and $WS+$ states to a stable $S+$ state, the other linking the $US-$ and $WS-$ states to a stable $S-$ state.  The $S\pm$ states only exist at zero clamping angle, because the pitchforks on the horizontal axis will be broken at any non-vanishing angle.  The unstable states on the low-shear side of these bifurcations connect back to the two unstable second-mode \emph{elastica} shapes shown at zero shear.
  Note that at zero clamping angle, the $US$ and $WS$ states are equivalent.  Upon symmetric change in the clamping angle, they will be distinct, and the connectivity described here will change.  In this study, we don't consider asymmetric changes in clamping angle, which observation suggests will stabilize either the second-mode \emph{elastica} or a pair of $S$-like planar shapes, depending on the value of the compression.
By shearing the unstable third-mode \emph{elastica} and following the branch to high shear, we encounter three branch points, none of which appear to create any stable states.  Between the second and third of these, there is a steep, but not yet folded back, section of the curve that will, upon a small change of clamping angle, become a stable section in between two folds.  We plot the companion curve below the horizontal axis in grey, as it will never acquire a stable segment.
Of the many twisted states at low shear, only two pairs of twisted states $TU\pm$ and $TW\pm$ (equivalent at zero clamping angle) are observed experimentally.  The two loops upon which they lie are complicated pretzel-like curves, each of which provides four (pairs of) states at zero shear, of which only one is experimentally observed in narrow bands.

   \begin{figure}[h!]
	\centering
	\includegraphics[width=0.85\textwidth]{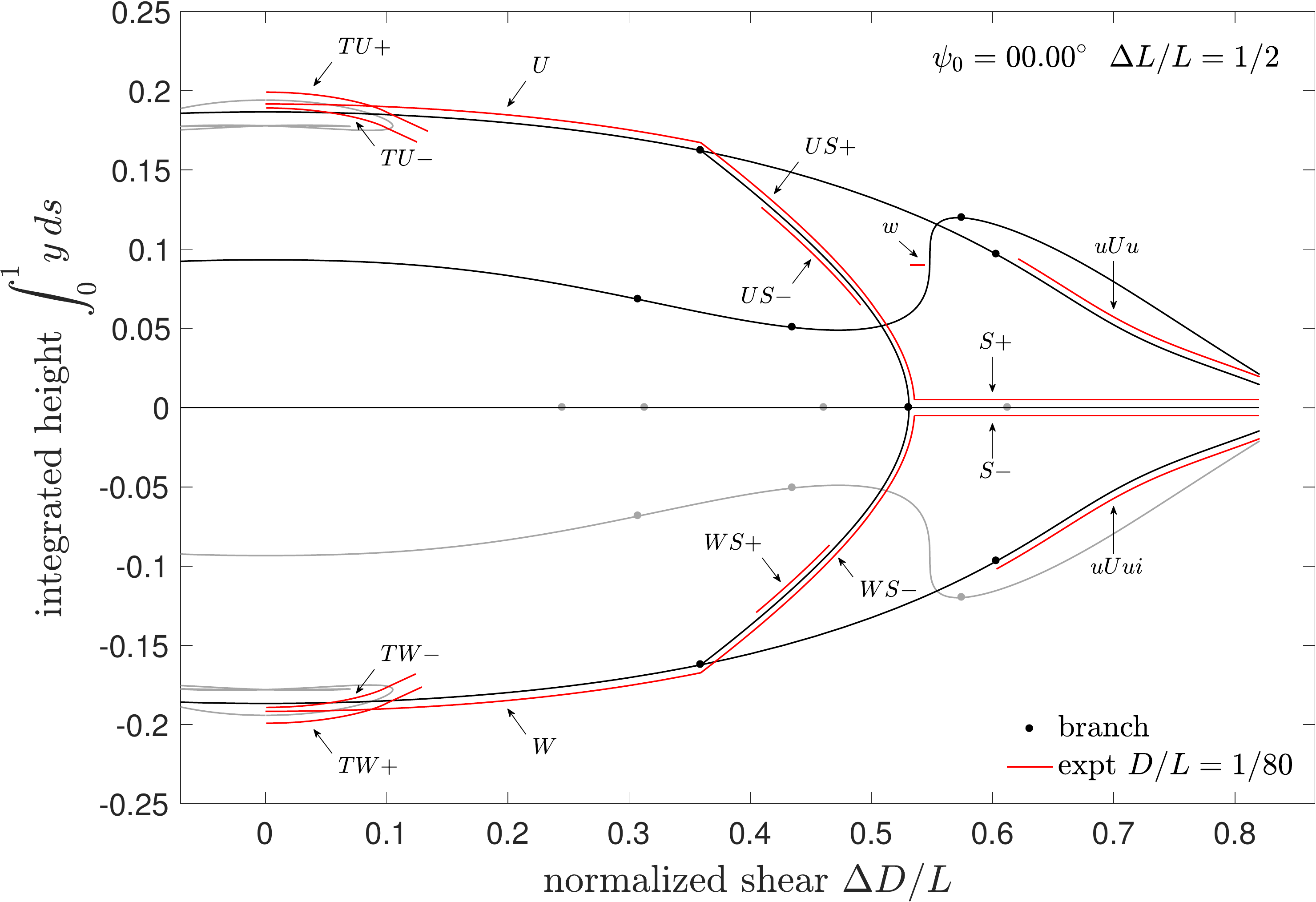}
	\caption{Experimental data (red curves) from narrow bands for clamping angle $\psi_0=0^{\circ}$, compared with numerical solutions of the anisotropic rod equations (black and grey curves).  Some solution curves branching from bifurcation points have been removed from the diagram for clarity. 
	The horizontal extent of the red curves is the range of stability (typical variation $\approx \pm 0.01\, \Delta D/L$), while the vertical position of the curves is not measured data, but is made to follow near the numerical curves whenever a comparison is possible.  No data was taken for $\Delta D/L < 0$.
Gravity causes asymmetry between $\pm$ data.  A stable $w$ state does not theoretically appear until $\psi_0 \ge\, \approx 0.135^{\circ}$, but is observed in experiments, likely due to error in clamping or alignment.
There is a smooth path from the first mode of planar \emph{elastica}, through the $U$, $US+$, and $S+$ states or the $W$, $WS-$, and $S-$ states, to approach the limiting shear.}
	\label{fig:Experi0}
\end{figure}

 In Figure \ref{fig:Experi0}, we compare the solutions of the anisotropic rod equations with experimental narrow band stability data for $\psi_0=0^{\circ}$, shown using red curves.  No experimental data was obtained for negative shear, so the data is truncated at the $\Delta D/L = 0$ axis.   Many solution curves have been removed from the figure for clarity.  Only the horizontal extent of the red experimental curves has any meaning.  The vertical position of these curves follows the corresponding solutions for ease of comparison, with $\pm$ pairs separated by a small gap, or the data is plotted as a horizontal line if no corresponding solution exists.  For example, the very short red line representing the $w$ state is observed experimentally, although in theory it should not appear until the clamping angle is slightly increased to $\psi_0 \approx 0.135^{\circ}$.  This discrepancy is likely due to some error in clamping or in vertical alignment in the presence of gravity; the system can be quite sensitive to boundary conditions close to a bifurcation. Similar comments can be made about any other asymmetries about the horizontal axis at $\psi_0 = 0^{\circ}$.  In this and subsequent figures, gravity is responsible for observed asymmetries between $\pm$ pairs, breaking pitchfork bifurcations like that between the $U$ and $US\pm$ states, such that $U$ always connects with $US+$.  The $US-$ and $WS+$ states are thus isolated states in between two fold bifurcations, and observed only by manual manipulation of the band, only because of the presence of gravity in a particular orientation.
This qualitative behavior was confirmed by augmenting the rod equations with a gravity term (see equations \ref{Gravity} in Appendix \ref{bvp}).
We observe that a first-mode \emph{elastica} will smoothly deform through the $U$, $US+$, and $S+$ states or through the $W$, $WS-$, and $S-$ states, to approach the limiting shear, with the $\pm$ choices being results of gravitational bias in this orientation.  This process, and many other bifurcations corresponding to Figure \ref{fig:Experi0}, are illustrated in the supplementary video {\texttt{transition0.mp4}} \cite{videos}.
 We will see that the numerical $W$-$WS$-$S$ path will be broken by any nonzero clamping angle, while the $U$-$US$-$S$ path will become a $U$-$US$ path at nonzero angle, and will eventually be broken at higher angles, with two merging events leading to a new smooth $U$-$w$-$uUui$ path.   At $\psi_0=0^{\circ}$, the $uUu$ and $uUui$ states also approach the limiting shear, but are not smoothly connected to planar configurations.

\begin{figure}[h!]
        \centering
        \includegraphics[width=0.85\textwidth]{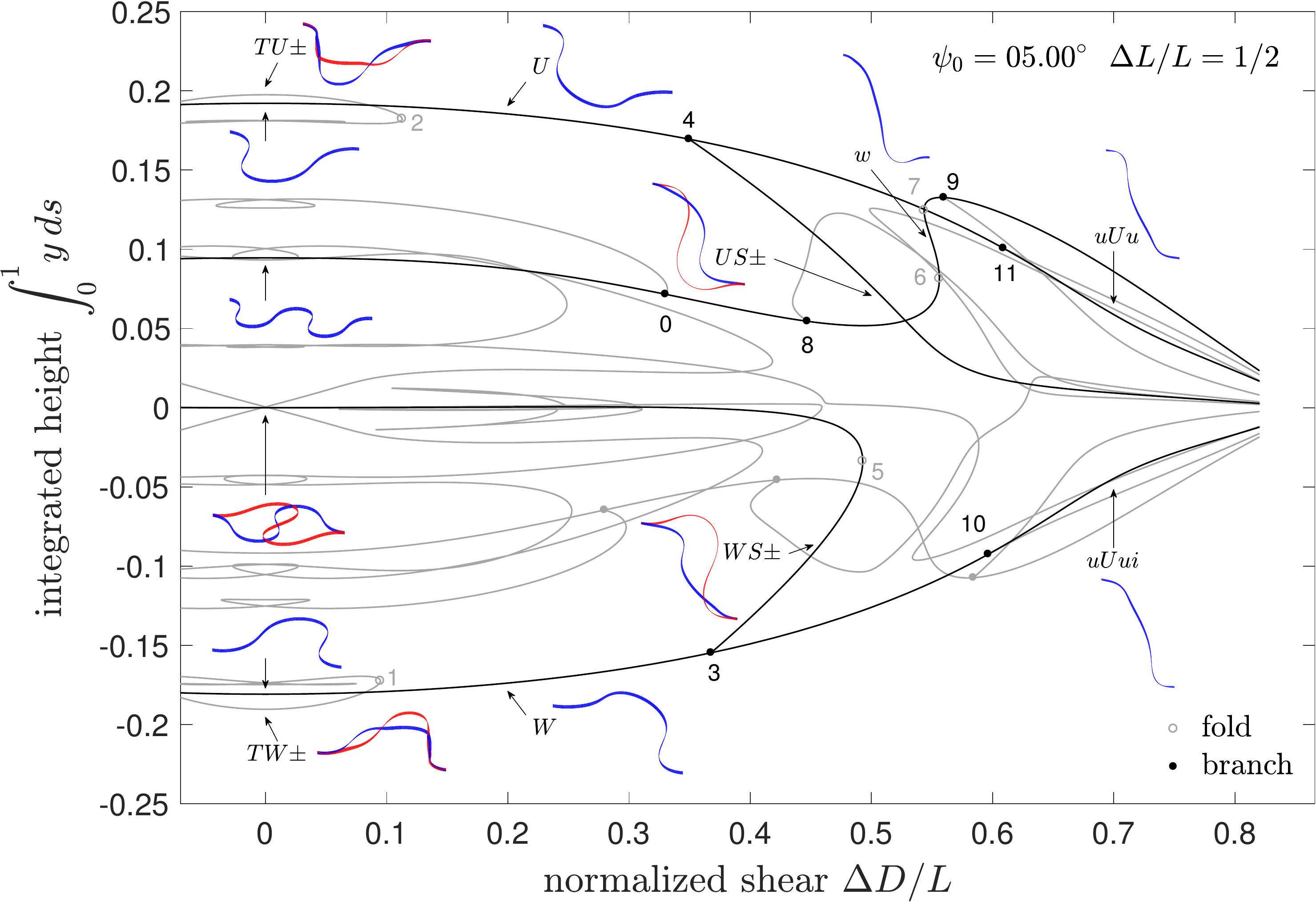}
        	\caption{from narrow bands for
	Some solutions (curves and renderings from the arrowed locations) and bifurcations (open and closed disks, some numbered) of the perfectly anisotropic Kirchhoff rod for clamping angle $\psi_0=5^{\circ}$.  Many bifurcations have been broken, and paths approaching the limit have been affected.  Two folds 6 and 7 and the $w$ state have been created.}
        \label{fig:Config5}
\end{figure}

\begin{figure}[h!]
        \centering
        \includegraphics[width=0.85\textwidth]{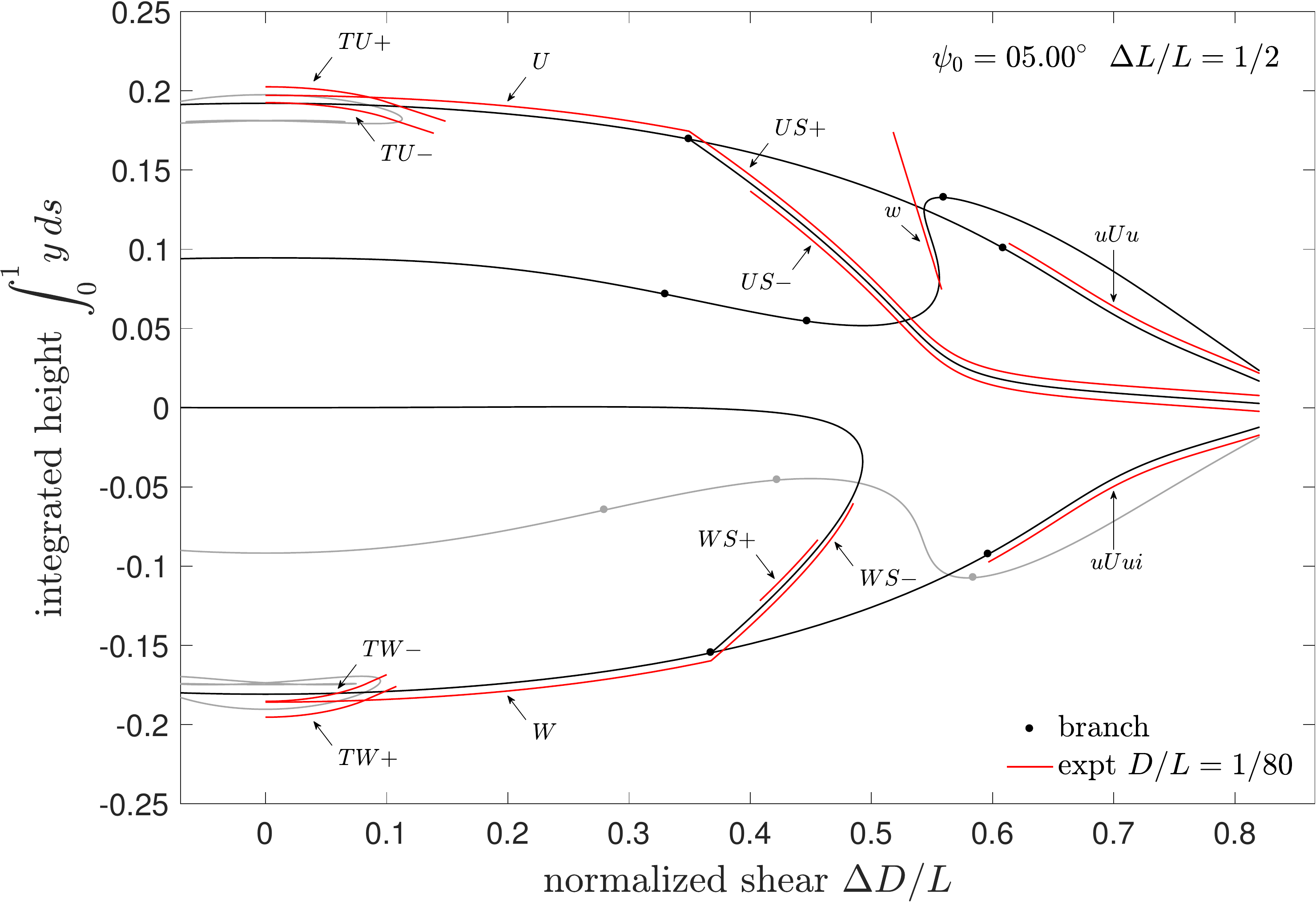}
        \caption{Experimental data (red curves) from narrow bands for clamping angle $\psi_0=5^{\circ}$, compared with numerical solutions of the anisotropic rod equations (black and grey curves).  Many solution curves have been removed from the diagram for clarity. 	The horizontal extent of the red curves is the range of stability (typical variation $\approx \pm 0.01\, \Delta D/L$), while the vertical position of the curves is not measured data, but is made to follow near the numerical curves whenever a comparison is possible. No data was taken for $\Delta D/L < 0$.  	
There is still a smooth path from the first mode of planar \emph{elastica}, through the $U$ and $US+$ states, to approach the limiting shear, but the corresponding path through $WS-$ has been broken; $WS\pm$ states will now jump to $US\pm$ states at intermediate shears.}
        \label{fig:Experi5}
\end{figure}

Figures \ref{fig:Config5} and \ref{fig:Experi5} show some solution manifolds, renderings, and experimental stability data for a small clamping angle, $\psi_0=5^{\circ}$.  In Figure \ref{fig:Config5} and some subsequent figures, some bifurcations have been numbered for convenient description and for further discussion in Section \ref{loci} and Appendix \ref{poisson}.  The nonzero clamping angle has broken the symmetry between the $U$ and $W$ families that live primarily above and below the horizontal axis in the figures.  At low shear, we can think of $U$ as the primary first-mode \emph{elastica} state, and $W$ as the corresponding inverted state (as the clamping angle increases, its  shape will more closely resemble its name, or perhaps an $M$ depending on one's orientation).  All the branch bifurcations on this axis, and some off of the axis, have been broken, creating numerous folds, and revealing the complex asymmetric connectivity of the curves.  
The breaking of the primary black pitchforks on the horizontal axis leads to the overlapping fold bifurcation pair 5 on the $WS\pm$ branch, now separated by a jump from the $US\pm$ branch, which branch has now merged with $S\pm$ and smoothly approaches the limiting shear.
Further increases in angle will shorten the stable range of the $WS\pm$ branch.
 Two folds 6 and 7 and an intermediate $w$ state have been created on an upper black branch.  The creation occurs at $\psi_0 \approx 0.135^{\circ}$ and corresponds to a cusp in $\psi_0$-$\Delta D/L$ space, which will be seen later in Figure \ref{fig:phasediagram}.  Increasing the clamping angle extends the stable range of the $w$ state; the corresponding inverted grey branch becomes shallower in slope, and will never produce a stable state under our choice of clamping path.
All of these features of the rod equation solutions are consistent with the experimental data.

In subsequent figures, we remove many of the complicated grey solution curves, indicating their existence with small stretches of dashed lines.

\begin{figure}[h!]
	\centering
	\includegraphics[width=0.85\textwidth]{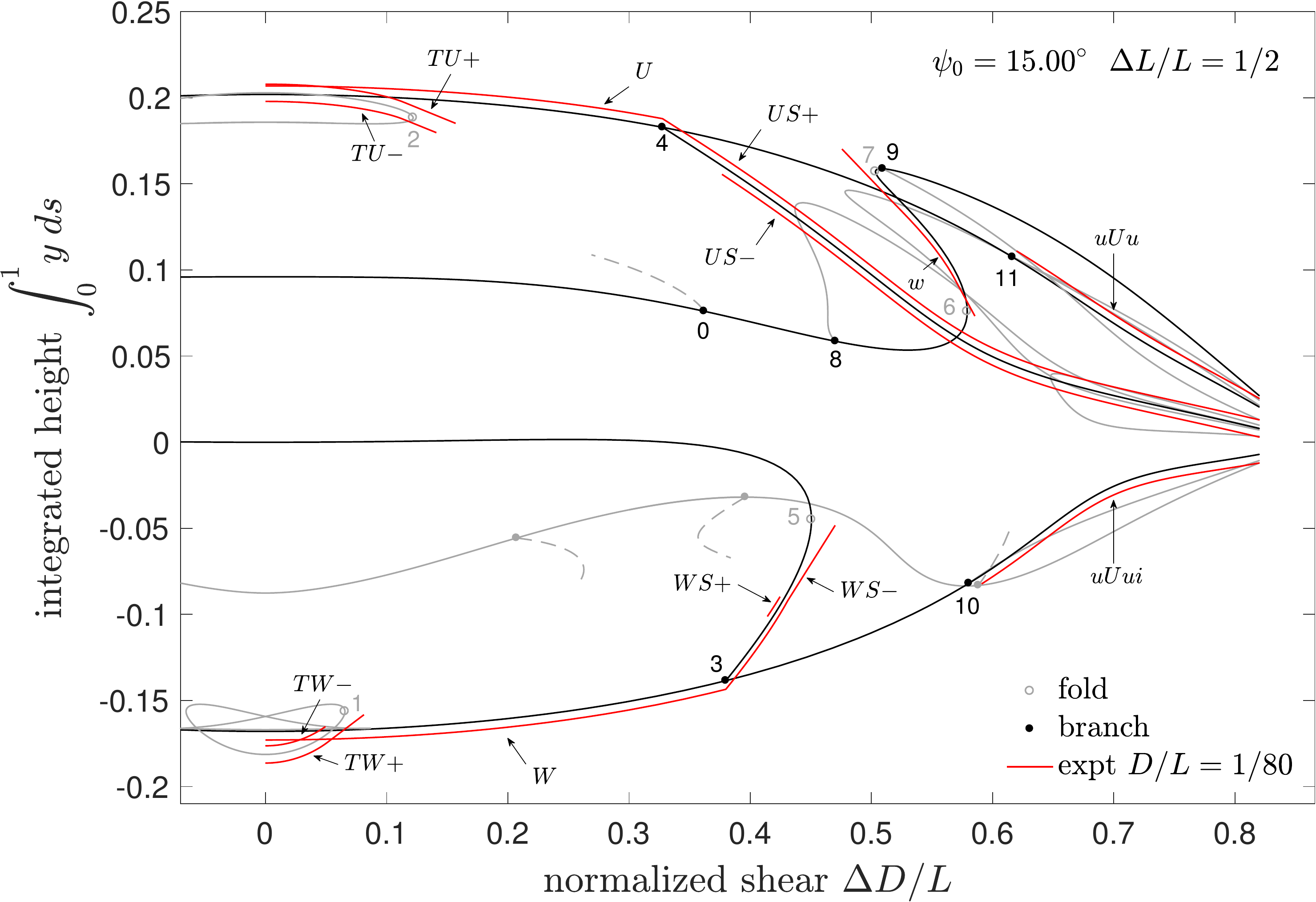}
	\caption{Some solutions (black and grey curves) and bifurcations (open and closed disks, some numbered) of the perfectly anisotropic Kirchhoff rod equations for clamping angle $\psi_0=15^{\circ}$, along with experimental data (red curves).  
		The horizontal extent of the red curves is the range of stability (typical variation $\approx \pm 0.01\, \Delta D/L$), while the vertical position of the curves is not measured data, but is made to follow near the numerical curves whenever a comparison is possible.  No data was taken for $\Delta D/L < 0$.    
		Numerically, the $w$ branch has been elongated, and the $WS\pm$ and $TW\pm$ branches have been shortened.  Experimentally, the $WS+$ state has nearly disappeared, in part due to the action of gravity.}
	\label{fig:Solution15}
\end{figure}

Figure \ref{fig:Solution15} shows some solution manifolds and experimental stability data for a clamping angle $\psi_0=15^{\circ}$.  Paths from zero to limiting shear have not changed.  However, the stable extent of the $w$ state has increased, while that of the $WS\pm$ states has decreased, with $WS+$ nearly disappearing due in part to the gravity-induced asymmetry.
 The primary twisted states $TU\pm$ are relatively unaffected by the clamping angle change, but their inverted partners, the $TW\pm$ states, now exist over a shorter extent.  In experiments, we observe that the $WS\pm$ states jump to the corresponding $US\pm$ states upon increasing shear, while the $WS+$ state jumps to the $WS-$ state, and the $US-$ state jumps to the $US+$ state, upon decreasing shear, due to gravity-induced folds.  The $w$ state jumps to the limiting $uUui$ state upon increasing shear, and to the $US+$ branch upon decreasing shear. 
Many of these transitions are illustrated in the supplementary video {\texttt{transition15.mp4}} \cite{videos}.
We note that much of the complexity of the solution manifolds arises from the anisotropy of the rod; Appendix \ref{anisotropy} shows relatively simple solution manifolds for isotropic rods that may be compared with Figures \ref{fig:Config0} and \ref{fig:Solution15}.

As we increase the clamping angle, there are many complicated changes to the solution structure.  Among these, the isolated $TW\pm$ loop partially merges with some of the complicated grey twisted curves (which we have already removed from the figures for clarity). 
The stable $TW\pm$ states
disappear at around $\psi_0 \approx 26.89^{\circ}$.   We don't show these, and many other, transitions here.

\begin{figure}[h!]
	\centering
	\includegraphics[width=0.85\textwidth]{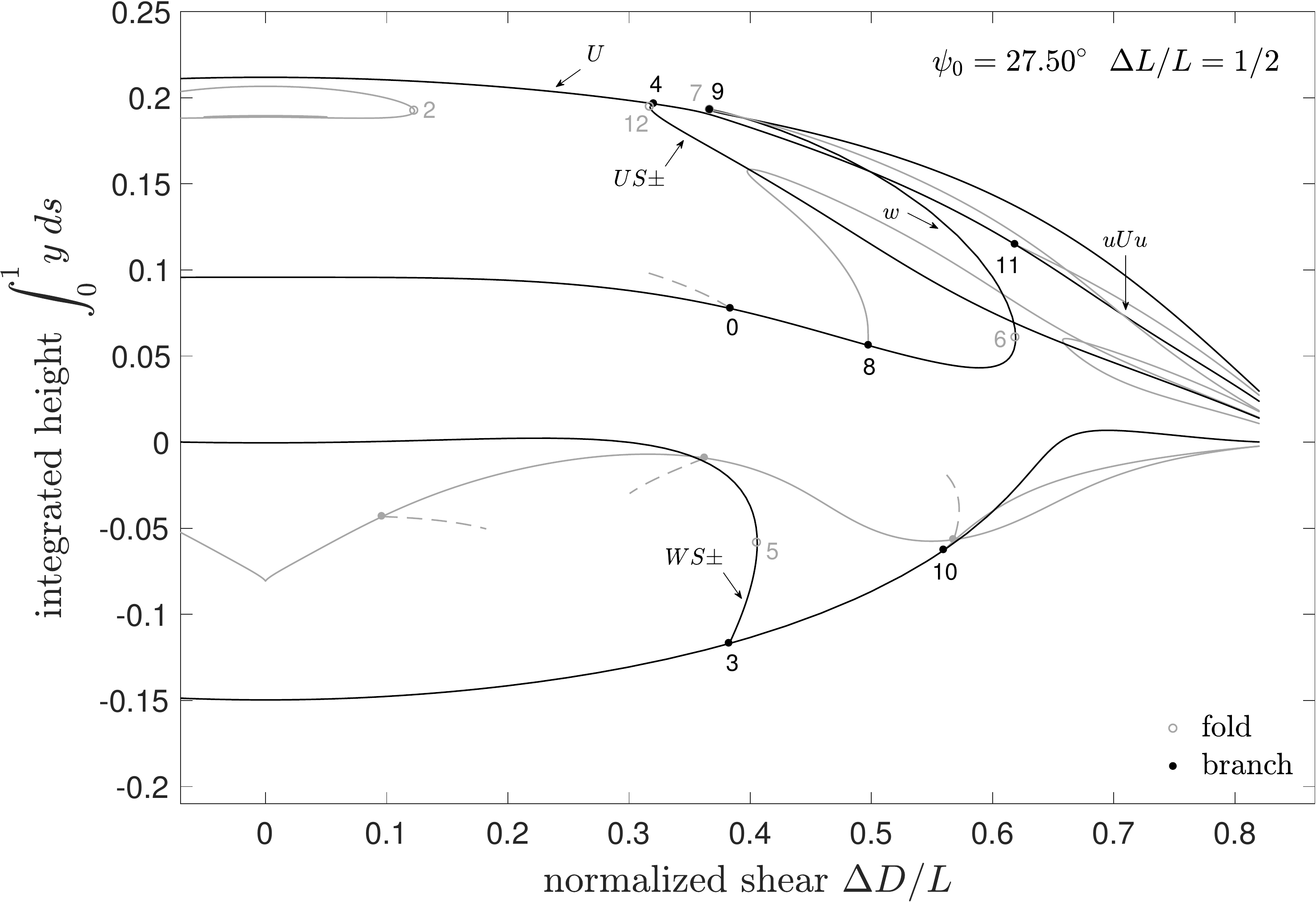}
	\caption{Some solution curves and bifurcations (open and closed disks, some numbered) of the perfectly anisotropic Kirchhoff rod for clamping angle $\psi_0=27.5^{\circ}$.  Pitchfork 4 has transformed from super- to sub-critical, and a new fold 12 has appeared.  The $U$ state will now (weakly) jump to a $US$ branch, which can be followed to the limit.  Fold 7 of the $w$ branch is approaching the $U$ branch.  Details of this region and subsequent transitions are shown in Figure \ref{fig:Transition1}.}
	\label{fig:Solution27p5}
\end{figure}

\begin{figure}[h!]
\centering
\includegraphics[width=0.9\textwidth]{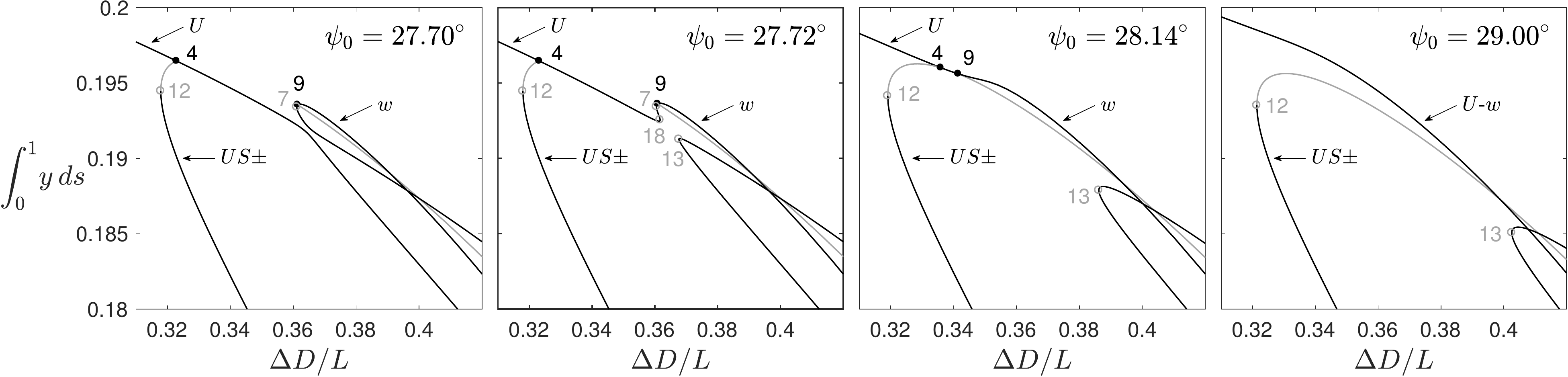}
\caption{Details of several transitions that change the connectivity of solution curves.  Closed and open disks are branch and turning points, respectively.  First, the $w$ branch collides with the $U$ branch, and the high-shear portion that includes the $uUu$ branch detaches from $U$ and attaches to part of $w$.  Then, folds 18 and 7 annihilate.  Then, subcritical pitchforks 4 and 9 annihilate, detaching the $US\pm$ branches from the $U$ branch, while $U$ attaches smoothly to $w$.}
\label{fig:Transition1}
\end{figure}

Some solution manifolds at $\psi_0=27.5^{\circ}$ are shown in Figure \ref{fig:Solution27p5}.  The $WS\pm$ branch has nearly disappeared.  At $\psi_0 \approx 26.29^{\circ}$, pitchfork bifurcation 4 transforms from super- to sub-critical, and a new fold bifurcation 12 appears.  This means that the $U$-$US$ path is no longer smooth.
  Fold bifurcation 7 of the $w$ branch is approaching the $U$ branch, and will merge through a complicated sequence shown in detail in Figure \ref{fig:Transition1}.
 At $\psi_0 \approx 27.71^{\circ}$, the branch containing the $w$ state touches the $U$ branch, and then splits to form two new folds 13 and 18.  This causes the $uUu$ branch to detach from the $U$ branch and attach to the unstable part of the $w$ branch to form an isolated branch that emerges from and loops back to the limiting shear.
   At $\psi_0 \approx 27.75^{\circ}$, folds 18 and 7 annihilate each other through a cusp in the $\Delta D/L - \psi_0$ plane.  Subcritical pitchfork 4 and branch point 9 annihilate each other at $\psi_0\approx28.15^{\circ}$; it appears that we can identify point 9 as a subcritical pitchfork at least for the small window of angles preceding this annihilation event, although stability information inferred from experiments does not allow us to make this identification in general.    This annihilation process detaches the $US\pm$ branches from the $U$ branch, while $U$ attaches smoothly to $w$ so that they are no longer distinct states, and the $U$-$US$ transition no longer occurs.  Now the primary first and third modes of planar \emph{elastica} are on the same curve, separated by a fold 6 and two branch points 8 and 0.

\begin{figure}[h!]
        \centering
        \includegraphics[width=0.85\textwidth]{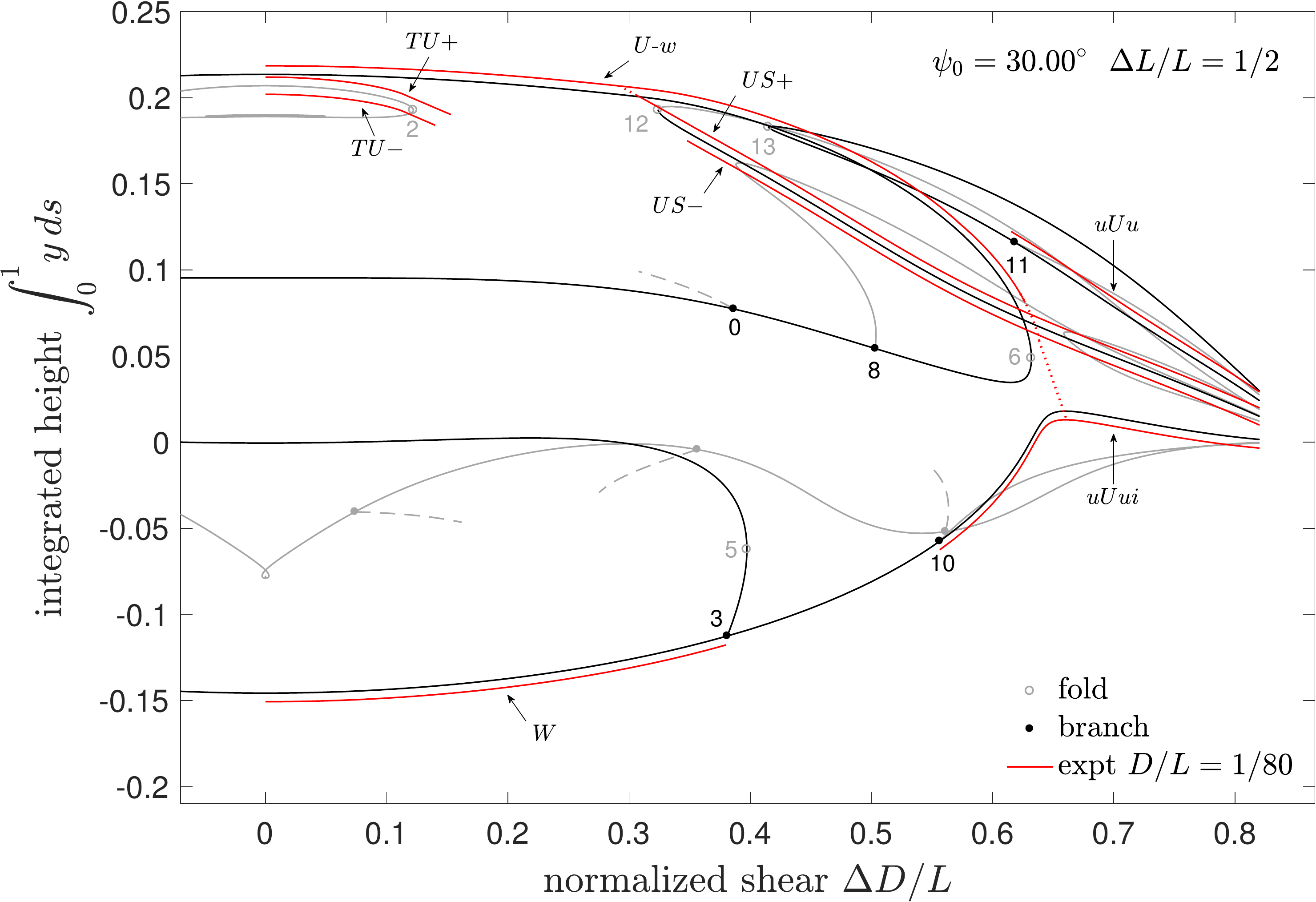}
        \caption{
        Some solutions (black and grey curves) and bifurcations (open and closed disks, some numbered) of the perfectly anisotropic Kirchhoff rod equations for clamping angle $\psi_0=30^{\circ}$, along with experimental data (red curves).  
		The horizontal extent of the red curves is the range of stability (typical variation $\approx \pm 0.01\, \Delta D/L$), while the vertical position of the curves is not measured data, but is made to follow near the numerical curves whenever a comparison is possible.  No data was taken for $\Delta D/L < 0$.    
		Numerically, the newly formed $U$-$w$ branch, which terminates in a fold rather than reaching the limit, is approaching the $uUui$ branch.  Red dotted lines indicate that sometimes smooth transitions from $U$-$w$ to $US+$ and from $uUui$ to $U$-$w$ are observed, which is inconsistent with the rest of the data (red solid lines) and the connectivity of the numerical solutions.
}
        \label{fig:Solution30}
\end{figure}

\begin{figure}[h!]
	\centering
	\includegraphics[width=0.8\textwidth]{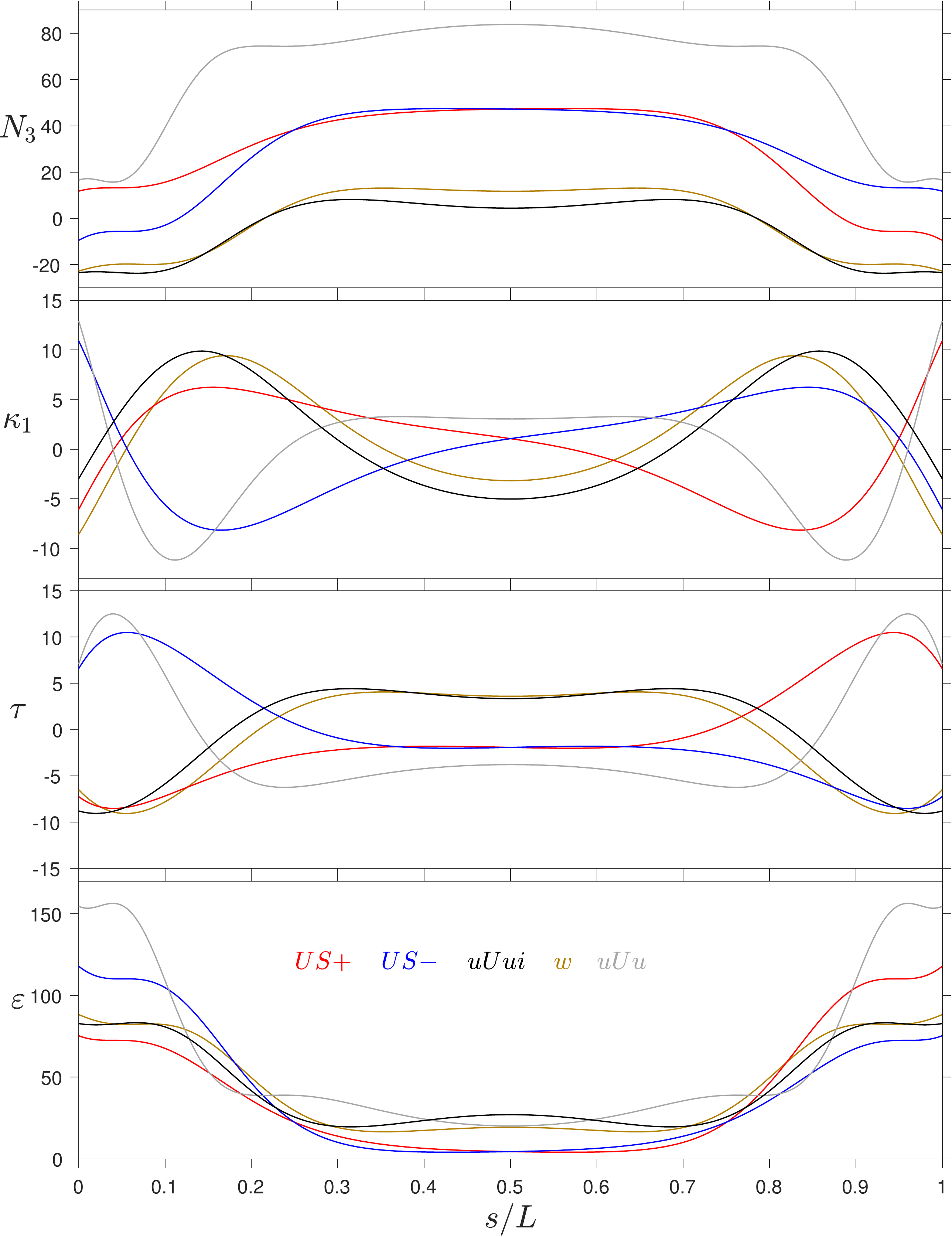}
	\caption{For five stable configurations at $\psi_0=30^{\circ}$, $\Delta L/L=1/2$, and $\Delta D/L=0.625$, we plot the axial force $N_3$, the curvature $\kappa_1$, the twist $\tau$, and the energy density $\varepsilon =\frac{1+\nu}{2} \kappa_1 ^2 + \tau ^2$, with $\nu = 0.25$.  The $uUui$, $w$ and $uUu$ states are reversibly symmetric about their midpoint, while the $US\pm$ states are a reversibly symmetric pair.  Increasing $s/L$ corresponds to moving from right to left on any curve renderings in the text.}\label{fig:30states}
\end{figure}

Figure \ref{fig:Solution30}  shows some solution manifolds and experimental stability data for a clamping angle $\psi_0=30^{\circ}$.  Both the $US\pm$ and $uUu$ branches are clearly detached from what is now the $U$-$w$ branch, which is approaching and will soon collide with the $uUui$ branch. 
Due to proximity to bifurcations that change connectivity of the solutions, the experimental data is not all consistent.  We draw red dotted lines to indicate that we often observe smooth transitions from $U$-$w$ to $US+$ and from $uUui$ to $U$-$w$, though at least half of the time these transitions do not occur and the data follow qualitatively with the numerical solutions.  The solid red lines correspond to these ``correct'' data.  Clearly, the system is sensitive to the presence of two nearby bifurcations in parameter space.  
The jump from $U$-$w$ to $uUui$ is weak and hard to observe.   This is now the main path to the limiting shear.

At $\Delta D/L=0.625$, there are five stable states.  Figure \ref{fig:30states} shows the numerically determined axial force $N_3$, the curvature $\kappa_1$, the twist $\tau$, and the energy density $\varepsilon =\frac{1+\nu}{2} \kappa_1 ^2 + \tau ^2$, with $\nu = 0.25$, for these states.  At this value of shear, all states have a higher energy density near the clamps than in the middle.  Interestingly, we observe that the $uUu$ state (primarily above the clamping plane) is purely tensile ($N_3 > 0$), while the corresponding $uUui$ state (primarily below the clamping plane) is compressive towards its ends and slightly tensile in the middle, although the depression at the center can be compressive at lower values of shear.  In general, higher shear will lead to increased tension as the limiting states are approached.

\begin{figure}[h!]
	\centering
	\includegraphics[width=0.85\textwidth]{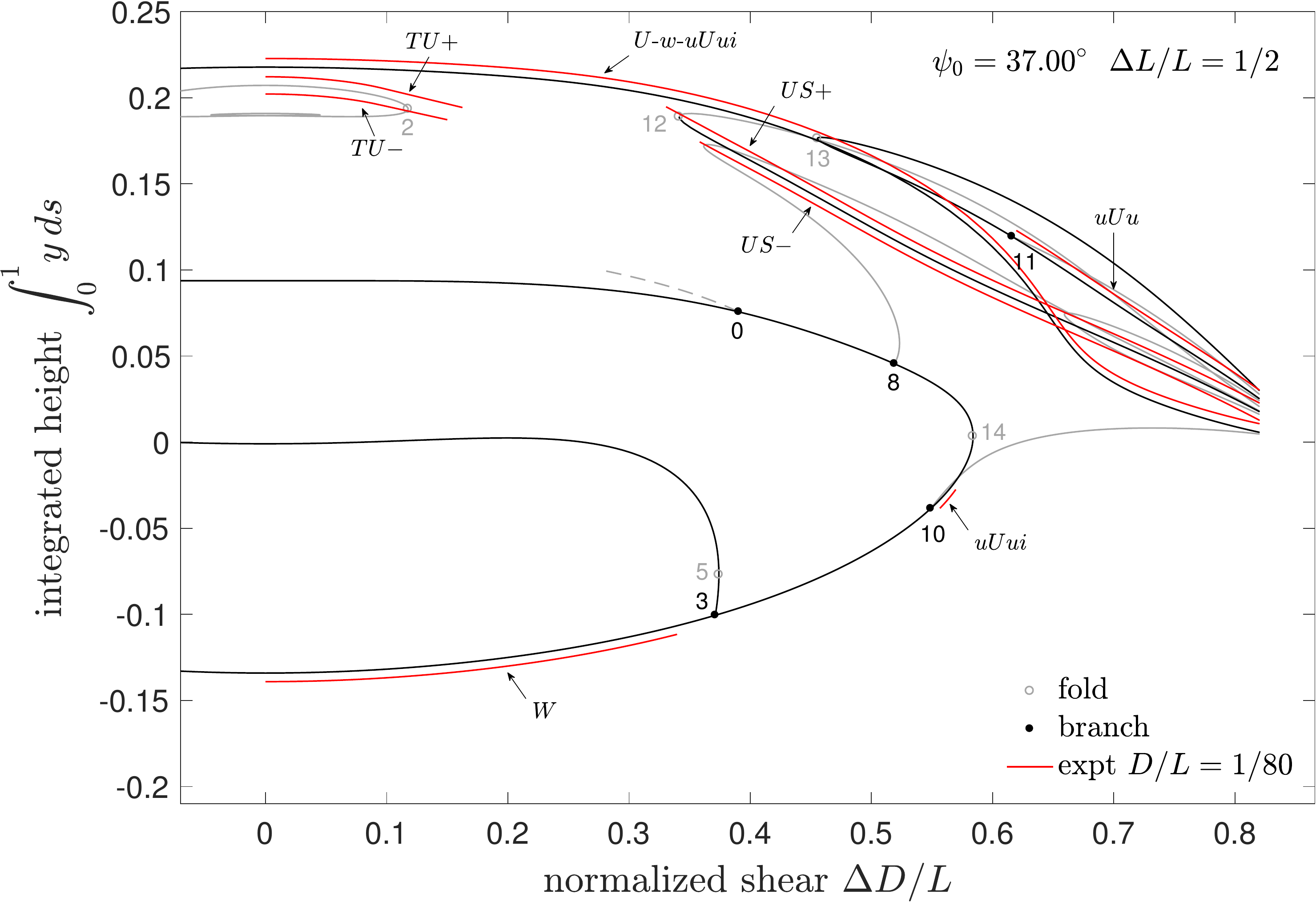}
	\caption{
	Some solutions (black and grey curves) and bifurcations (open and closed disks, some numbered) of the perfectly anisotropic Kirchhoff rod equations for clamping angle $\psi_0=37^{\circ}$, along with experimental data (red curves).  Some of the grey curves from prior diagrams have been removed. 
		The horizontal extent of the red curves is the range of stability (typical variation $\approx \pm 0.01\, \Delta D/L$), while the vertical position of the curves is not measured data, but is made to follow near the numerical curves whenever a comparison is possible.  No data was taken for $\Delta D/L < 0$.    
		Numerically, the $U$-$w$ branch has collided with the $uUui$ branch, creating a $U$-$w$-$uUui$ branch and a small residual $uUui$ branch.  These are also observed experimentally.  There is now a continuous $U$-$w$-$uUui$ path from first mode planar \emph{elastica} to approach the limiting shear.  The $WS\pm$ states are close to disappearing numerically, and are not observed experimentally.}
	\label{fig:Solution37}
\end{figure}

\begin{figure}[h!]
	\centering
	\includegraphics[width=0.85\textwidth]{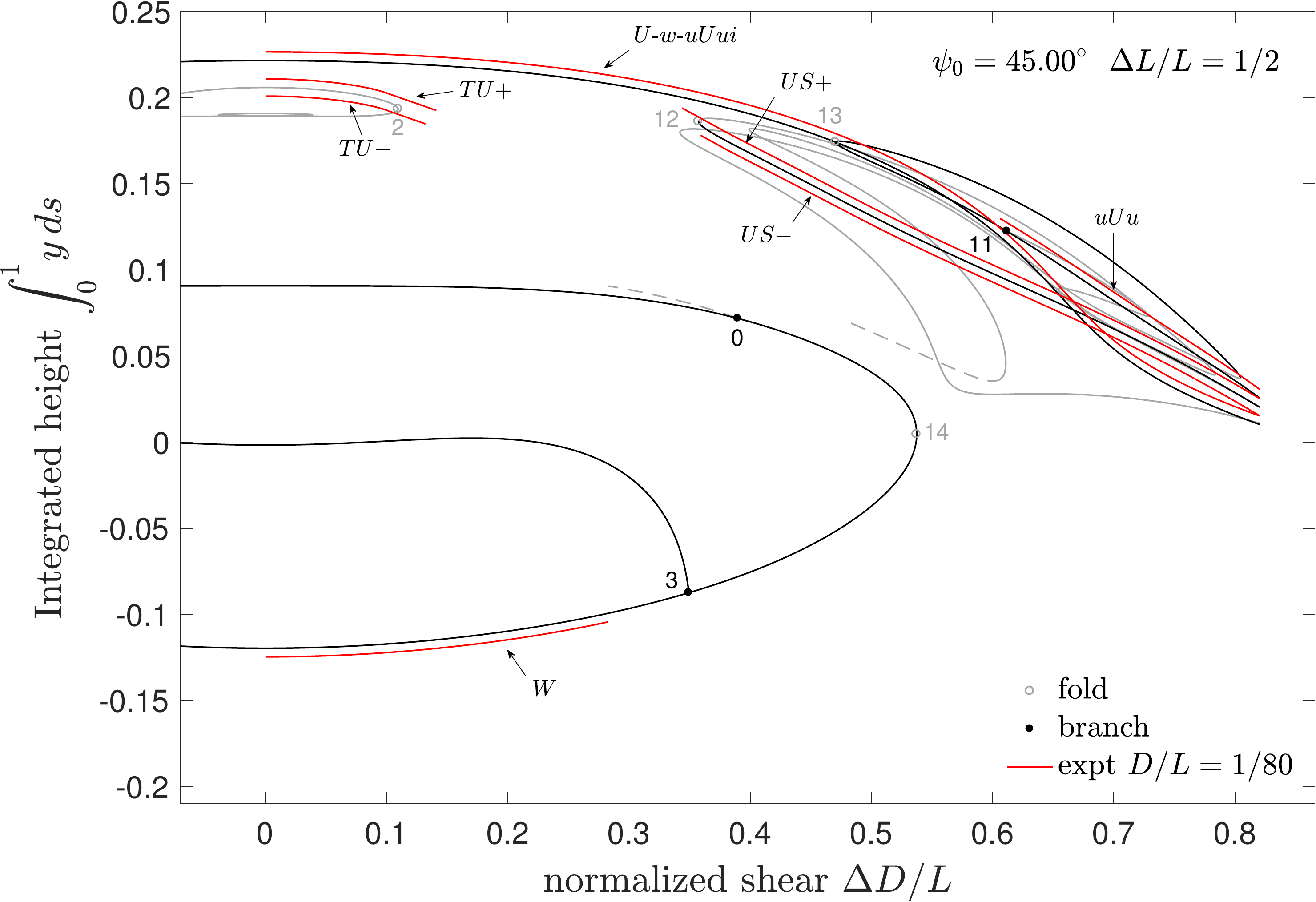}
	\caption{
Some solutions (black and grey curves) and bifurcations (open and closed disks, some numbered) of the perfectly anisotropic Kirchhoff rod equations for clamping angle $\psi_0=45^{\circ}$, along with experimental data (red curves).  One grey curve has been truncated with a dashed line.
		The horizontal extent of the red curves is the range of stability (typical variation $\approx \pm 0.01\, \Delta D/L$), while the vertical position of the curves is not measured data, but is made to follow near the numerical curves whenever a comparison is possible.  No data was taken for $\Delta D/L < 0$.    
		The $WS\pm$ states have become unstable, and the residual $uUui$ branch has disappeared. }
	\label{fig:Solution45}
\end{figure}

At $\psi_0 \approx 30.62^{\circ}$, the $U$-$w$ branch collides with the $uUui$ branch.  This merge-split event leads to a continuous $U$-$w$-$uUui$ path from first mode planar \emph{elastica} to approach the limiting shear.  However, part of the original $uUui$ branch remains, and can be seen between pitchfork 10 and a new fold 14 in Figure  \ref{fig:Solution37}.  Figure \ref{fig:Solution37}   shows some solution manifolds and experimental stability data for a clamping angle $\psi_0=37^{\circ}$.  
Now the primary third mode and the inverted first mode of planar \emph{elastica} are on the same curve, separated by several branch points 3, 10, 8, and 0, and a fold 14.  
The $WS\pm$ states have a very narrow extent, and will soon disappear (at $\psi_0 \approx 41.74^{\circ}$) as the supercritical pitchfork 3 absorbs the fold 5 and becomes subcritical, a process we will clearly see later in Figure \ref{fig:phasediagram}.  Already we do not observe them in the experiments.
Some of the grey curves in the lower half of previous figures, including the inverted third mode counterpart to the upper branch that contains the $w$ state, have been removed, as they have collided with other very complicated states that we have already removed.  We retain a small grey hairpin curve near the limiting shear, as it will eventually link up with one of the black curves at higher angles.
The experiments confirm the changes in connectivity, including the presence of a residual $uUui$ branch with short extent at intermediate shear.
Interestingly, this $uUui$ shape continues to jump to the $U$-$w$-$uUui$ branch upon increasing shear, and to the $US-$ branch upon decreasing shear, even as the range of stability shrinks and these bifurcations (fold and subcritical pitchfork) approach the same value of shear.  This, along with the relatively steep slope of the branch, imply that the shapes change significantly over a narrow range of shear.
Many transitions are illustrated in the supplementary video {\texttt{transition37.mp4}} \cite{videos}.

\begin{figure}[h!]
\centering
\includegraphics[width=0.9\textwidth]{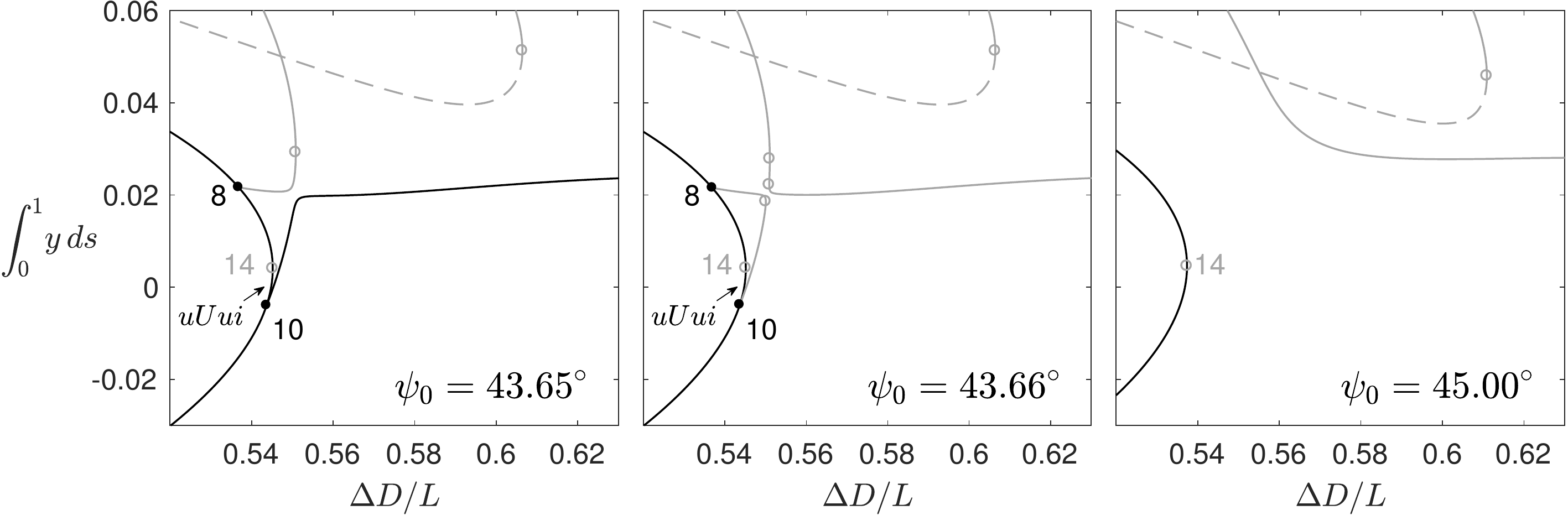}
\caption{Details of several transitions that change the connectivity of solution curves.  The annihilation of branch points 8 and 10 occurs at fold point 14 and leads to the disappearance of the residual $uUui$ branch.
Closed and open disks are branch and turning points, respectively.}
\label{fig:Transition2}
\end{figure}

Figure \ref{fig:Solution45}  shows some solution manifolds and experimental stability data for a clamping angle $\psi_0=45^{\circ}$, after a collision between pitchfork 3 and fold 5 at $\psi_0 \approx 41.74^{\circ}$ has made $WS\pm$ states unstable, and a collision-annihilation of branch points 8 and 10 at $\psi_0 \approx 44.27^{\circ}$ has eliminated the residual $uUui$ branch.  Details of the latter process are shown in Figure \ref{fig:Transition2}.
Some complicated changes in connectivity of unstable (unobserved) states have occurred near the limiting shear, and one curve has turned back to link up with states we have already removed; we use a dashed line to truncate this curve.  This connection is actually short-lived and will soon be lost again, so this dashed line will not appear again in subsequent figures.

\begin{figure}[h!]
	\centering
	\includegraphics[width=0.85\textwidth]{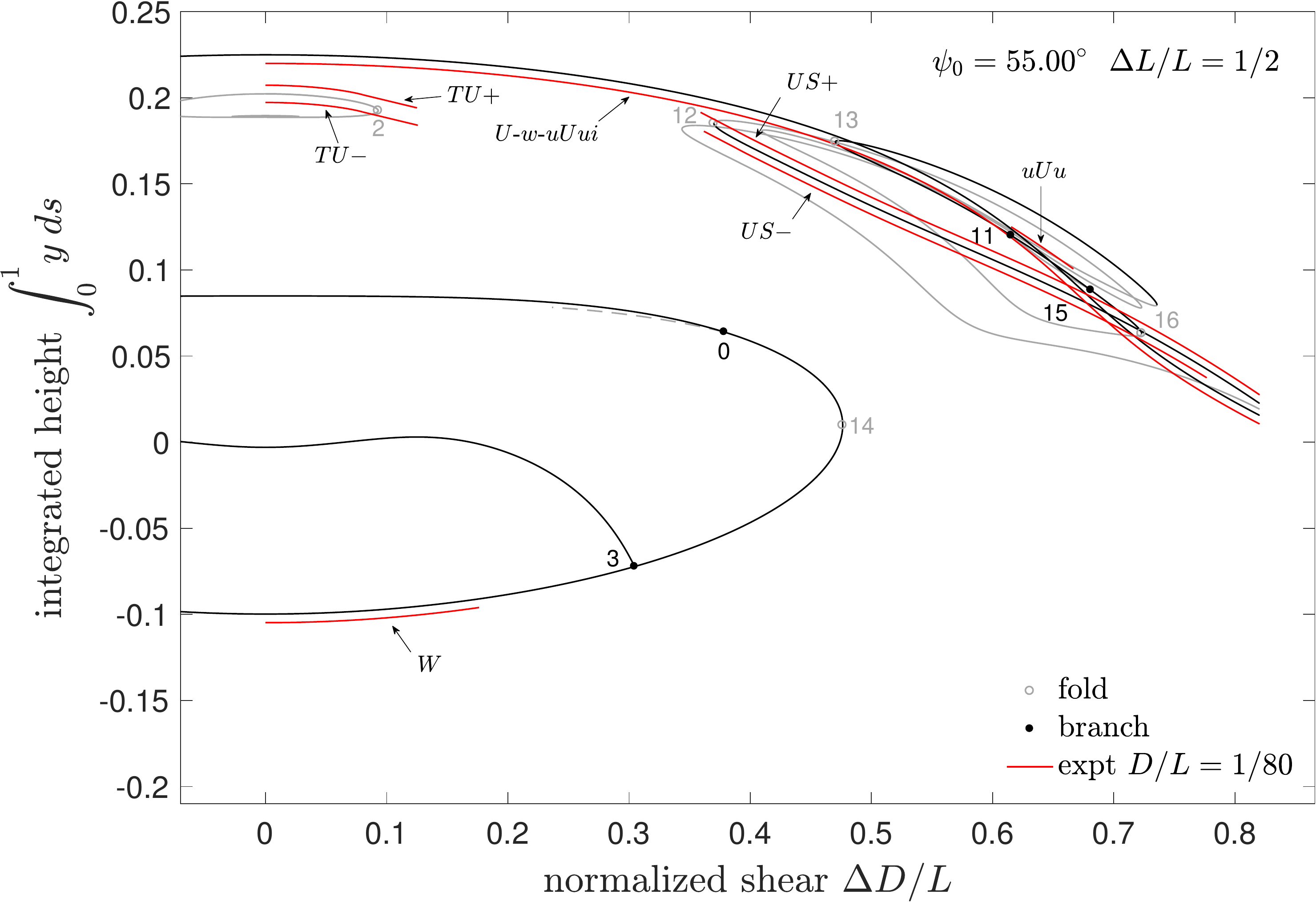}
	\caption{Some solutions (black and grey curves) and bifurcations (open and closed disks, some numbered) of the perfectly anisotropic Kirchhoff rod equations for clamping angle $\psi_0=55^{\circ}$, along with experimental data (red curves).  
		The horizontal extent of the red curves is the range of stability (typical variation $\approx \pm 0.01\, \Delta D/L$), while the vertical position of the curves is not measured data, but is made to follow near the numerical curves whenever a comparison is possible.  No data was taken for $\Delta D/L < 0$.    
		Several changes have occurred and are described in the text.  The $US\pm$ branch has a fold at higher shear than what we show here, and connects back to the loopy grey curves.  The $uUu$ branch is not always observed experimentally.	}
	\label{fig:Solution55}
\end{figure}

Figure \ref{fig:Solution55}  shows some solution manifolds and experimental stability data for a clamping angle $\psi_0=55^{\circ}$.  At $\psi_0 \approx 52.09^{\circ}$, a pair of subcritical pitchforks is born on the $uUu$ branch at $\Delta D/L \approx 0.789$ .  By $\psi_0 = 55^{\circ}$, one pitchfork 15 remains, the other has exited to the right at high shear, while a fold 16 has entered from the right.  The details of this process are not known, but from the loci shown later in Figure \ref{fig:phasediagram}, it seems that there is a curve splitting at high shear that we don't observe.
Other curves have also developed folds that move in from the limiting shear, and additional complicated changes in connectivity of unstable (unobserved) states have occurred.
At this angle, the $US\pm$ branch turns around at $\Delta D/L \approx 0.856$ and connects to the loopy grey curves.  We experimentally observe that both $US\pm$ states lose stability before this value of shear, although for the $US+$ state this is at a value of shear higher than what we show in the figures.
Bifurcation 3 should now take the $W$ state to the $U$-$w$-$uUui$ branch instead of the $US\pm$ branch.  This change in path actually begins to happen at lower angles due to the destabilizing effects of gravity on the $W$ state, as can be seen in Figure \ref{fig:Solution45}.  Therefore, starting with any planar state at this high clamping angle and simply applying shear, we will approach the limit through a $U$-like state, and not an $S$-like state.
Experimental results on the short $uUu$ branch are inconsistent, in that sometimes the state is not observed.  The data shown are for ``correct'' observations.  The system is sensitive due to its proximity to an event at 
$\psi_0 \approx 56.29^{\circ}$, when subcritical pitchforks 11 and 15 annihilate each other, leading to the disappearance of the $uUu$ state.

\begin{figure}[h!]
	\centering
	\includegraphics[width=0.85\textwidth]{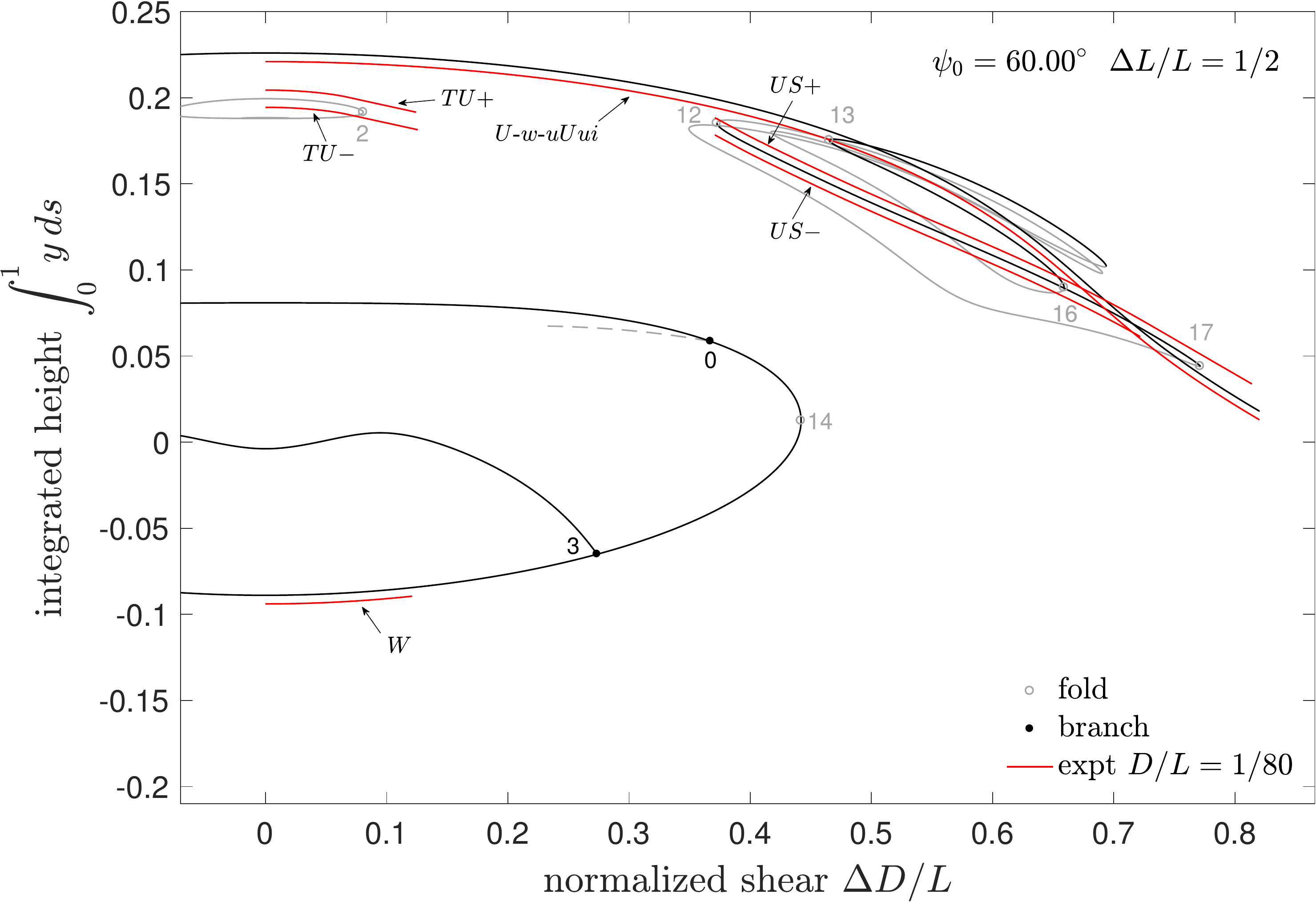}
	\caption{Some solutions (black and grey curves) and bifurcations (open and closed disks, some numbered) of the perfectly anisotropic Kirchhoff rod equations for clamping angle $\psi_0=60^{\circ}$, along with experimental data (red curves).  
		The horizontal extent of the red curves is the range of stability (typical variation $\approx \pm 0.01\, \Delta D/L$), while the vertical position of the curves is not measured data, but is made to follow near the numerical curves whenever a comparison is possible.  No data was taken for $\Delta D/L < 0$.    
		The $uUu$ branch has disappeared, and the fold 17 on the $US\pm$ branch appears at a lower shear.  Experimentally, the $W$ state is significantly destabilized by gravity.	}
	\label{fig:Solution60}
\end{figure}

\begin{figure}[h!]
	\centering
		\includegraphics[width=0.8\textwidth]{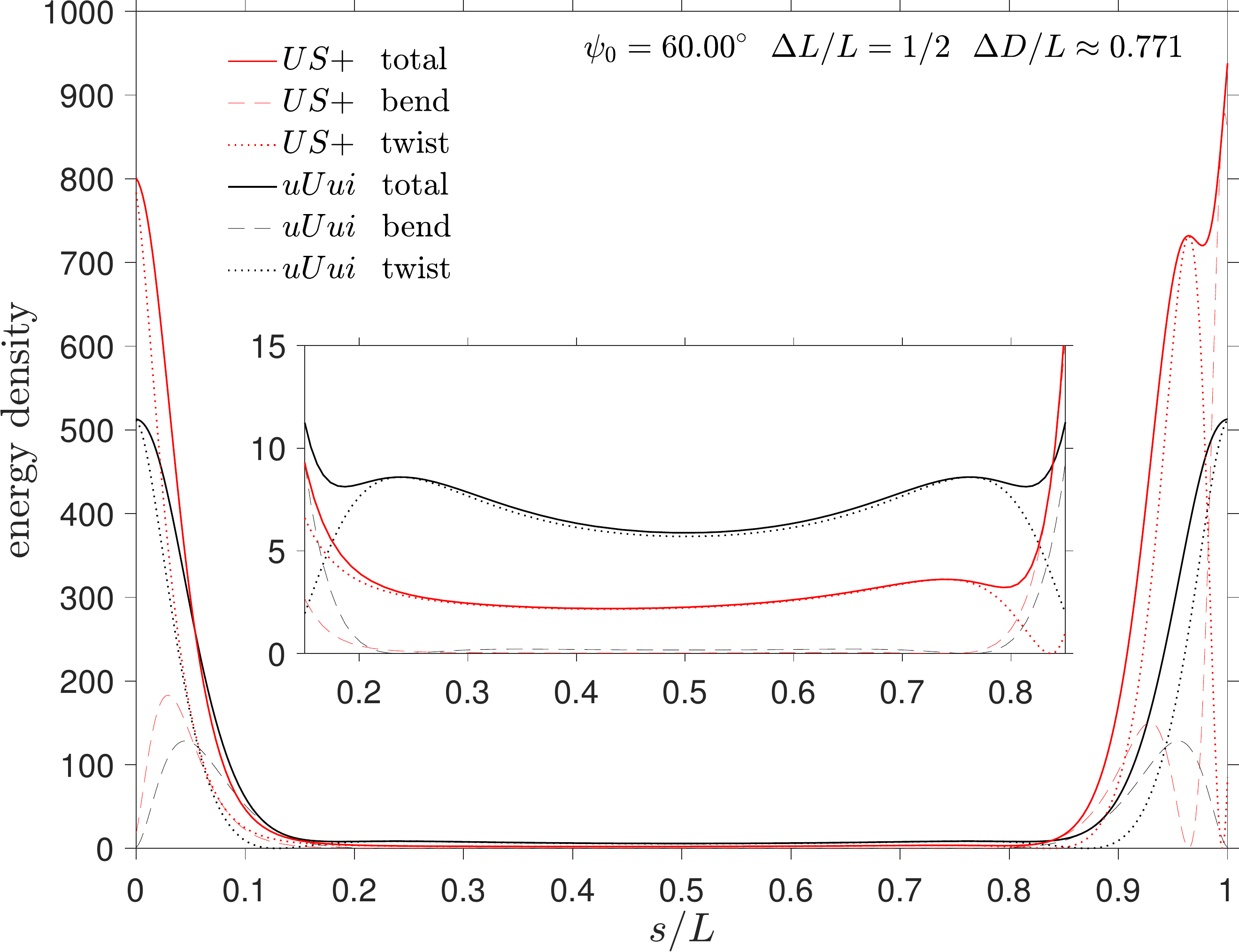}
		\caption{
	The bending energy density $\varepsilon =\frac{1+\nu}{2} \kappa_1 ^2$ with $\nu = 0.25$, the twist energy density $\tau ^2$, and the total energy density for the two states before and after the $US+$ to $uUui$ transition, the latter being the high shear portion of the $U$-$w$-$uUui$ branch.
	The jump relieves a high concentration of bending energy at the $s=1$ end of the $US+$ state.  The central expanse of the rod, shown in the inset, stores relatively little elastic energy, most of it in twist.
Increasing $s/L$ corresponds to moving from right to left on any curve renderings in the text.}
	\label{fig:jumpenergy}
\end{figure}

Figure \ref{fig:Solution60}  shows some solution manifolds and experimental stability data for a clamping angle $\psi_0=60^{\circ}$.
The terminal fold 17 of the $US\pm$ branch now appears at a lower shear, $\Delta D/L \approx 0.771$.  Experiments are qualitatively consistent with the solutions, with gravity significantly destabilizing the $W$ state at this high angle and asymmetrizing the $US\pm$ transitions with respect to fold 17.

 Transitions at $\psi_0=55^{\circ}$ and $\psi_0=60^{\circ}$ are shown in the supplementary video {\texttt{transition5560.mp4}}  \cite{videos}.   It can be seen that the $US+$ to $U$-$w$-$uUui$ transition through fold 17 at $\psi_0=60^{\circ}$ involves a rapid rotation of one end.  A similar rapid rotation is seen during local snap-through events in the tensile loading of slit sheets \cite{marcelopersonal}. This transition is explored further in Figure \ref{fig:jumpenergy}, which plots the bending energy density $\varepsilon =\frac{1+\nu}{2} \kappa_1 ^2$, with $\nu = 0.25$, the twist energy density $\tau ^2$, and the total energy density for the states just before and after this transition.  It can be seen that the jump relieves a high concentration of bending energy at one end and partially relieves some twisting energy near that end while shifting its maximum to the end, and partially relieves some twisting energy at the other end. Some of the energy has moved into the central expanse of the rod, but this region stores relatively little elastic energy either before or after the transition, most of it in twist.

If we continue to increase the clamping angle above $\psi_0  = 60^{\circ}$, the twisted $TU\pm$ loops will shrink and disappear through the annihilation of two folds at $\psi_0 \approx 74.03^{\circ}$.  The loopy structure of $US\pm$ also shrinks and disappears after undergoing some complicated transitions which we do not investigate here.
Additionally, the subcritical pitchfork 3 delimiting the stability of the $W$ state approaches the zero-shear axis and annihilates with its negative-shear twin to eliminate the stable $W$ state at $\psi_0 \approx 76.95^{\circ}$--- this is the classic snap-through of an inverted \emph{elastica} arch under end rotations \cite{plaut2009vibration}.  After this, the only remaining stable configuration is $U$-$w$-$uUui$.  We did not proceed past clamping angles of 
$\psi_0=80^{\circ}$.

To conclude this section, we remark that, despite some variable results due to sensitivity of the system near bifurcations, all of our experimental observations for narrow bands seem to be explained by the anisotropic Kirchhoff model, with allowance for the effects of gravity.

\section{Loci of bifurcations related to stable states}\label{loci}

\begin{figure}[h!]
	\centering
	\includegraphics[width=0.85\textwidth]{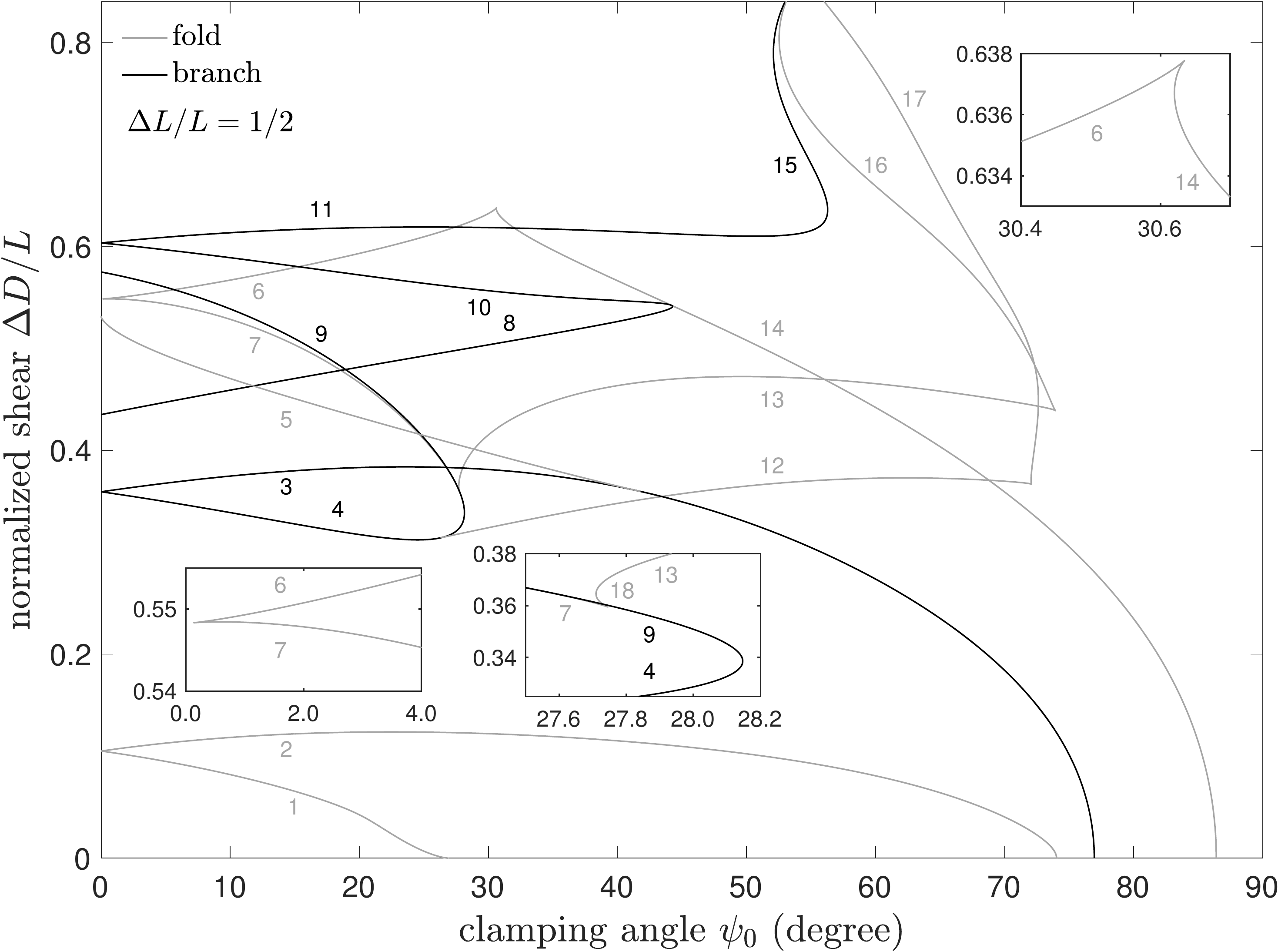}
	\caption{Loci of various bifurcations for the perfectly anisotropic Kirchhoff rod in the plane spanned by clamping angle $\psi_0$ and shear $\Delta D/L$, at fixed compression $\Delta L/L=1/2$.  Numbers correspond to those on figures in Section \ref{results}. }
	\label{fig:phasediagram}
\end{figure}

The complicated landscape of connectivity changes surveyed in the previous section can be better understood by tracing the loci of bifurcation points of the perfectly anisotropic Kirchhoff equations in a higher-dimensional parameter space.  For our present study at fixed compression $\Delta L/L=1/2$, this is the two-dimensional space spanned by normalized shear $\Delta D/L$ and clamping angle $\psi_0$.
Figure \ref{fig:phasediagram} shows the paths traced in this space by many fold and branch points, numbered as on figures in Section \ref{results}.  Most of these are connected in some way with states observed in experiments, and thus delineate regions of stability for various configurations.
The leftmost inset shows the cusp that gives rise to folds $6$ and $7$ and the $w$ state at the small value of clamping angle $\psi_0 \approx 0.135^{\circ}$.
The middle inset corresponds to the complicated series of transitions shown in Figure \ref{fig:Transition1}.  The upper right inset corresponds to the merge-split event between the $w$ and $uUui$ branches at $\psi_0 \approx 30.62^{\circ}$.
The turning points connecting pitchfork 4 and branch point 9, and pitchforks 11 and 15, represent the annihilation events between these two pairs of bifurcations at $\psi_0\approx28.15^{\circ}$ (Figure \ref{fig:Transition1}) and $\psi_0\approx56.29^{\circ}$, respectively.
 Continuing along the 15 curve, there is another turning point at higher shear, which implies that 15 and another pitchfork appear together on the $uUu$ branch at $\psi_0\approx52.09^{\circ}$, with the other moving off to higher shears, as discussed earlier with respect to Figure \ref{fig:Solution55}.
Between folds 12 and 17, there are two turning points and a cusp, indicating some complicated behavior involved in the disappearance of the $US\pm$ states at clamping angles above $70^{\circ}$.
The diagram is symmetric about zero shear; note the asymmetry between folds 1 and 2 that govern the disappearance of twisted states, such that 1 has a cusp on the zero-shear line, while 2 has a smooth turning point.  The diagram also has a symmetry about zero clamping angle, but in a pairwise sense; for example, curves 1 and 2 will exchange their identities upon crossing this axis.

The loci provide some information about regions of stability.  For example, with reference to the positive shear and clamping angle quadrant shown here, the $TU \pm$ and $TW \pm$ states are stable below curves 2 and 1, respectively, and the $W$ state is stable below curve 3.  The $WS\pm$ states are stable in the region between curves 3 and 5.  The $US\pm$ states are stable above and to the left of a curve connecting the loci of 4, 12, and 17.

 It is clear that clamping at large angles reduces the number of available states, and thus the occurrence of jump events.  We can use a diagram like Figure \ref{fig:phasediagram} to avoid such violent events.  For example, we might wish to transform a large clamping angle, large shear $US+$ state to a large clamping angle, small shear $U$ state, without experiencing the jump that would occur upon simply reducing the shear.  Instead, we can decrease the clamping angle, decrease the shear, and increase the clamping angle again.  Thus we avoid crossing line 12, corresponding to a fold-induced jump, and instead cross line 4, corresponding to a supercritical pitchfork.
For another example, we can transform a small clamping angle, large shear $w$ state to either a small clamping angle, small shear $U$ state or a small clamping angle, large(r) shear $uUui$ state by first increasing the clamping angle, then shearing back or forward, and finally decreasing the clamping angle again.  This avoids crossing lines $7$ or $6$, which are fold-induced jumps, and makes use of the continuous $U$-$w$-$uUui$ branch available at large clamping angles.

\section{Width effects: from rods to ribbons to plates}\label{widtheffect}

\begin{figure}[h!]
	\captionsetup[subfigure]{labelformat=empty}
	\centering
	\begin{subfigure}[t]{0.13\textwidth}
		\centering
		\includegraphics[width=0.8in]{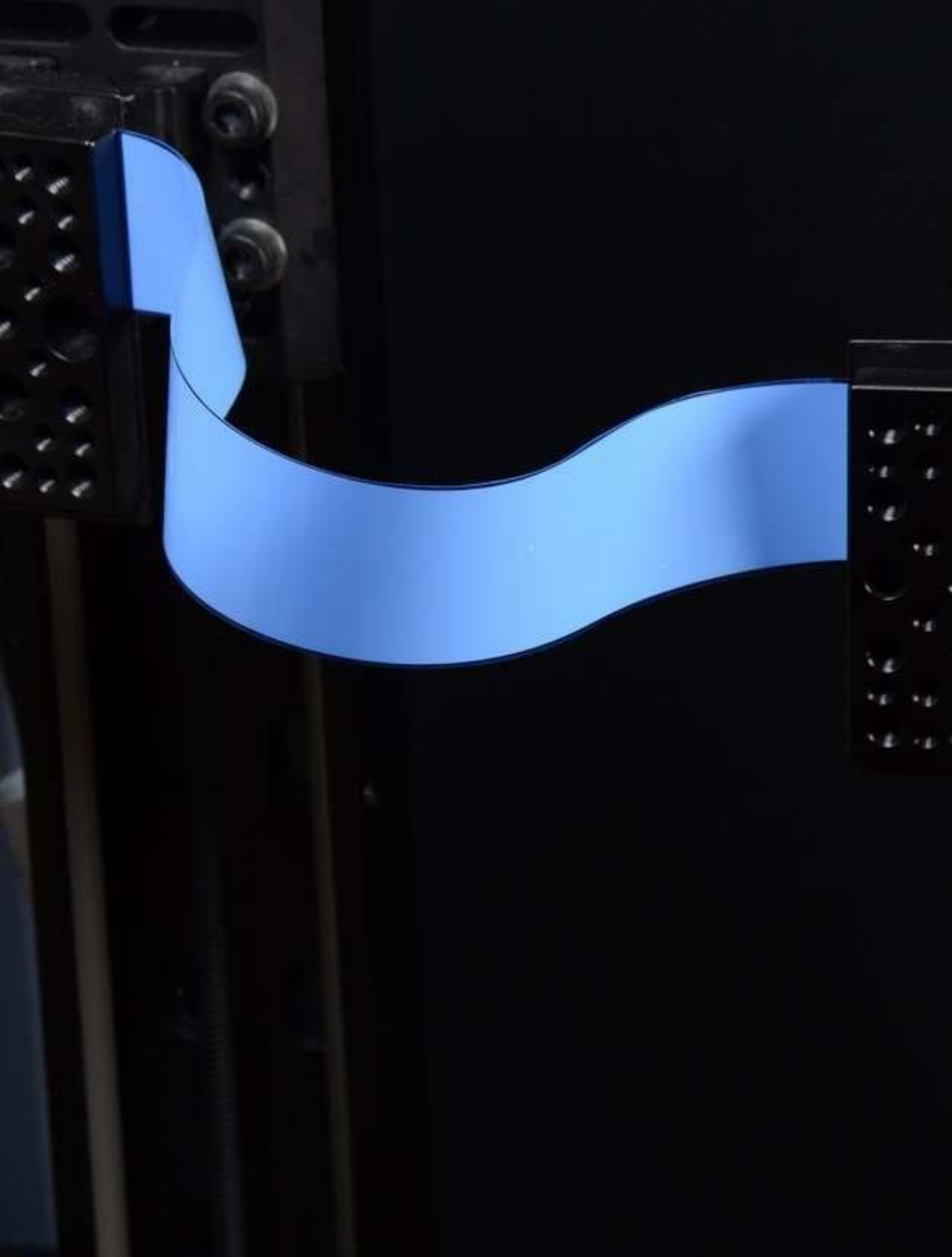}
		\caption{$U1$}
	\end{subfigure}%
	\hspace{1pt}
	\begin{subfigure}[t]{0.13\textwidth}
		\centering
		\includegraphics[width=0.8in]{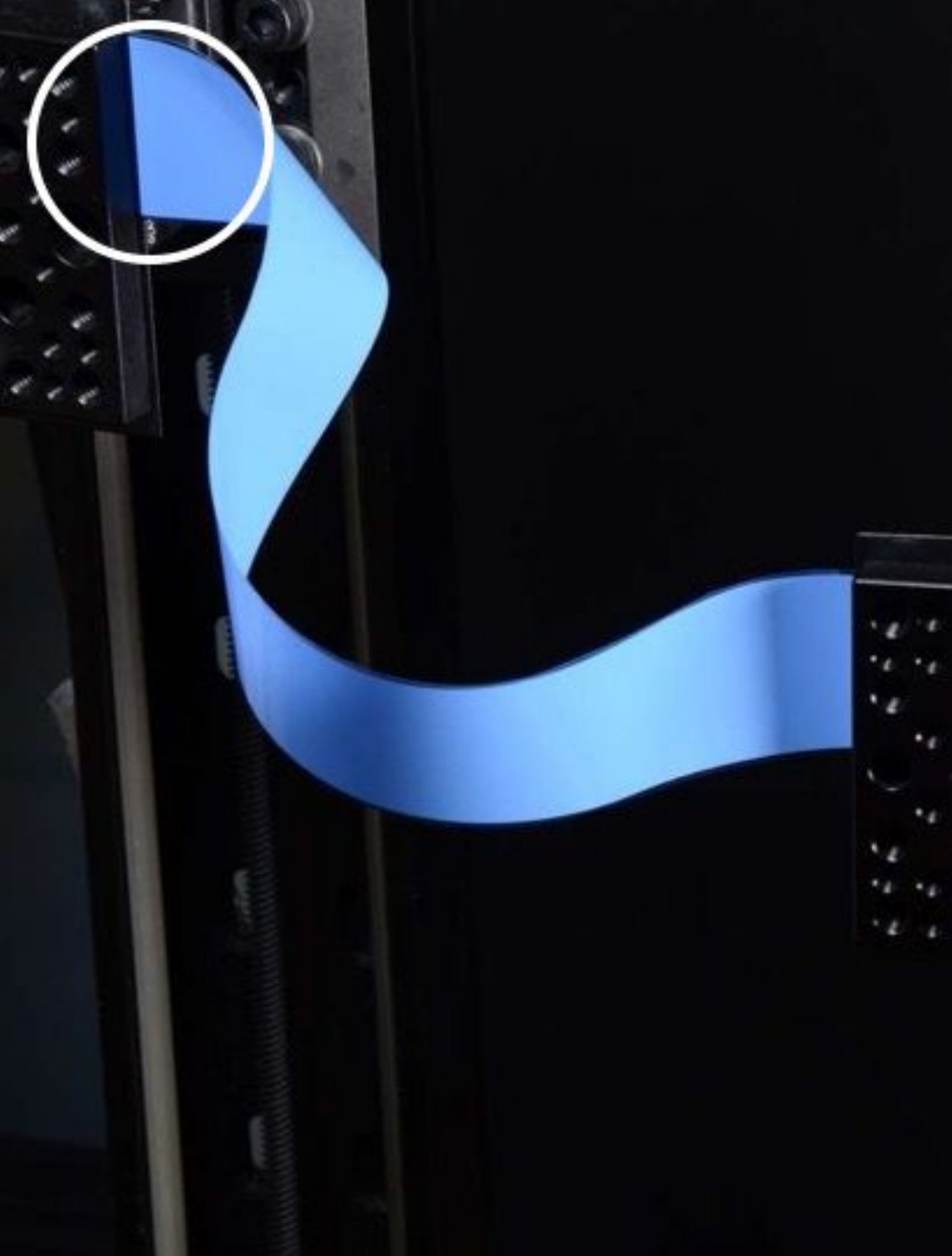}
		\caption{$US1+$}
	\end{subfigure} 
	\begin{subfigure}[t]{0.13\textwidth}
		\centering
		\includegraphics[width=0.8in]{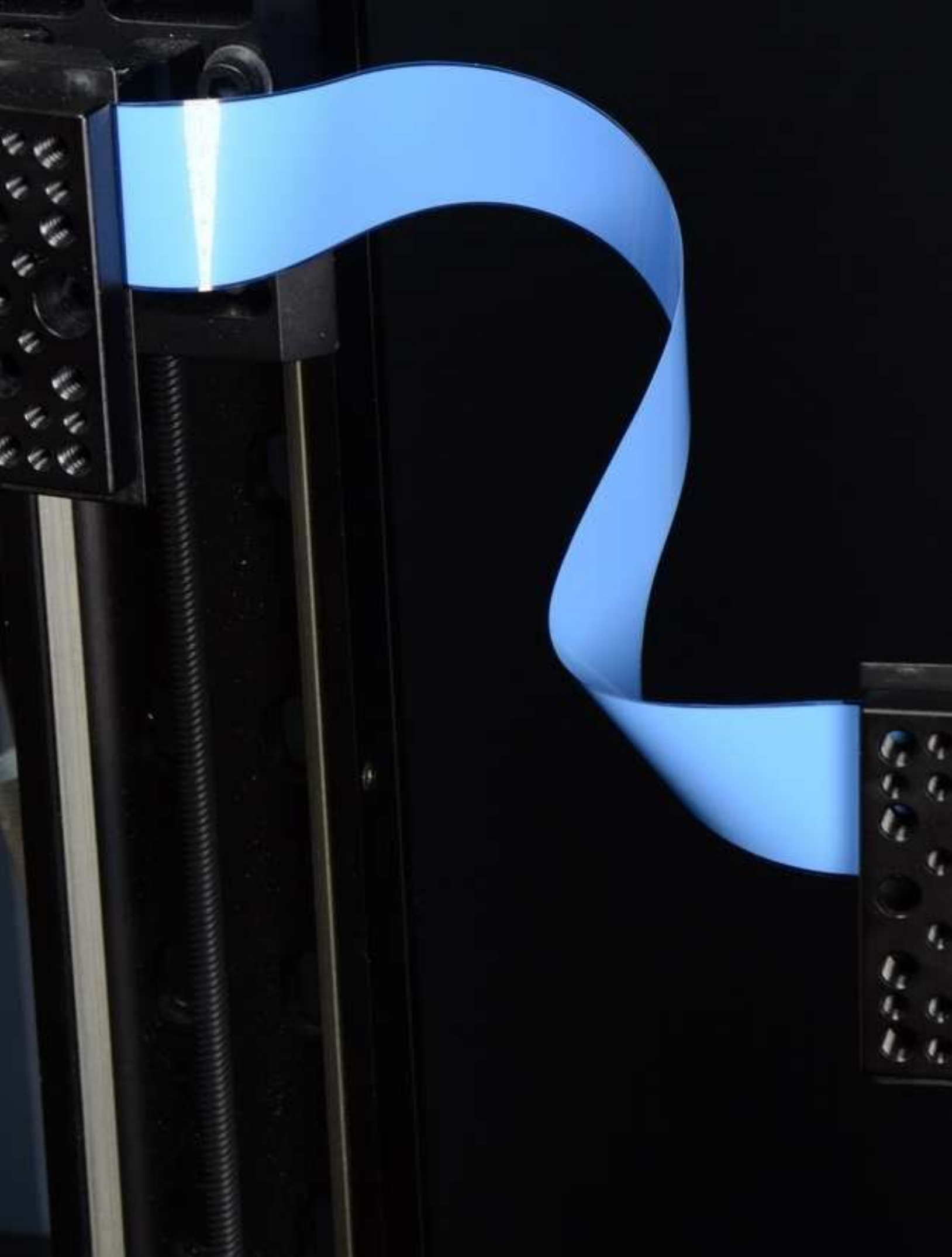}
		\caption{$WS+$}
	\end{subfigure}
	\begin{subfigure}[t]{0.13\textwidth}
		\centering
		\includegraphics[width=0.8in]{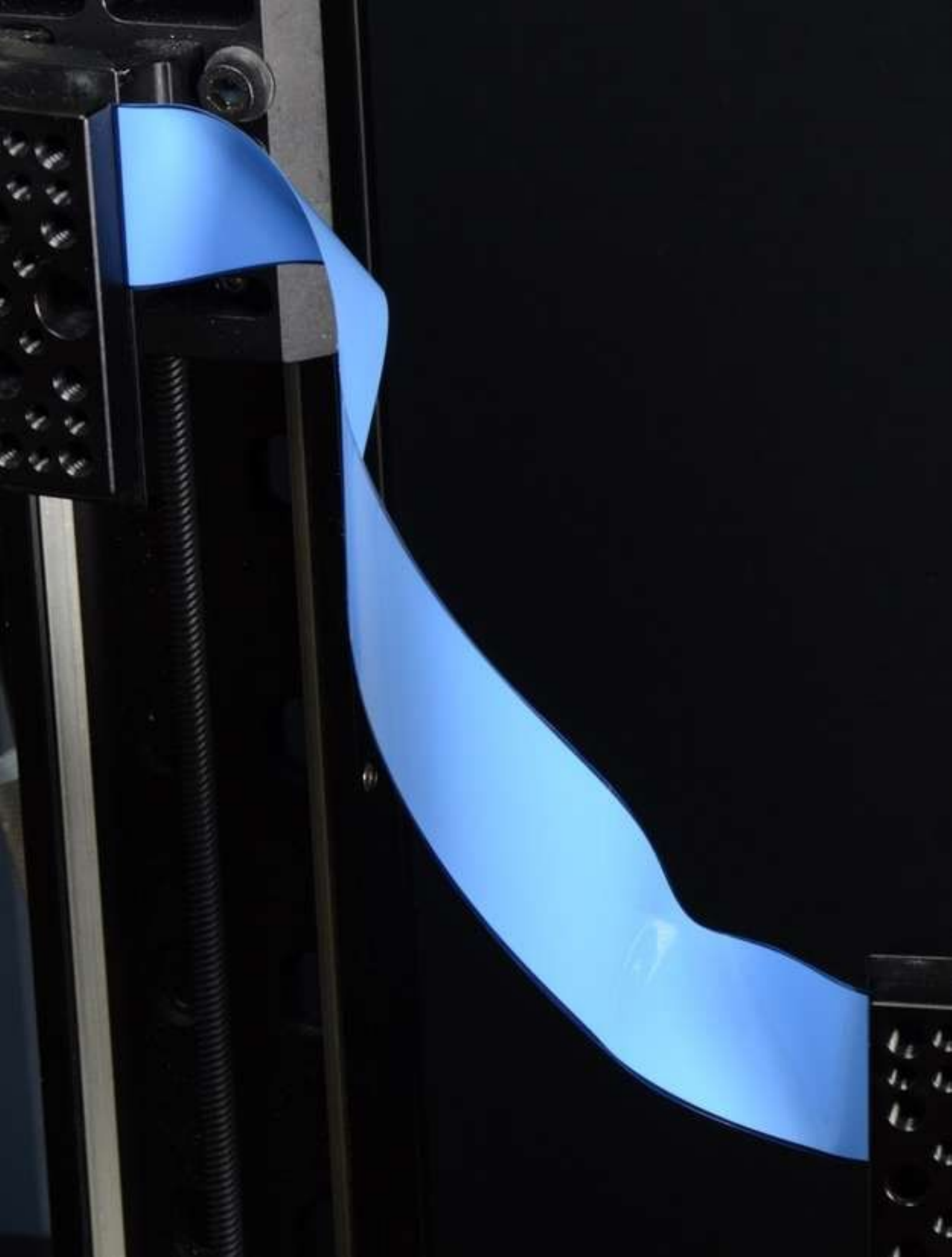}
		\caption{$uUu$}
	\end{subfigure}
	\begin{subfigure}[t]{0.13\textwidth}
		\centering
		\includegraphics[width=0.8in]{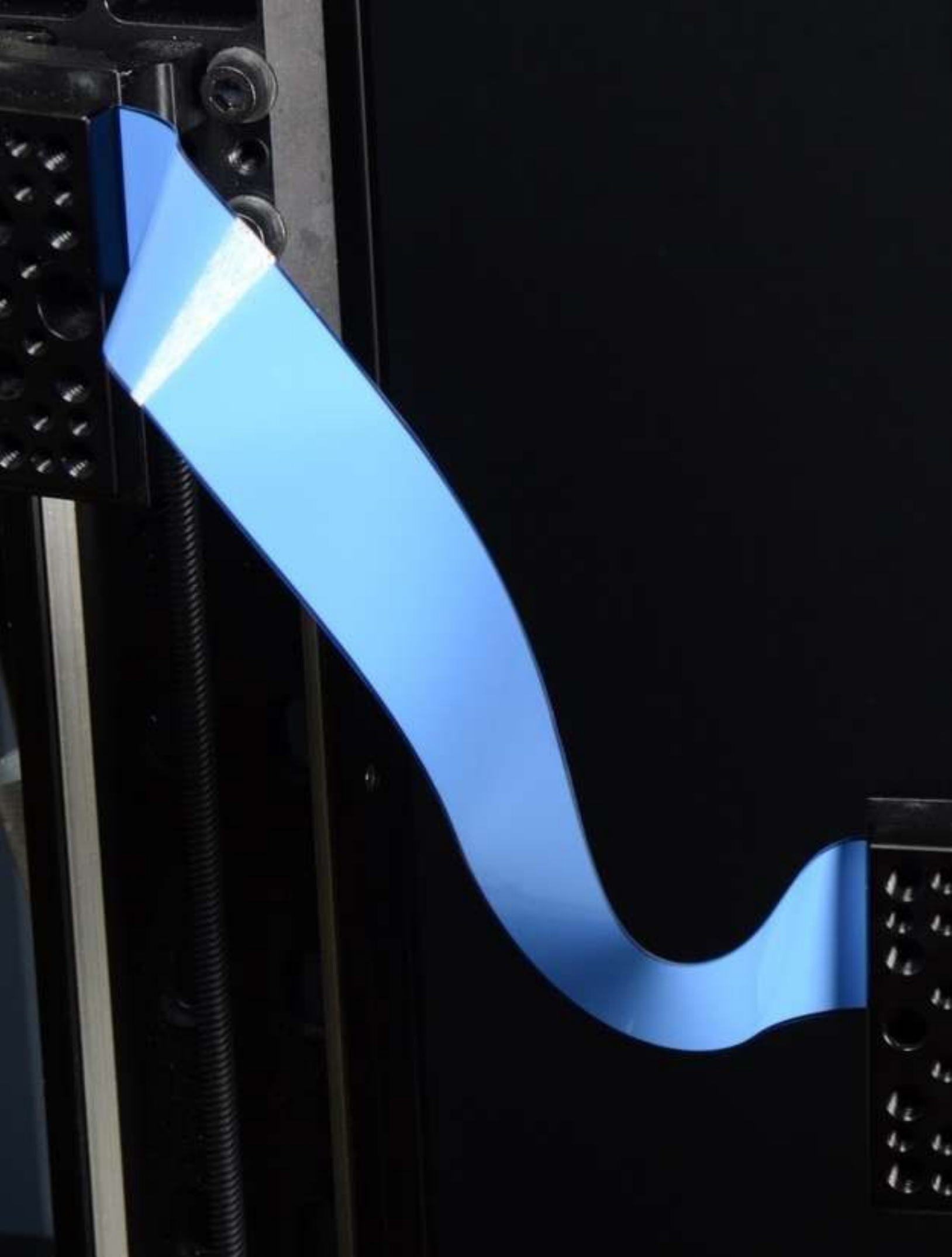}
		\caption{$w$}
	\end{subfigure}
	\begin{subfigure}[t]{0.13\textwidth}
	\centering
	\includegraphics[height=1.056in]{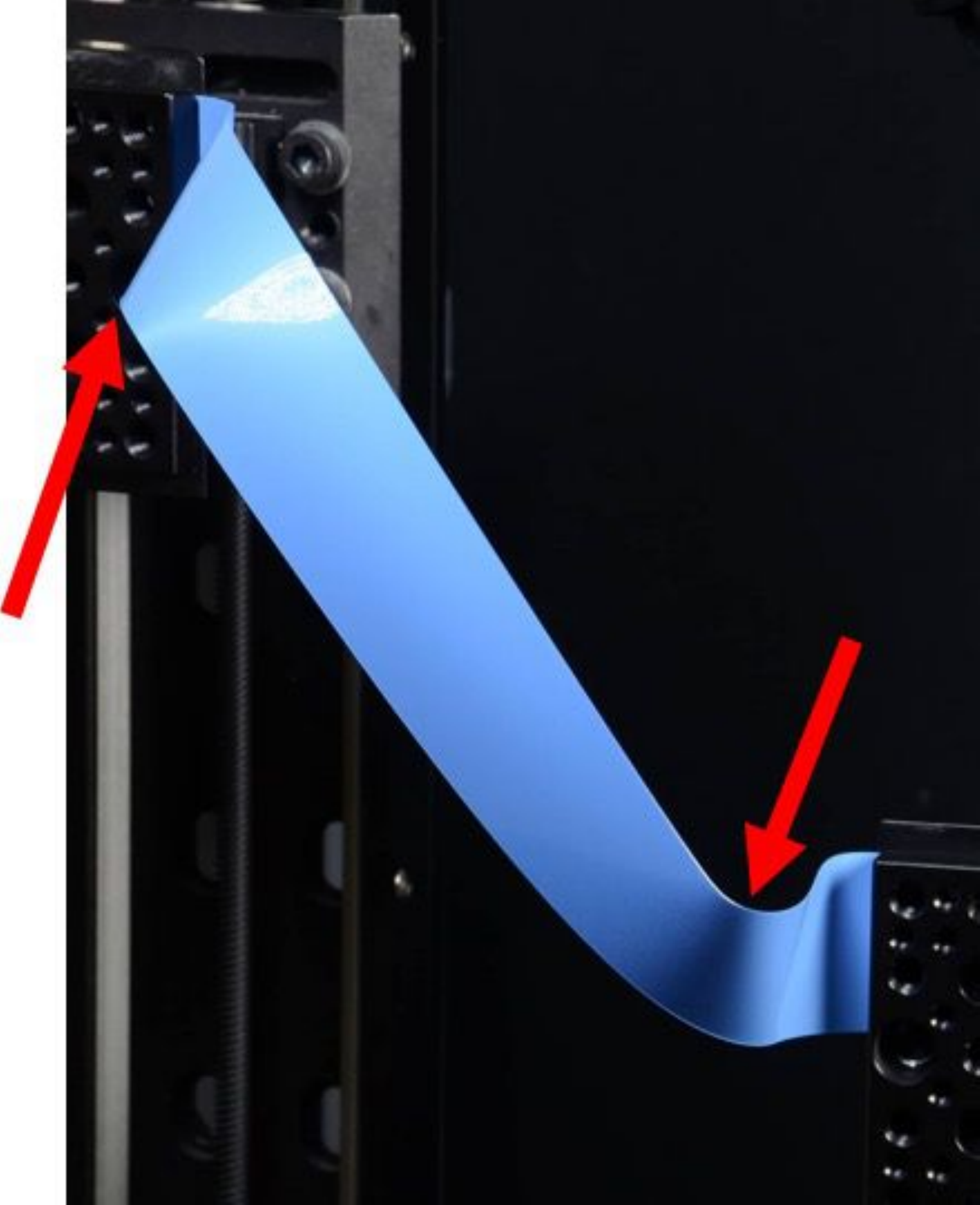}
	\caption{$U2$}
  \end{subfigure}
  \hspace{-1pt}
	\begin{subfigure}[t]{0.13\textwidth}
		\centering
		\includegraphics[width=0.8in]{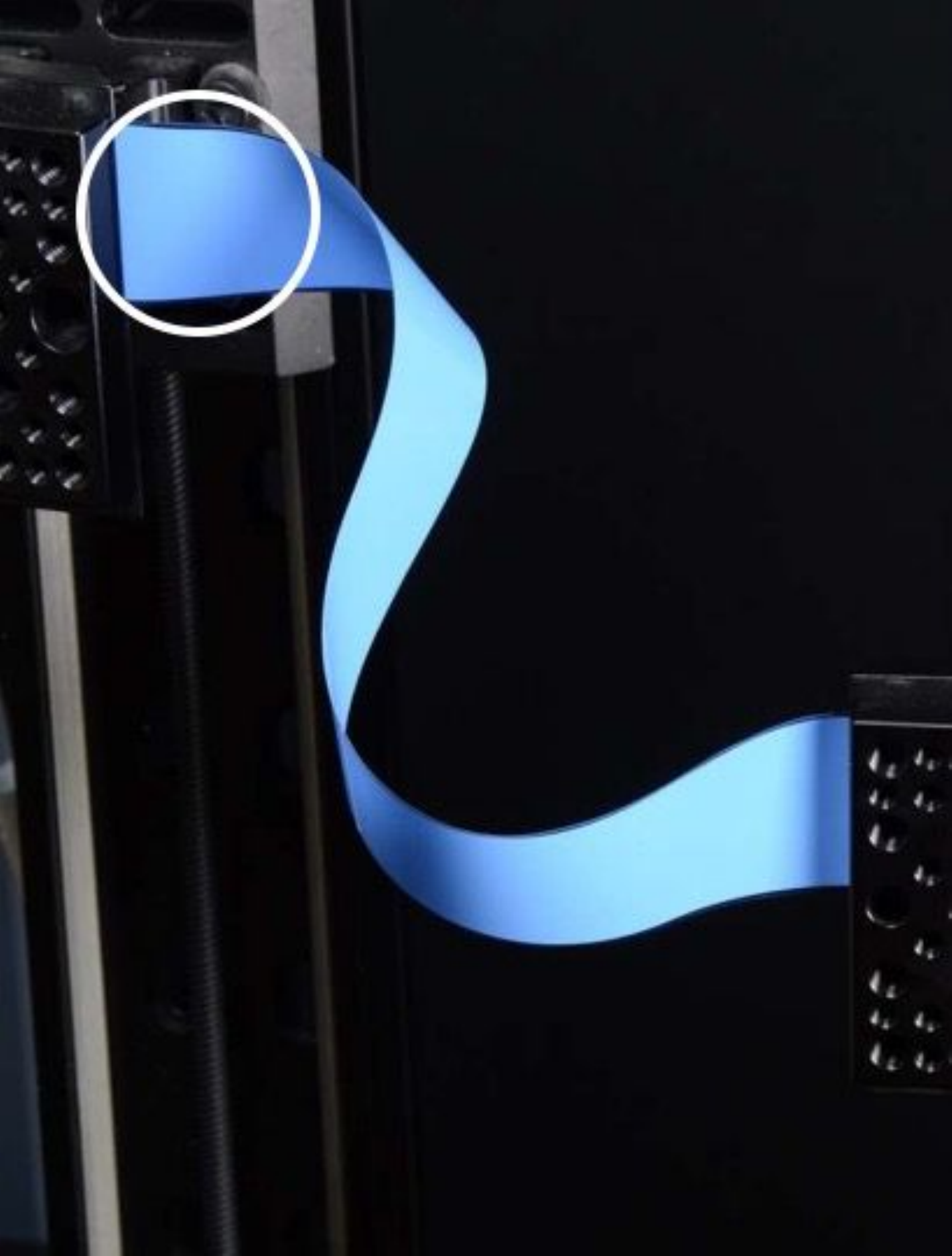}
		\caption{$US2+$}
	\end{subfigure}     \\
	\vspace{10pt}
	\begin{subfigure}[t]{0.13\textwidth}
		\centering
		\includegraphics[width=0.8in]{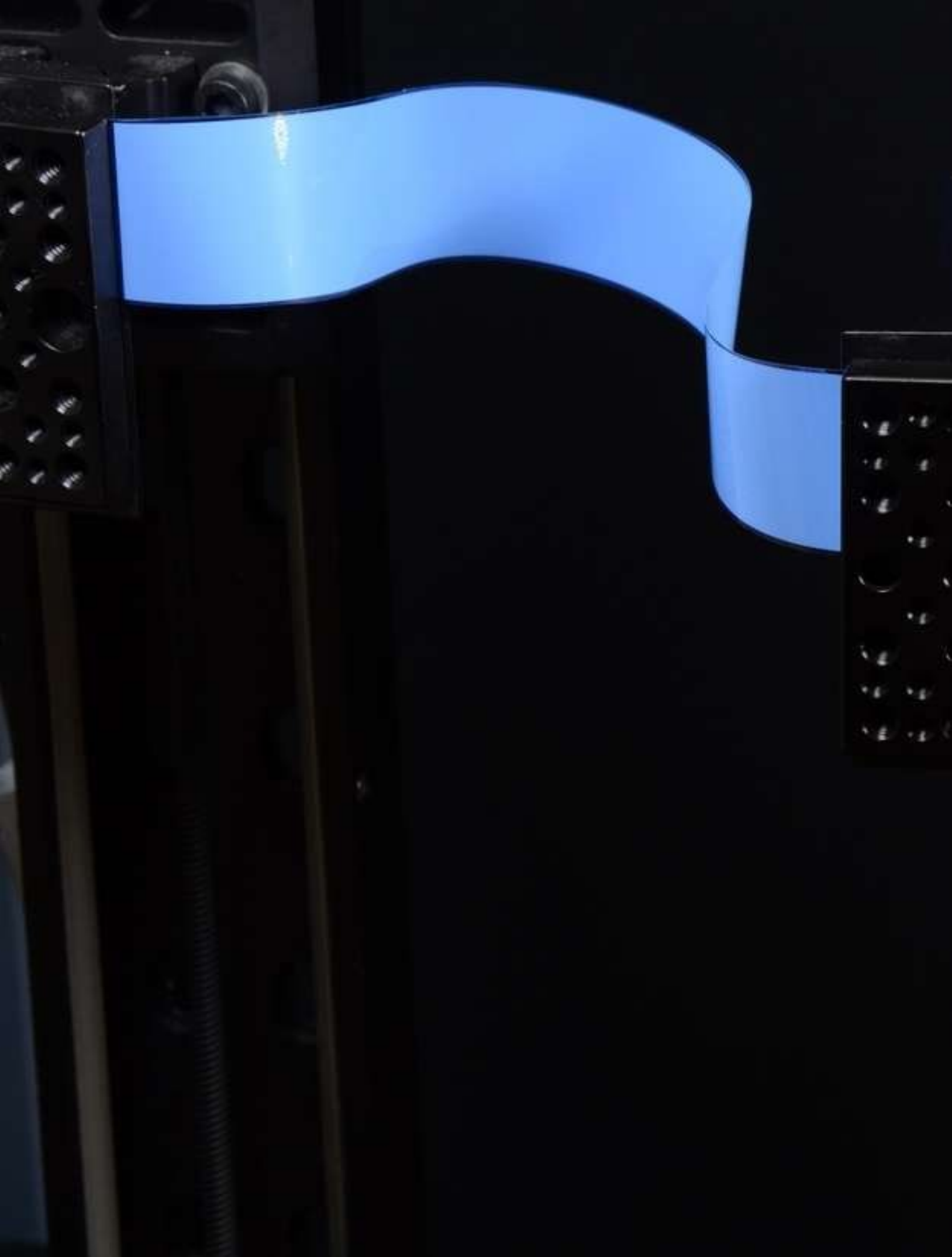}
		\caption{$W$}
	\end{subfigure}%
	\hspace{1pt}
	\begin{subfigure}[t]{0.13\textwidth}
		\centering
		\includegraphics[width=0.8in]{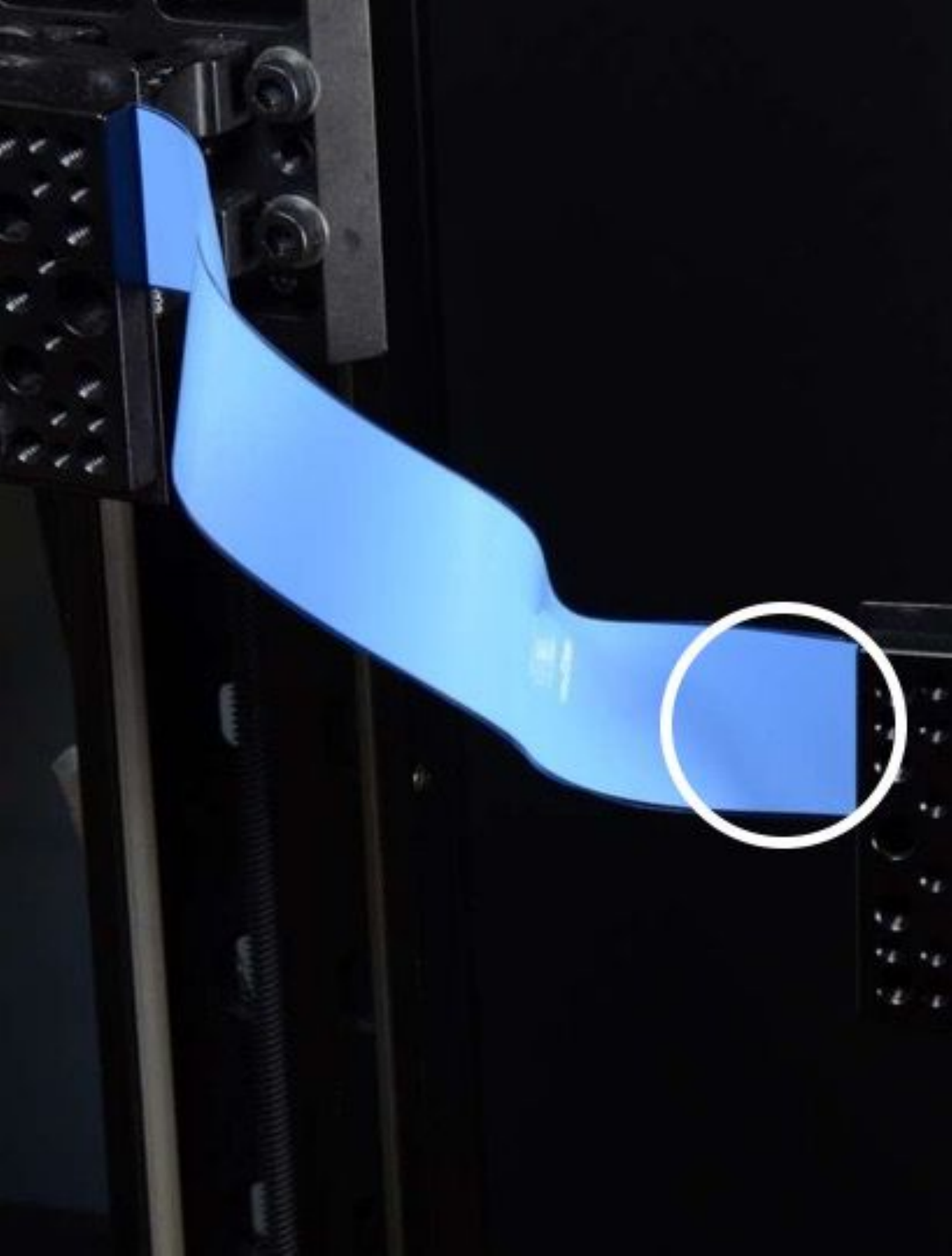}
		\caption{$US1-$}
	\end{subfigure} 
	\begin{subfigure}[t]{0.13\textwidth}
		\centering
		\includegraphics[width=0.8in]{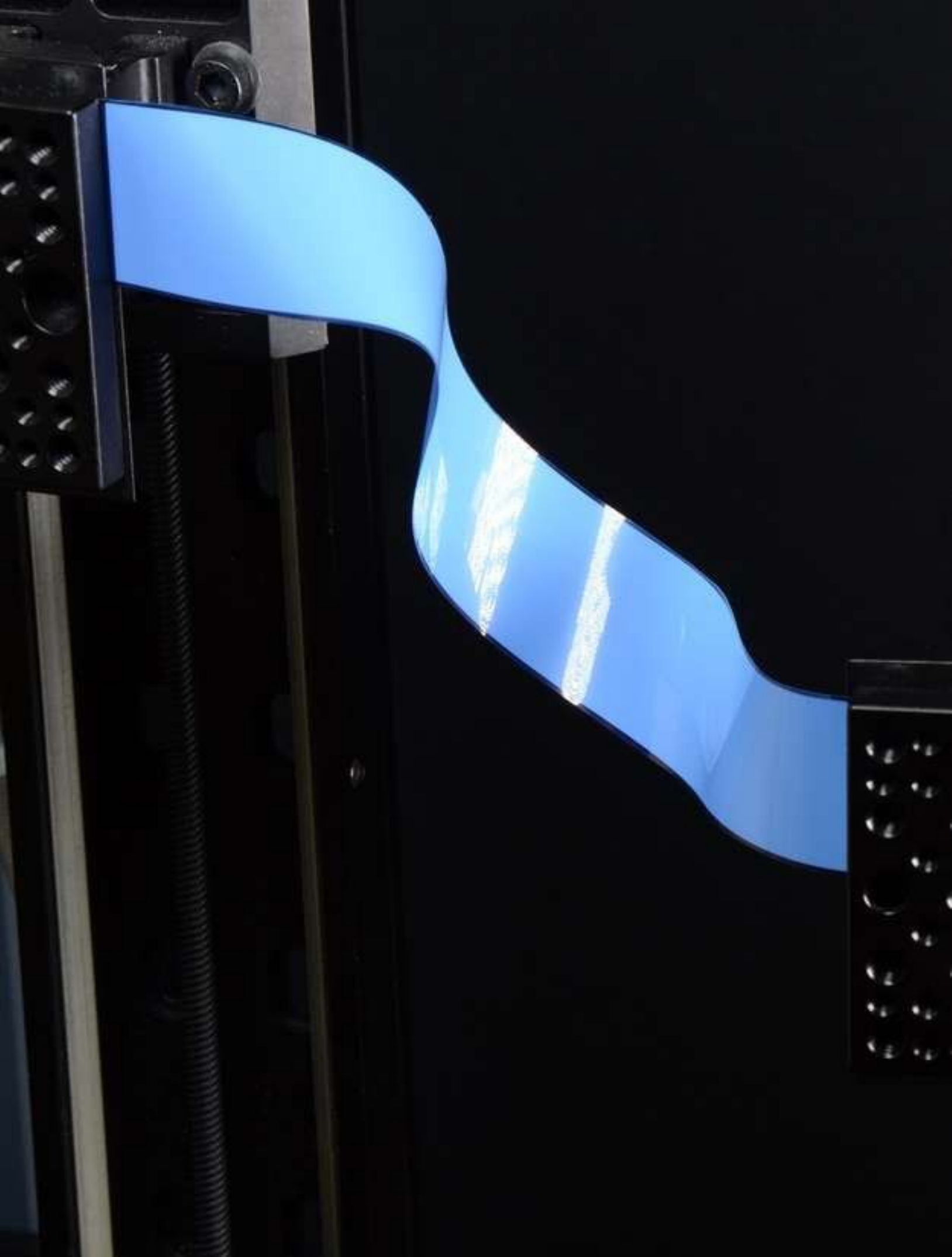}
		\caption{$WS-$}
	\end{subfigure}
	\begin{subfigure}[t]{0.13\textwidth}
		\centering
		\includegraphics[width=0.8in]{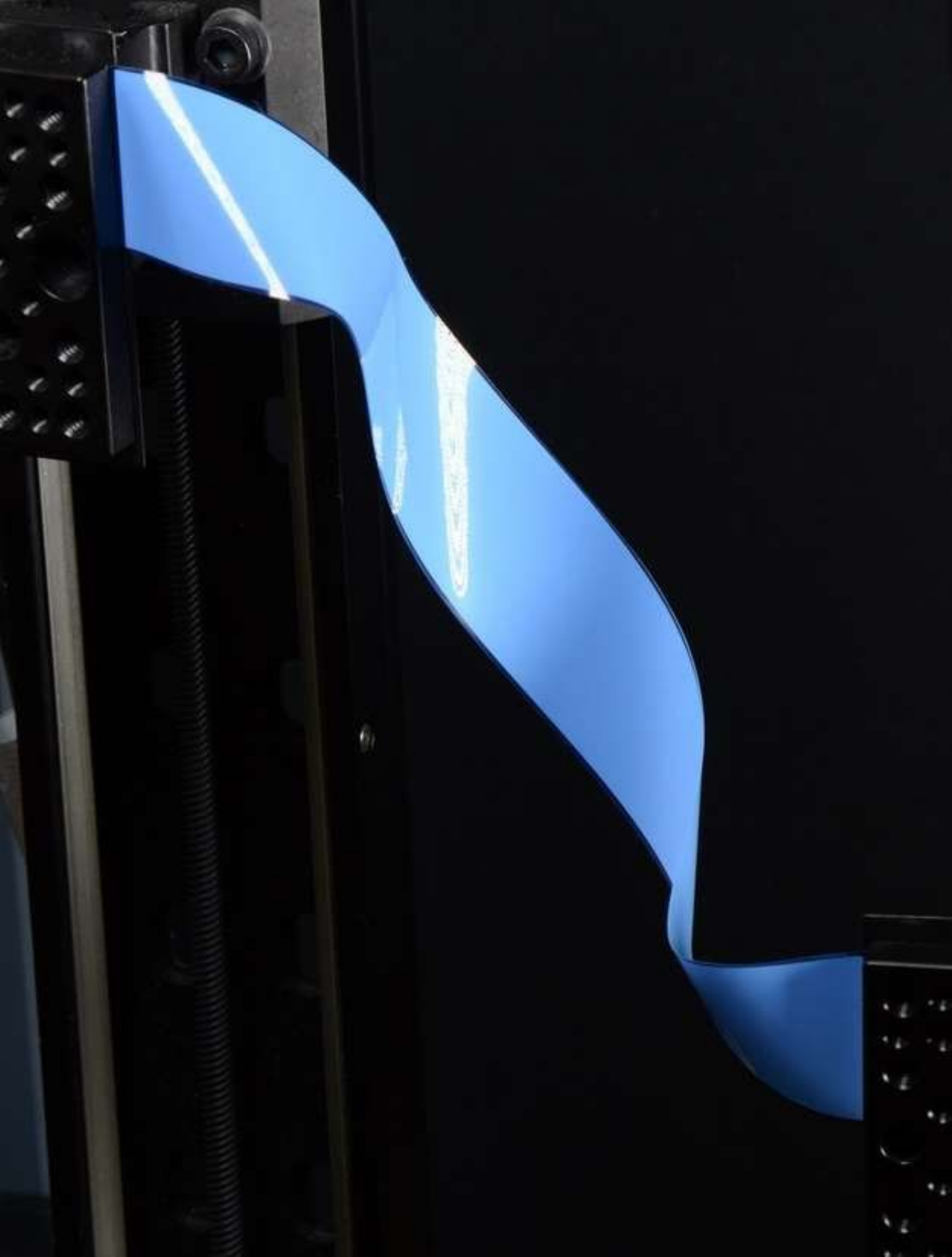}
		\caption{$uUui$}
	\end{subfigure}
	\begin{subfigure}[t]{0.13\textwidth}
		\centering
		\includegraphics[width=0.8in]{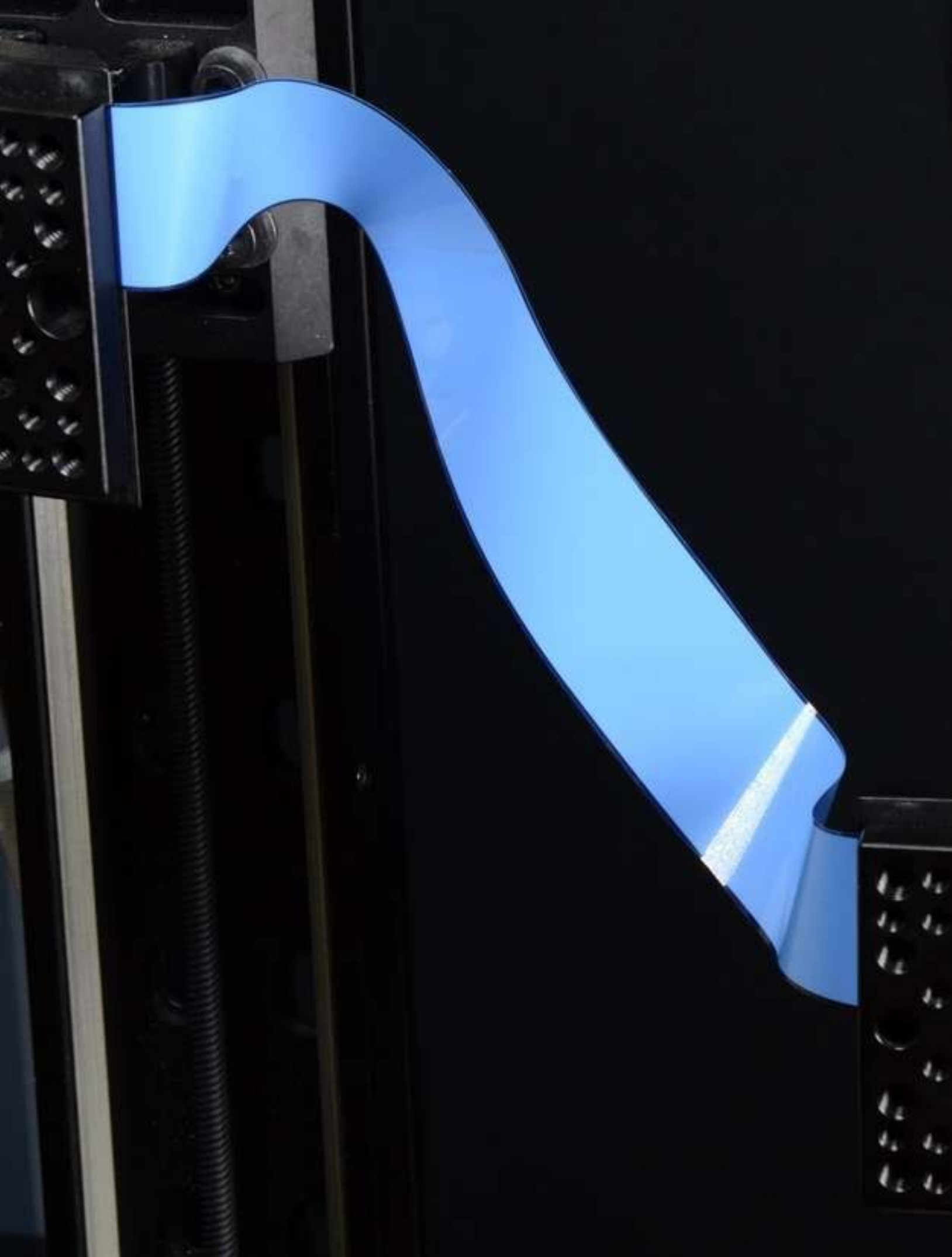}
		\caption{$wi$}
	\end{subfigure}
\hspace{0.13\textwidth}
	\begin{subfigure}[t]{0.13\textwidth}
		\centering
		\includegraphics[width=0.8in]{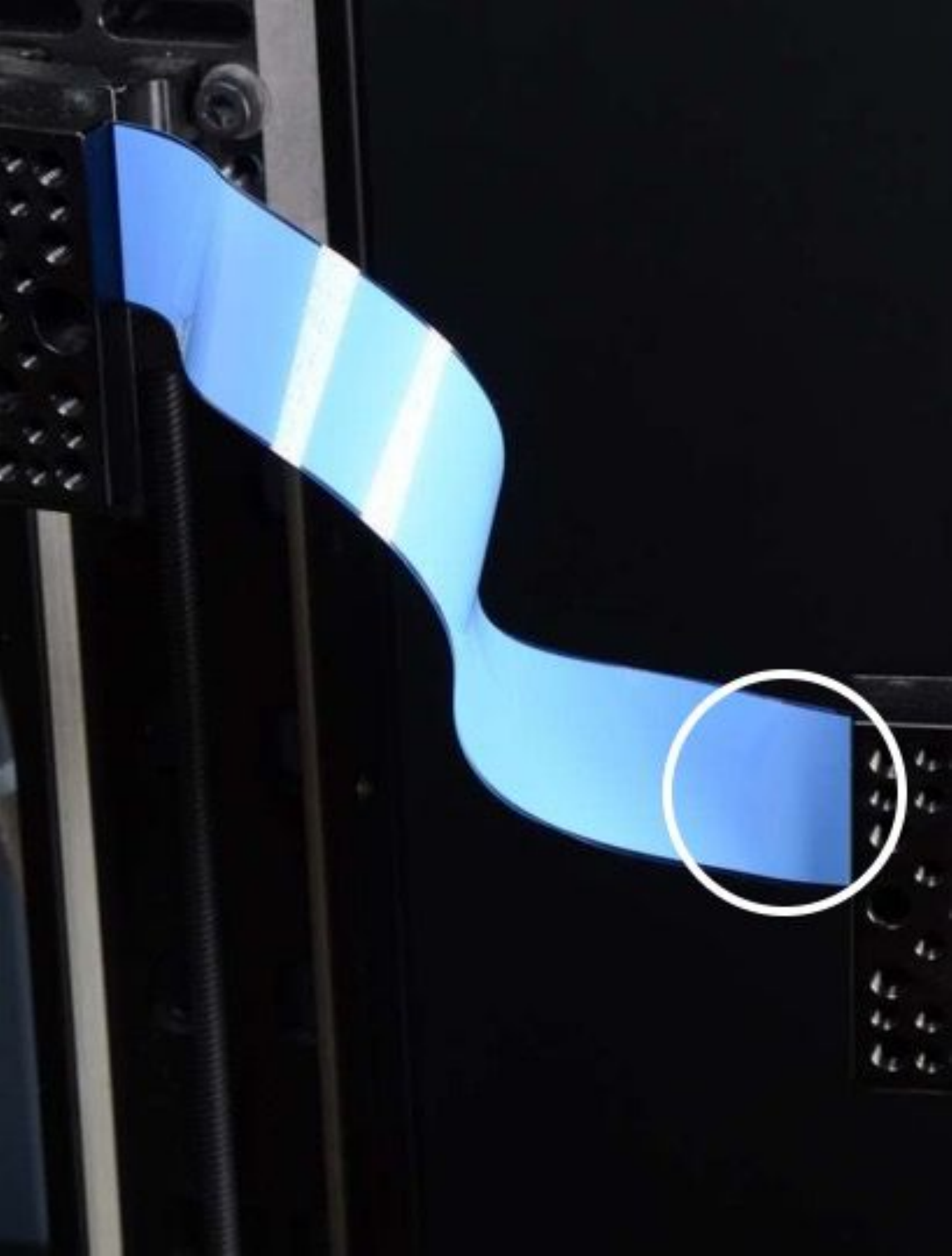}
		\caption{$US2-$}
	\end{subfigure}
	\caption{Some experimental configurations observed in wide bands, with aspect ratio $D/L=1/8$,  compression $\Delta L/L=1/2$, clamping angle $\psi_0 = 5^{\circ}$, and various values of shear $\Delta D / L$.  Twisted states are not shown.
	 Gravity is roughly vertical in these images.
The $US1\pm$ and $US2\pm$ states are separated by weak local jumps in the circled regions.  At very low clamping angles, there is a similar separation of $WS\pm$ into $WS1\pm$ and $WS2\pm$ states, but these have already merged at this clamping angle. 
The $U1$ and $U2$ states will become connected at higher clamping angles. The red arrows on the $U2$ state point at regions of focused curvature.  }
	\label{fig:WidExpConfiguration}
\end{figure}

In this section, we present experimental results on wide bands ($D/L = 1/8$, $1/4$) and compare them with the narrow band ($D/L = 1/80$) experiments and the anisotropic Kirchhoff rod model for a couple of choices of clamping angle ($\psi_0 = 0^{\circ}$, $15^{\circ}$) at the same compression $\Delta L/L = 1/2$.
The effects of gravity become less important as the width of the band increases.
The behavior of twisted states for intermediate width bands is quite complicated, including the appearance of new stable states and self-contact.  Reserving a deeper exploration for future study, we leave this behavior out of the present discussion, other than to present later a few examples of twisted states for various intermediate width bands in Figure \ref{fig:WidthEffect}.  Also shown there are indented states, which become possible for very wide bands, and are another topic we reserve for future study.  For the wider $D/L = 1/4$ bands, no twisted states are observed.
Aside from twisted states, several new states appear in wide bands, but these still appear to be related to the states we have already seen for narrow bands, and we can attempt to organize all the results around the Kirchhoff solutions.
The reversible symmetry properties of the Kirchhoff equations appear to persist in all of the experimentally observed wide band states.

\begin{figure}[h!]
	\centering
	\includegraphics[width=0.85\textwidth]{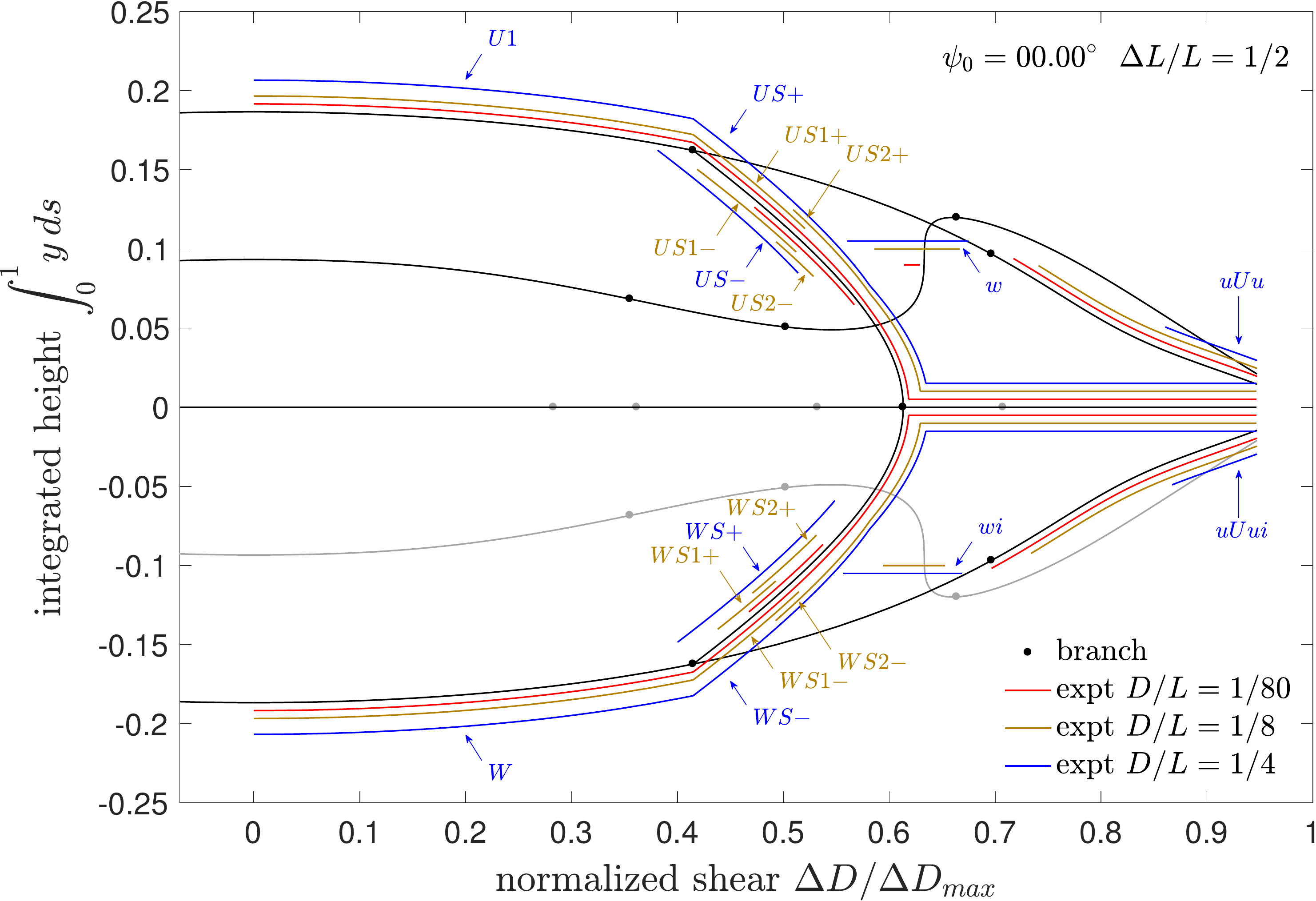}
	\caption{Experimental data (red, brown, and blue curves) from narrow and wide bands, normalized by limiting shear, for clamping angle $\psi_0=0^{\circ}$, compared with numerical solutions of the anisotropic rod equations (black and grey curves).  Twisted states are not included. The horizontal extent of the experimental curves is the range of stability (estimated error $\approx \pm 0.01\, \Delta D/L$), while the vertical position of the curves is not measured data, but is made to follow near the numerical curves whenever a comparison is possible.  No data was taken for $\Delta D/L < 0$.  The $US\pm$ states are split into two states for the intermediate-width bands.  For the widest bands, the pitchforks between $U1$ and $US\pm$ and $W$ and $WS\pm$ are only weakly broken by gravity, and transitions to either of the $\pm$ pair are observed.}
	\label{fig:WidExperi0}
\end{figure}

\begin{figure}[h!]
	\centering
	\includegraphics[width=0.85\textwidth]{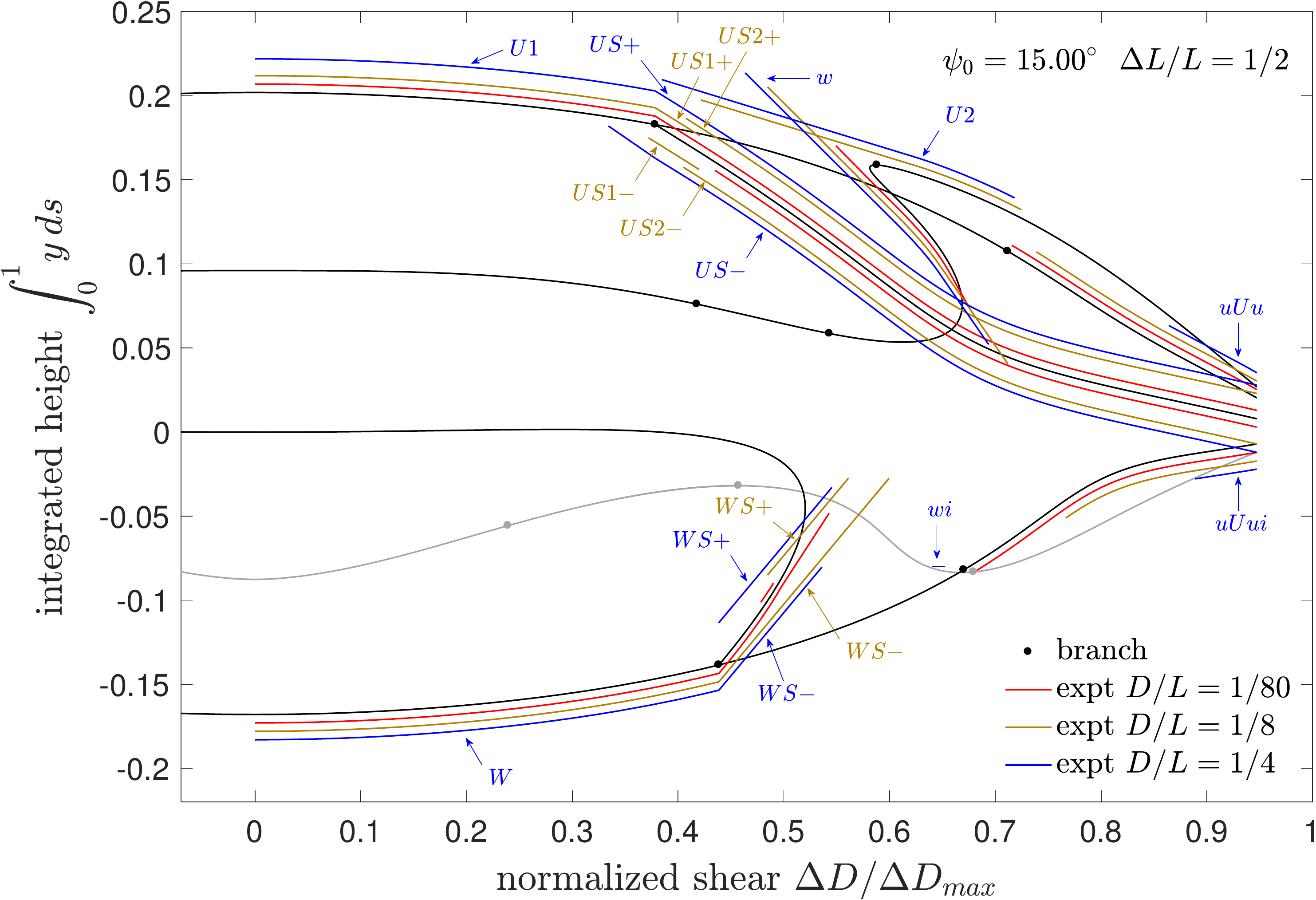}
	\caption{As in Figure \ref{fig:WidExperi0}, but with clamping angle $\psi_0=15^{\circ}$. The $U2$ state first appears at small nonzero clamping angles.  For the widest bands, the pitchforks between $U1$ and $US\pm$ and $W$ and $WS\pm$ are only weakly broken by gravity, and transitions to either of the $\pm$ pair are observed.}
	\label{fig:WidExperi15}
\end{figure}

Figure \ref{fig:WidExpConfiguration} shows some of the states we observe in wide bands for a band of aspect ratio $D/L = 1/8$, a shallow clamping angle $\psi_0 = 5^{\circ}$, and various values of shear $\Delta D / L$.  We name the states with reference to those found in narrow bands.  The $U$ state is now called $U1$, as there is a $U2$ branch stabilized at higher shears; the two will eventually connect at higher clamping angles.  Increasing the shear in the $U2$ state leads to focusing of generators and bending energy, as indicated by the two red arrows.  The intermediate width bands ($D/L = 1/8$) now feature two sets of $US\pm$ states, which are separated by weak local jumps in the circled regions.  At very low clamping angles, there is a similar separation of $WS\pm$ into $WS1\pm$ and $WS2\pm$ states, but these have already merged with each other into a single $WS\pm$ set at this small clamping angle.  Twisted states are also present, but not shown in this figure.  The inverted version of the $w$ state, the $wi$ state, is observed in wide bands.

 To compare bands of different width, we now normalize the shear $\Delta D$ using the limiting shear for a band of given width, as discussed in Appendix \ref{limitstates}.  We have $\Delta D_{max} \approx 207.85$ mm for $D/L=1/80$, $\Delta D_{max}=180.00$ mm for $D/L=1/8$ and $\Delta D_{max} \approx 156.33$ mm for $D/L=1/4$.
Figures \ref{fig:WidExperi0} and \ref{fig:WidExperi15} show some solution manifolds for the perfectly anisotropic Kirchhoff rod, along with experimental data for narrow and wide bands for clamping angles $\psi_0=0^{\circ}$ and $\psi_0=15^{\circ}$, respectively.  Twisted states are not included.  While the $U2$ state is not present at zero clamping angle, it begins to exist at small nonzero clamping angles, and can be seen in Figure \ref{fig:WidExperi15}.  Based on its shape, it seems to correspond to a state observed in the Kirchhoff solutions but which is experimentally unstable for narrow bands.  For wide bands, the $w$ state can be reached by gently poking the stable $U2$ state so that it buckles inwards near its midpoint.  At both clamping angles, it is clear that increasing the width of the band stabilizes the $w$ and $wi$ states and destabilizes the $uUu$ and $uUui$ states.  Recall that the $w$ state does not appear for the Kirchhoff solutions until a small nonzero clamping angle.  The states all seem to follow the Kirchhoff rod backbone, but there is clearly an additional jump (probably two folds?) separating the $US1\pm$ and $US2\pm$ states, and another separating the $WS1\pm$ and $WS2\pm$ states, for intermediate width bands ($D/L = 1/8$) at sufficiently low clamping angles.  This effect is shown in the supplementary video {\texttt{widtheffect15.mp4}} \cite{videos}.
 For the widest bands ($D/L=1/4$), the pitchforks between $U1$ and $US\pm$ and $W$ and $WS\pm$ are only weakly broken by gravity, and transitions to either of the $\pm$ pair are observed, in contrast to the consistently biased choices made by narrow bands.

\begin{figure}[h!]
	\captionsetup[subfigure]{labelformat=empty}
	\begin{subfigure}[t]{0.18\textwidth}
		\includegraphics[width=0.9\textwidth]{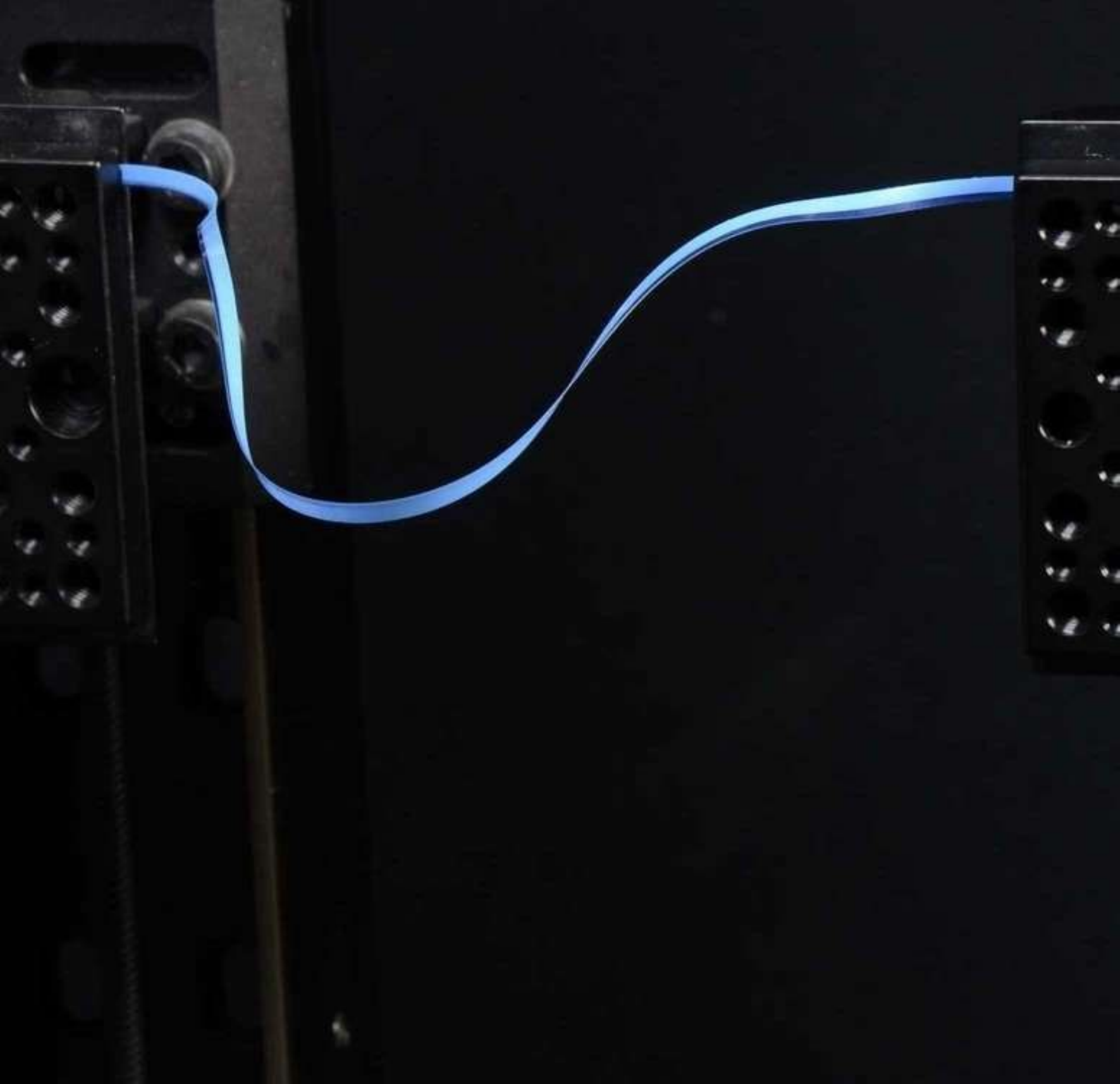}
		\caption{$D/L=1/80$}
	\end{subfigure}%
	\hspace{0.3pt}
	\begin{subfigure}[t]{0.18\textwidth}
		\includegraphics[width=0.9\textwidth]{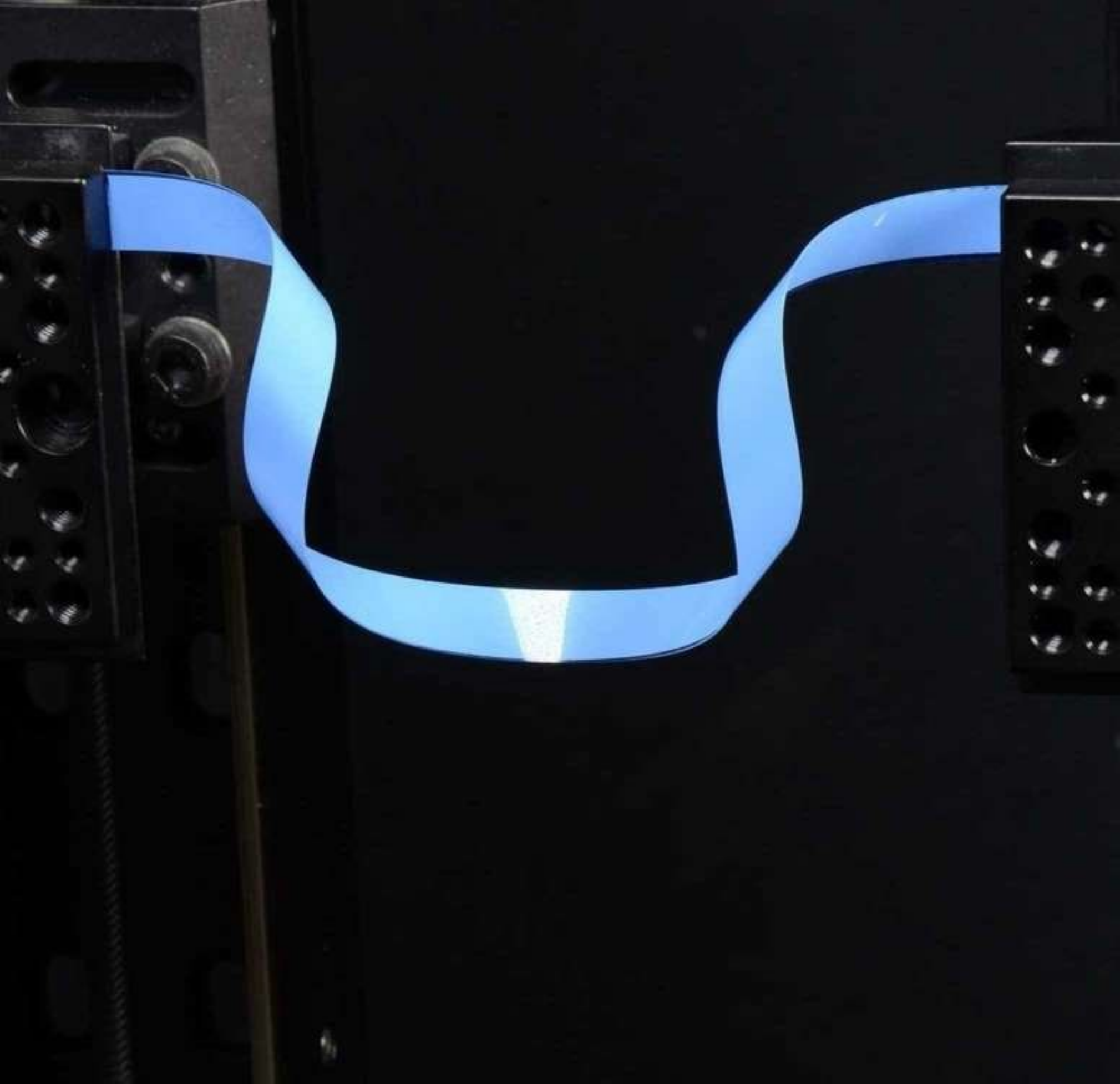}
		\caption{$D/L=1/24$}
	\end{subfigure}
	\begin{subfigure}[t]{0.18\textwidth}
		\includegraphics[width=0.9\textwidth]{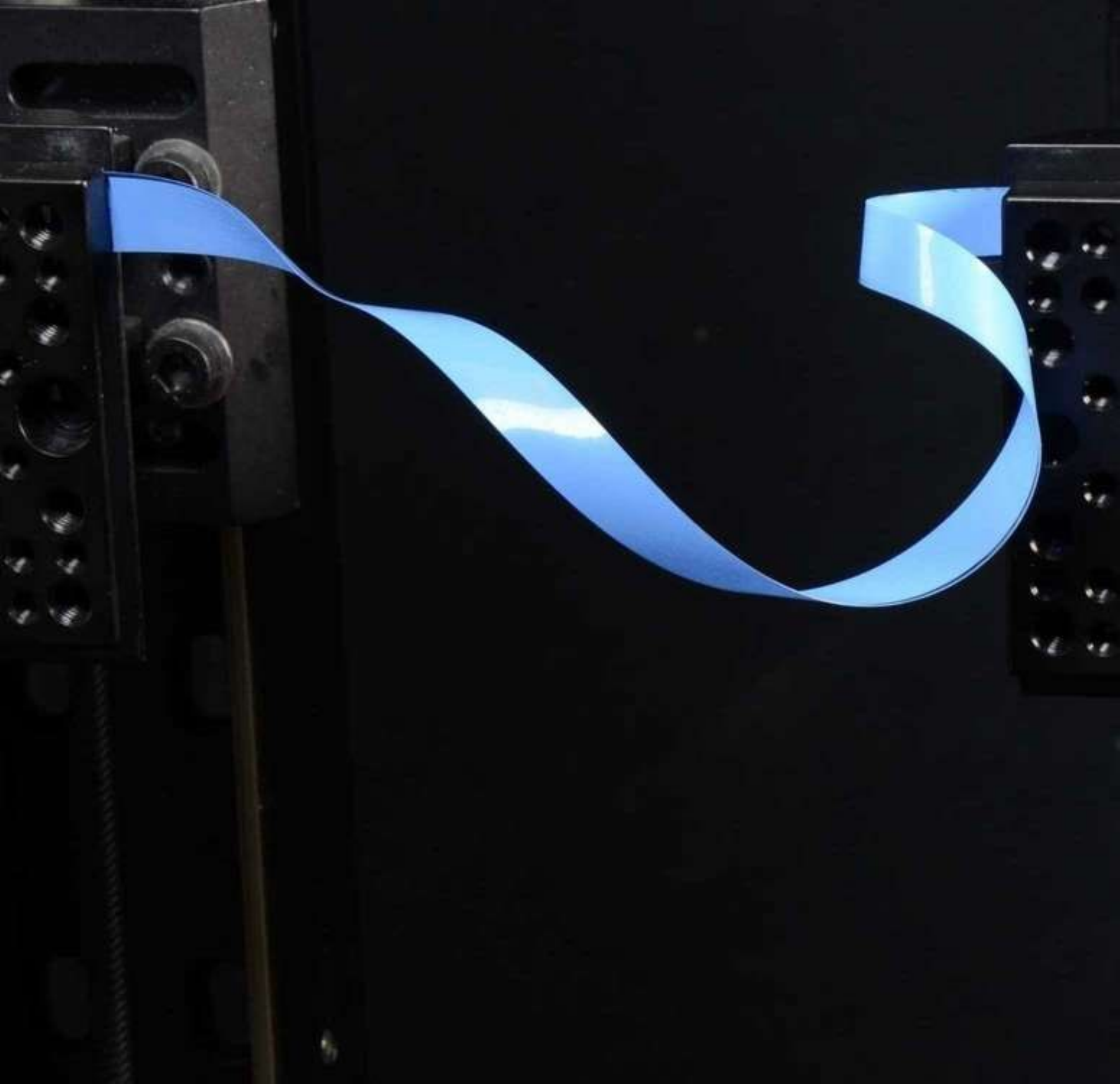}
		\caption{$D/L=1/24$}
	\end{subfigure}
	\begin{subfigure}[t]{0.18\textwidth}
		\includegraphics[width=0.9\textwidth]{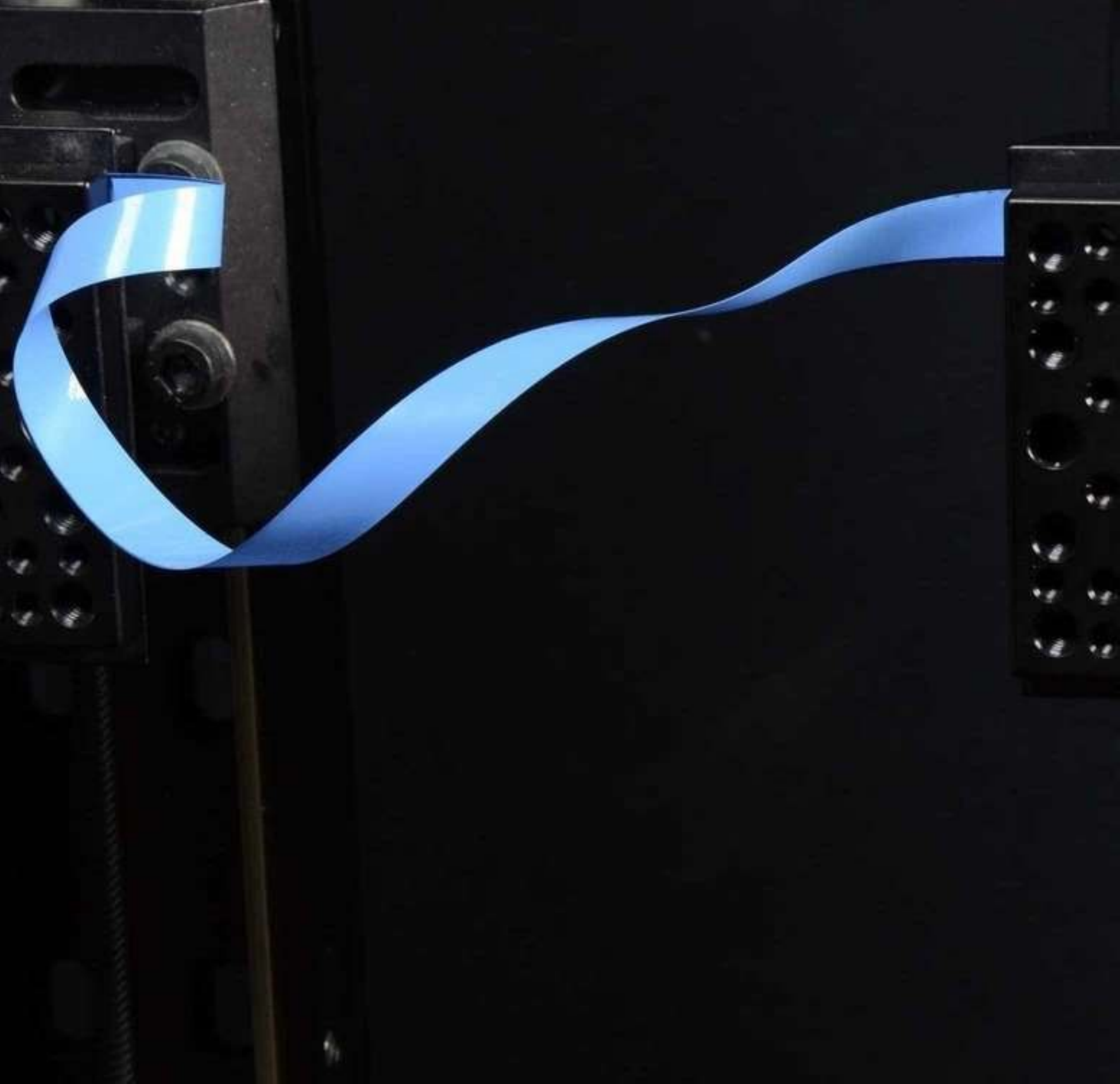}
		\caption{$D/L=1/24$}
	\end{subfigure}
	\begin{subfigure}[t]{0.18\textwidth}
		\includegraphics[width=0.9\textwidth]{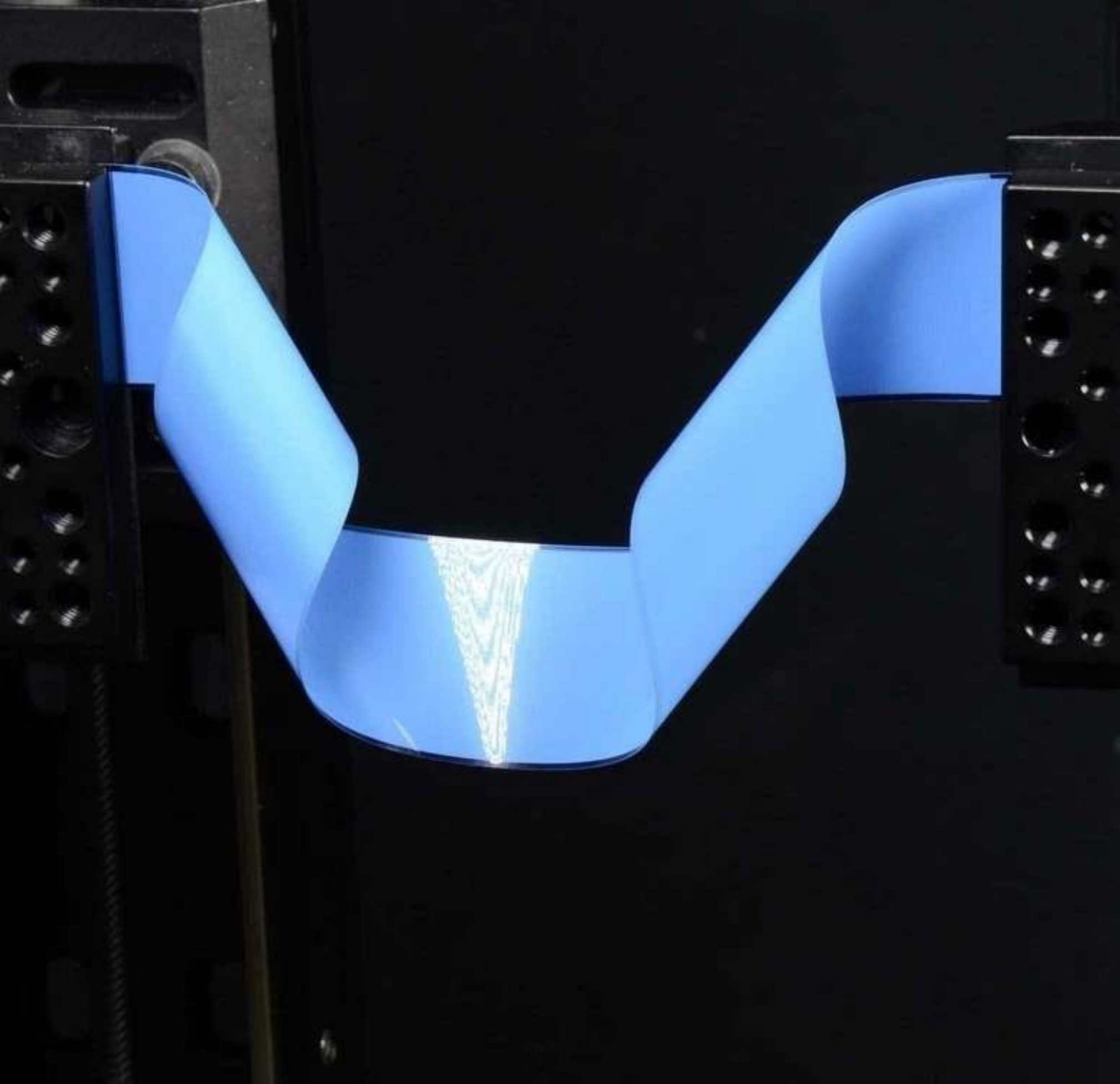}
		\caption{$D/L=1/8$}
	\end{subfigure} \\
		\vspace{10pt}
	\begin{subfigure}[t]{0.3\textwidth}
		\includegraphics[width=0.9\textwidth]{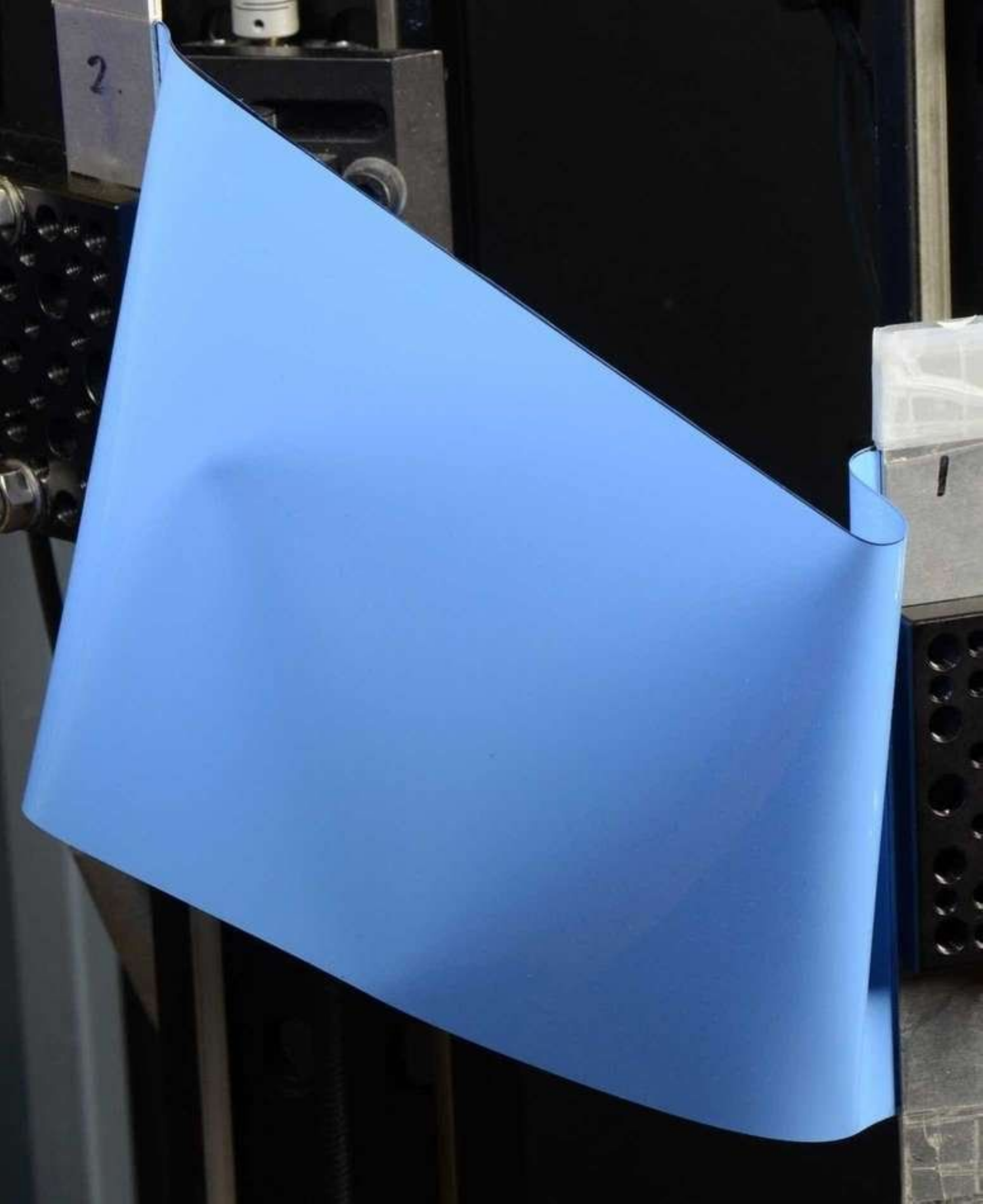}
		\caption{$D/L=1/2$}
	\end{subfigure}
	\begin{subfigure}[t]{0.3\textwidth}
		\includegraphics[width=0.9\textwidth]{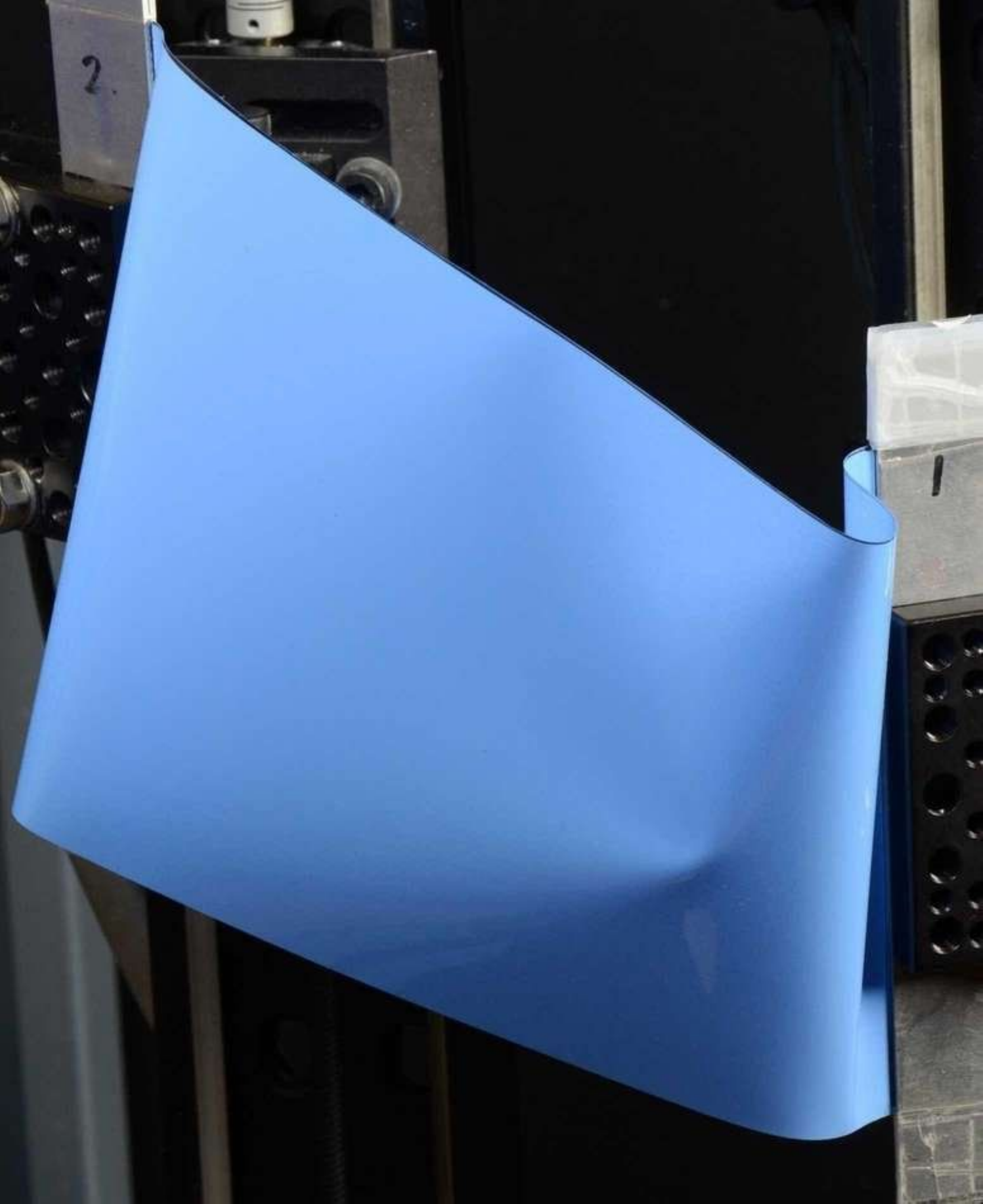}
		\caption{$D/L=1/2$}
	\end{subfigure}
	\caption{Examples of states stabilized by width effects.  Twisted states, different from the narrow band state shown in the upper left, are stabilized and then destabilized or eliminated by self-contact issues by increasing width.  Each twisted state shown is a $-$ state, and there exist corresponding $+$ states.  A pair of indented states exist for very wide bands; the elastic defects are stable only for a very narrow range of shear.  All bands are compressed to $\Delta L/L=1/2$; the bands with $1/80 \leq D/L \leq 1/8$ are clamped at $\psi_0=0^{\circ}$ and the very wide band with $D/L=1/2$ is clamped at $\psi_0=30^{\circ}$. }
	\label{fig:WidthEffect}
\end{figure}

At slightly higher clamping angles, the $U1$ and $U2$ states become connected for wider bands.  This is in contrast to the Kirchhoff solutions and narrow band data, for which the $U1$ state (called $U$ in the prior Section \ref{results}) merges with the $w$ state.  At higher clamping angles, shearing the $U1$-$U2$ state will lead to two successive snap-throughs to the $w$ and $uUui$ states.  
These transitions are shown in the supplementary video {\texttt{widtheffect45.mp4}}  \cite{videos}, where it can be clearly seen that the violence of the first snap-through increases with increasing width.  In general, greater width exaggerates the effect of energy focusing of the generators of a developable strip \cite{korte2010triangular,starostin2015equilibrium,chopin2016disclinations}; the snap-through transitions release some of this stored bending energy.   
The limiting shear is approached \emph{via} one of four possible limiting states, either the $S$-like $US\pm$ pair or the $U$-like $uUu$ and $uUui$ states.  These contain highly focused conical singularities near their ends, as shown in Appendix \ref{limitstates}.

No other new states were observed in experiments at $\psi_0=30^{\circ}$, $45^{\circ}$, or $60^{\circ}$, although interesting changes in connectivity do occur, one of which is shown in 
the supplementary video {\texttt{widtheffect45.mp4}}  \cite{videos}.

We briefly mention some other states that are stabilized by width effects, some of which are shown in Figure \ref{fig:WidthEffect}.  At a width $D/L = 1/24$, the narrow band $D/L = 1/80$ twisted state pair is replaced by three other $\pm$ pairs. These states are sensitive to boundary conditions; one can play by hand with a band of $D/L = 1/40$ and see all eight twisted states by changing compression and clamping angle.  Depending on the state, application of shear may lead to a snap-through, a looped structure, or self-contact.  At a width of $D/L = 1/8$, the intermediate-width $\pm$ pair has become unstable,
 and at higher widths the symmetric intermediate-width state requires self-contact.
For very fat bands, such as the $D/L = 1/2$ bands shown in in the figure, there is a pair of stably indented states, each of which contains a pair of ``d-cones''.  These elastic defects are only stable for a very narrow range of applied shear, and upon decreasing or increasing shear will respectively annihilate or propagate through the structure to emerge through the boundaries, enabling a snap-through transition.
Both the twisted and indented examples suggest rich avenues of research, which we reserve for future work.

\section{Further discussion}\label{discussion}

We have presented experimental results on the stability of thin elastic bands subject to compression, shear, and symmetric clamping.  The Kirchhoff equations for perfectly anisotropic rods serve as a surprisingly good guide to the behavior of these bands, particularly when they are narrow in width.  We have explored only a limited region of the parameter space of boundary conditions and band geometry, but have already stumbled on many new stable configurations and jump phenomena.  Here we briefly discuss some confusing issues and avenues for future work.

First we note again that there is a distinct difference between a Kirchhoff rod, whose cross section remains orthogonal to its centerline, and a developable strip.  This difference is quite obvious in an image like Figure \ref{fig:UlikeRl}, where the director $\bm{d_1}$ associated with the slice of material perpendicular to the rod centerline is clearly not aligned with a straight line (generator) on the strip.  This ``rod'''s cross section is actually bending in the width direction.  This is why our renderings of the Kirchhoff solutions as strips representing the rod frame are not equivalent to renderings of isometrically deforming elastic strips.  This point is reiterated in Appendix \ref{bvp}.  However, the assumption of developability is itself problematic, particularly when applied to narrow strips.  We refer the reader to Appendix \ref{stripmodel} for a demonstration of the limitations of strip models in the present context.

From the behavior of wider bands, we might have expected that branch point 9 in the Kirchhoff solutions would be a pitchfork bifurcation with a stable $U2$-like state on one side.  We do not observe any such stable state for narrow bands.  For wider bands, the $U2$ state is somewhat shell-like, with synclastic curvature, and one can pop back and forth between $U2$ and the slightly indented $w$ state.  In contrast, width appears to destabilize the highly bent and twisted $uUu$ and $uUui$ states.  The stability of twisted states at low shear also shows a complicated dependence on width.
These and more complicated stabilization effects, such as the narrow range of boundary conditions allowing one to create defects by indenting very wide plate-like bands such as those shown in Figure \ref{fig:WidthEffect}, raise interesting questions about the boundaries between rod-like, plate-like, and shell-like behavior in thin sheets.
In order to capture such width effects, one needs either a model of a two-dimensional plate or strip, or a Cosserat rod model with a more complicated structure derived from such a two-dimensional model \cite{starostin2007shape, dias2014non, audoly2015buckling}.
In some such models, the width of the band can appear as a potential continuation parameter.  However, strip models lead to difficulties in numerical implementation because of the singular way in which they handle inflection points, which must be added by hand and cannot arise spontaneously during continuation.  These models will not admit rod-like solutions such as those shown in Figures \ref{fig:30states} and \ref{fig:jumpenergy}, where the twist, which for strips is identified with the torsion, does not vanish simultaneously with the curvature.  More importantly, without some further modification, such models cannot be used to continue solutions such as those of the $U$-$w$-$uUui$ branch, in which inflection points smoothly appear during deformation.


The full space of boundary conditions includes positions and general tilts in all directions, such that the director frames at the end points may take arbitrary values in the space of rotations SO(3).  In addition to energy barriers leading to multi-stability and jump phenomena, the non-simply-connected nature of SO(3) can also create topological barriers to deformation.  This fact-- related to the famous ``belt trick'' of Dirac-- implies a lower bound on elastic energy for rods subject to boundary conditions in which the end tangents are parallel \cite{baez1991topological}.  It would be of great interest to expand these results to arbitrary boundary conditions, which would require some consideration of how to define an appropriate linking number for open rods \cite{alexander1982ambiguous, van2000helical, van2007end, prior2016extended}.

Our elastic system has an analogue in the geometric controls literature, in which the orientation of the frame of our perfectly anisotropic rod appears as the orientation of a vehicle that can pitch and roll, but cannot yaw, and the elastic energy appears as a quadratic cost function for the two allowed controls \cite{Baillieul78}. However, without constraints on path length in the controls problem, the analogy is only strictly correct for an elastic setup in which the rod is not clamped, but is allowed to vary its length by sliding in or out of sleeves.
The ability to satisfy boundary conditions that favor a forbidden bending or steering rotation through an indirect combination of other allowed rotations reflects the fact that the commutator of two infinitesimal rotations in $\mathbb{E}^3$ provides the third.

\section*{Acknowledgments}
We thank G H M van der Heijden for helpful discussions, particularly regarding the treatment of singularities in Appendix \ref{stripmodel}.  We also thank M A Dias and T J Healey for helpful comments, and P S Krishnaprasad for the reference \cite{Baillieul78}.

\newpage

\appendix

\section{Limiting states}\label{limitstates}

\begin{figure}[h!]
	\captionsetup[subfigure]{labelfont=normalfont,textfont=normalfont}
	\begin{subfigure}[t]{0.43\textwidth}
		\centering
		\includegraphics[height=2.9in]{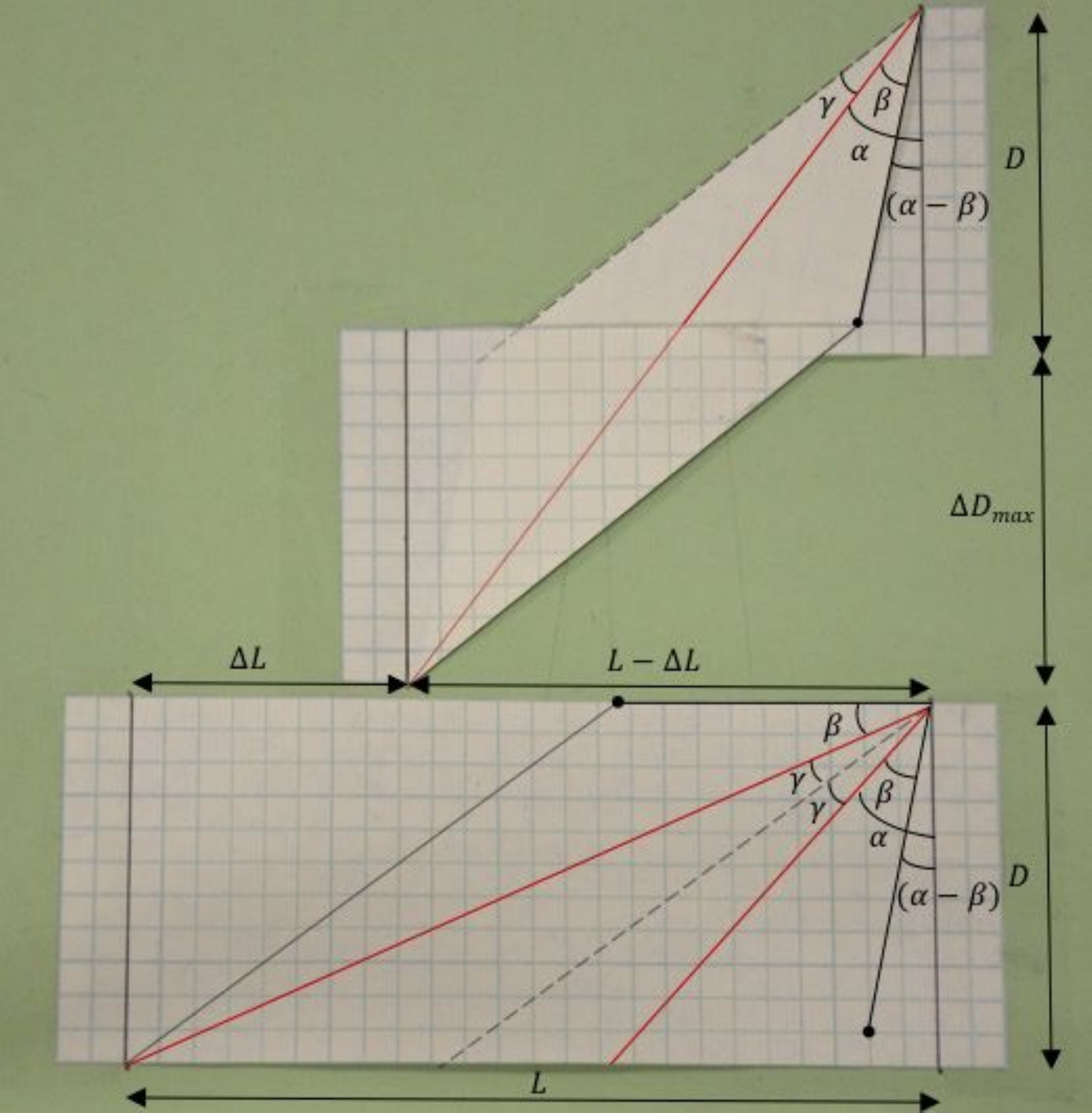}
		\caption{}\label{SlikePaper}
	\end{subfigure}
	\begin{subfigure}[t]{0.43\textwidth}
		\centering
		\includegraphics[height=2.9in]{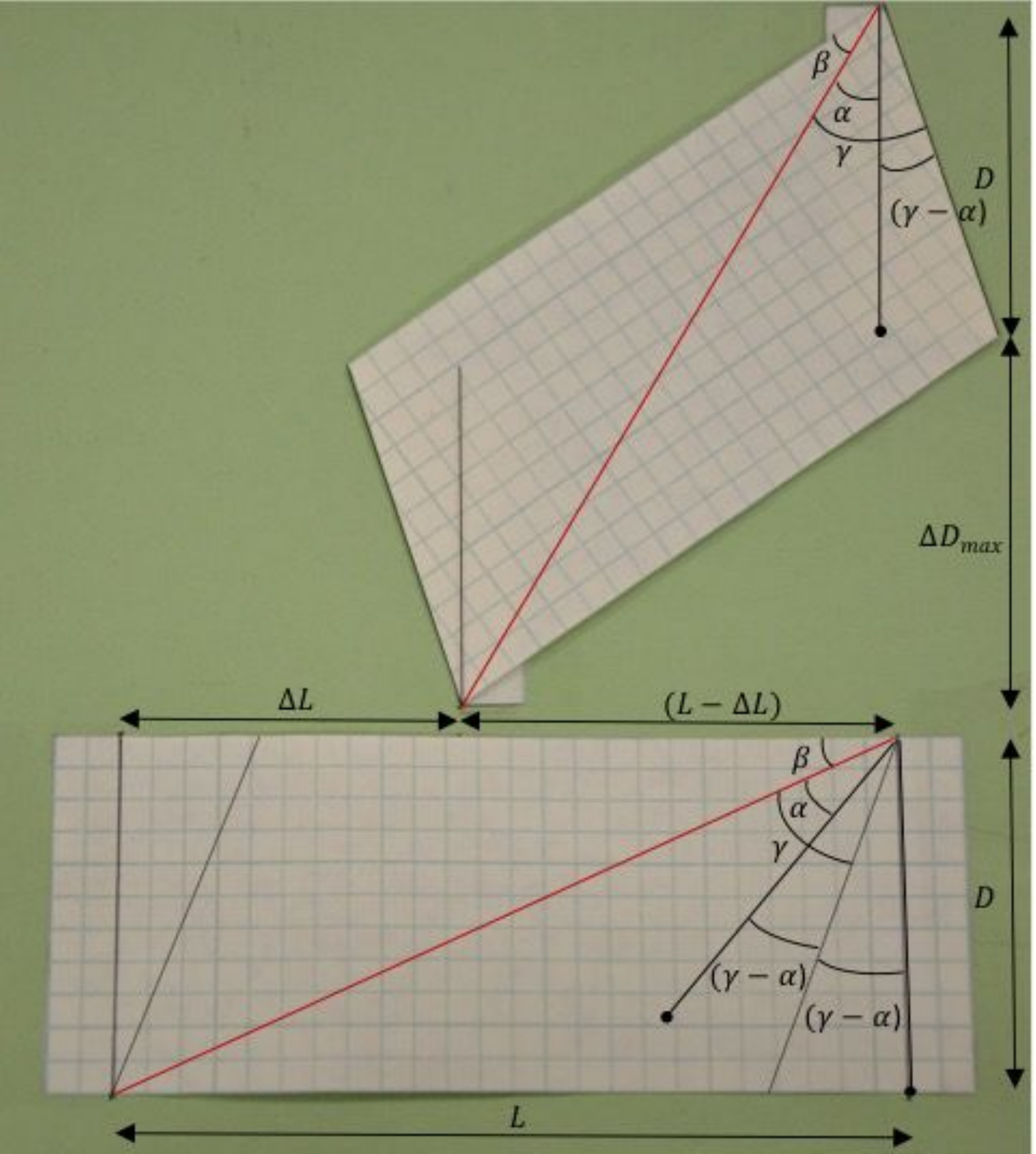}
		\caption{}\label{UlikePaper}
	\end{subfigure}
	\begin{subfigure}[t]{0.12\textwidth}
		\centering
		\includegraphics[height=1.5in]{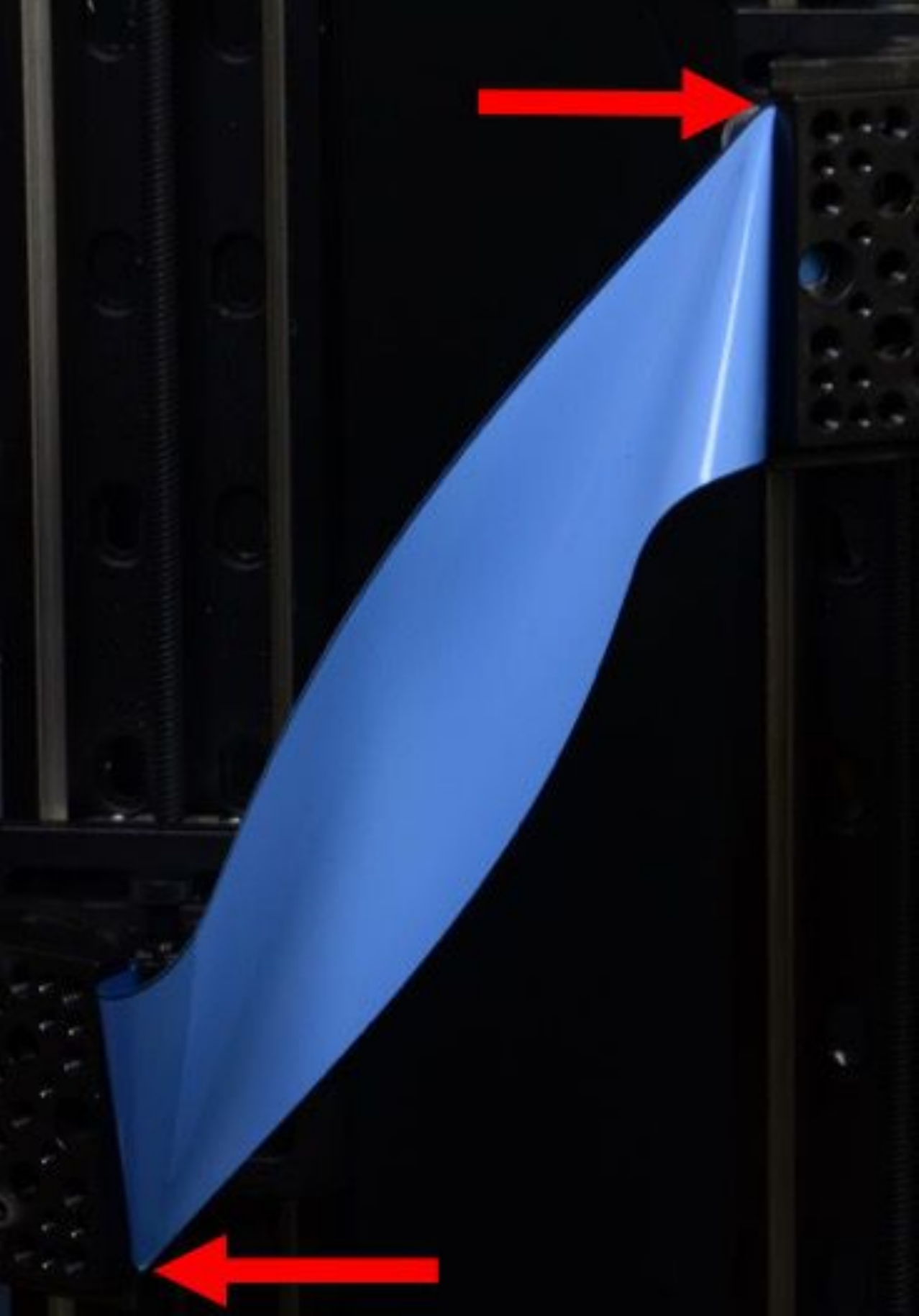}
		\caption{}\label{cone}
	\end{subfigure}
	\caption{Folded paper models representing (\subref{SlikePaper}) a limiting $S$-like state for $\psi_0=0^{\circ}$ and (\subref{UlikePaper}) a limiting $U$-like state for $\psi_0=180^{\circ}$.  Clamping occurs on the vertical lines located two squares in from the ends of the gridded paper. The folding process brings red lines and black dots on the bottom reference strips into coincidence (note that there is slight difference in scale between the left and right images).  (\subref{cone}) An experimental $uUui$ state for a clamping angle of $30^{\circ}$.  Red arrows point at incipient conical singularities.}\label{PaperModel}
\end{figure}

The shearing process tends towards limiting states, past which the sheet cannot deform without stretching somewhere.
Figures \ref{SlikePaper}-\subref{UlikePaper} show folded paper models representing a limiting $S$-like state for $\psi_0=0^{\circ}$ and a limiting $U$-like state for $\psi_0=180^{\circ}$.  The models are flat-foldable for these choices of clamping angle, but more general models can be made.  
Figure \ref{cone} shows an experimental $uUui$ state (a $U$-like state from the $W$ family) for a clamping angle $\psi_0=30^{\circ}$.  Red arrows point at incipient conical singularities forming near the clamps.  In the limit, these cones share a single straight line generator formed from a band diagonal (red lines) that sets the limiting shear to 
\begin{align}\label{limitshear}
\Delta D_{max}=\sqrt{L^2 + D^2 -(L-\Delta L)^2}-D \, .
\end{align}

\section{Numerical continuation of the Kirchhoff equations}\label{bvp}

\begin{figure}[h!]
	\captionsetup[subfigure]{labelfont=normalfont,textfont=normalfont}
	\centering
	\begin{subfigure}[t]{0.18\textwidth}
		\centering
		\includegraphics[height=1.3in]{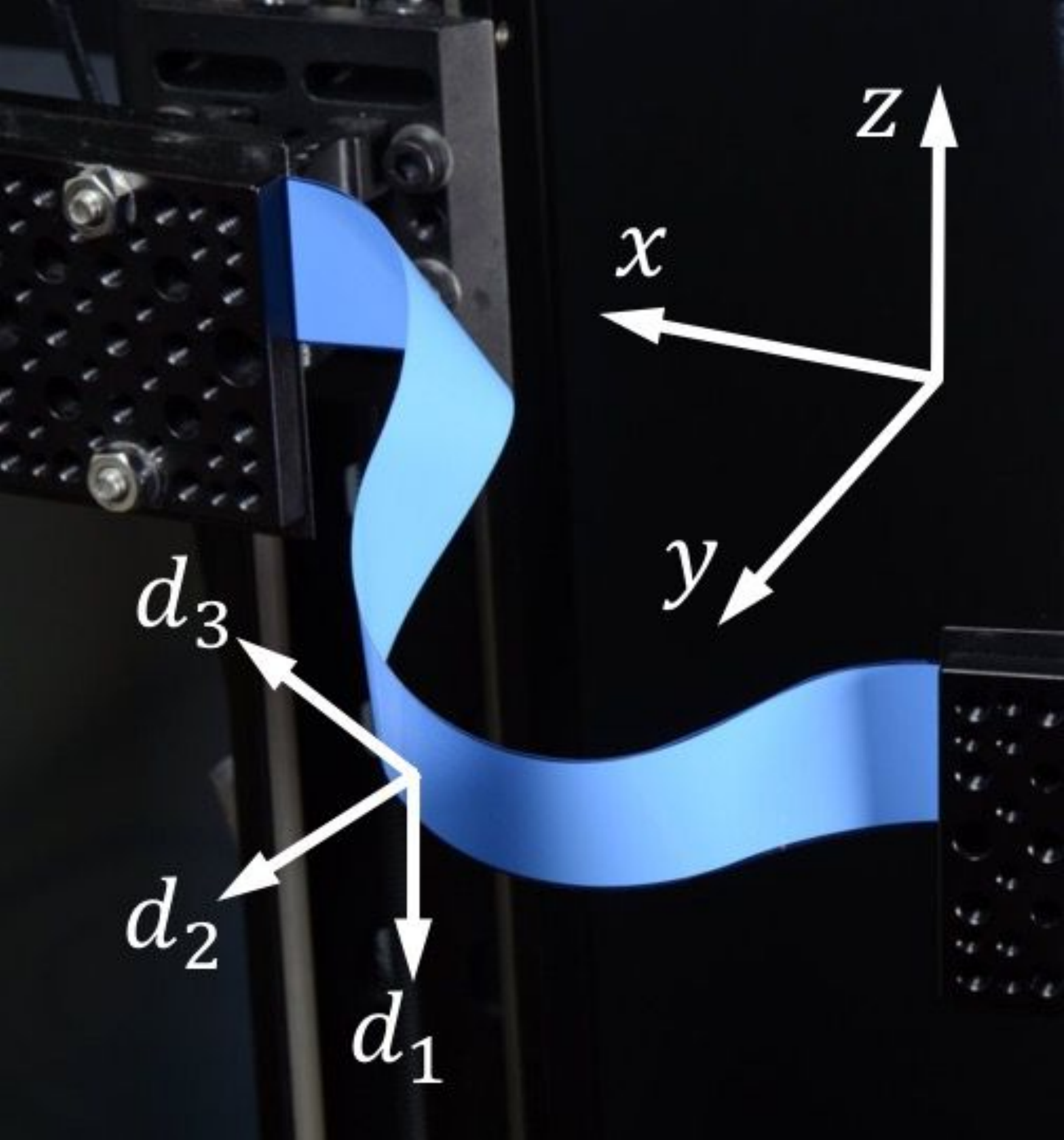}
		\caption{}\label{fig:framecoordinate}
	\end{subfigure}\hspace{.1in}
	\begin{subfigure}[t]{0.18\textwidth}
		\centering
		\includegraphics[height=1.3in]{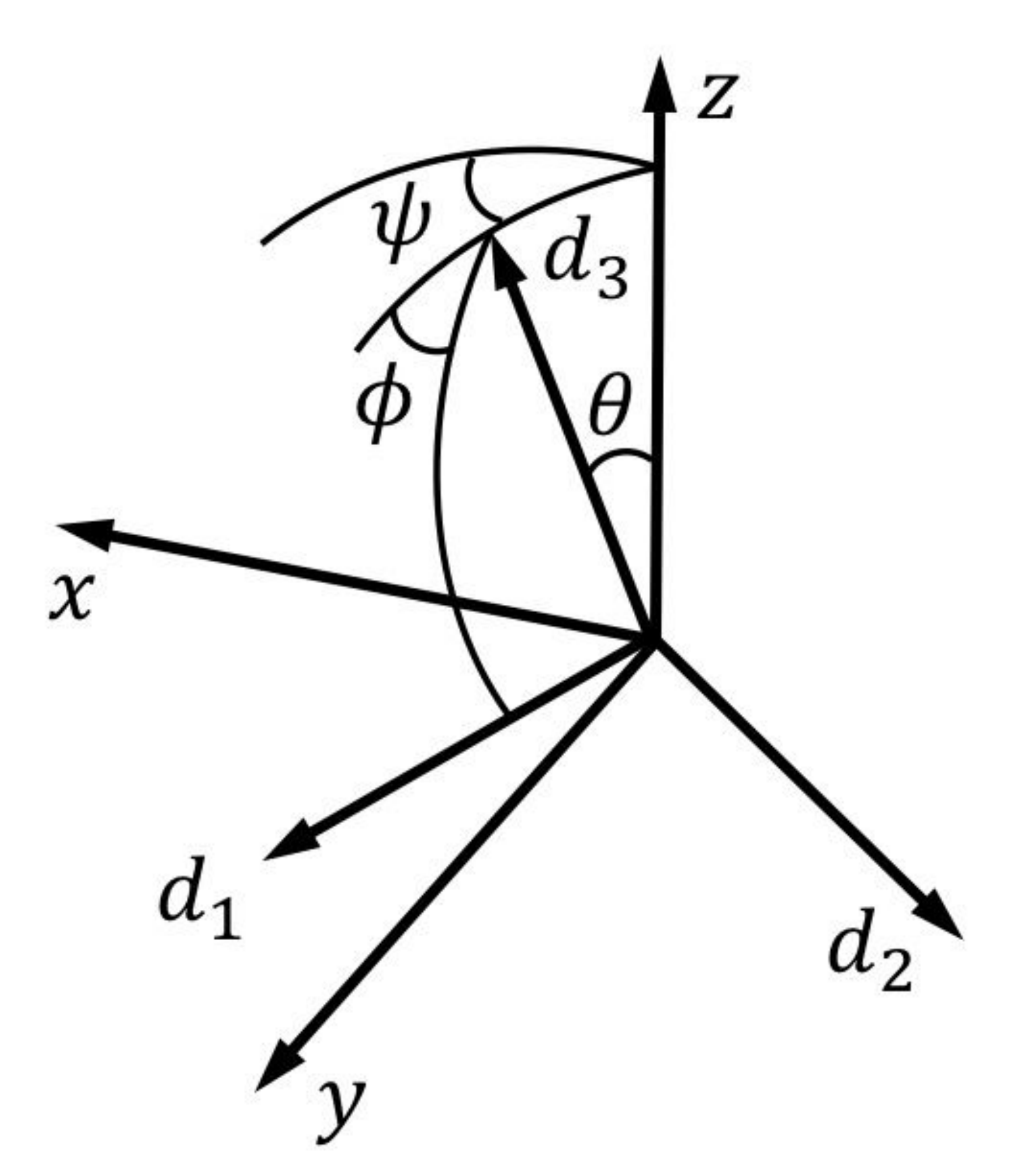}
		\caption{}\label{fig:eulerangle}
	\end{subfigure} \hspace{.1in}
	\begin{subfigure}[t]{0.54\textwidth}
	\centering
	\includegraphics[height=1.3in]{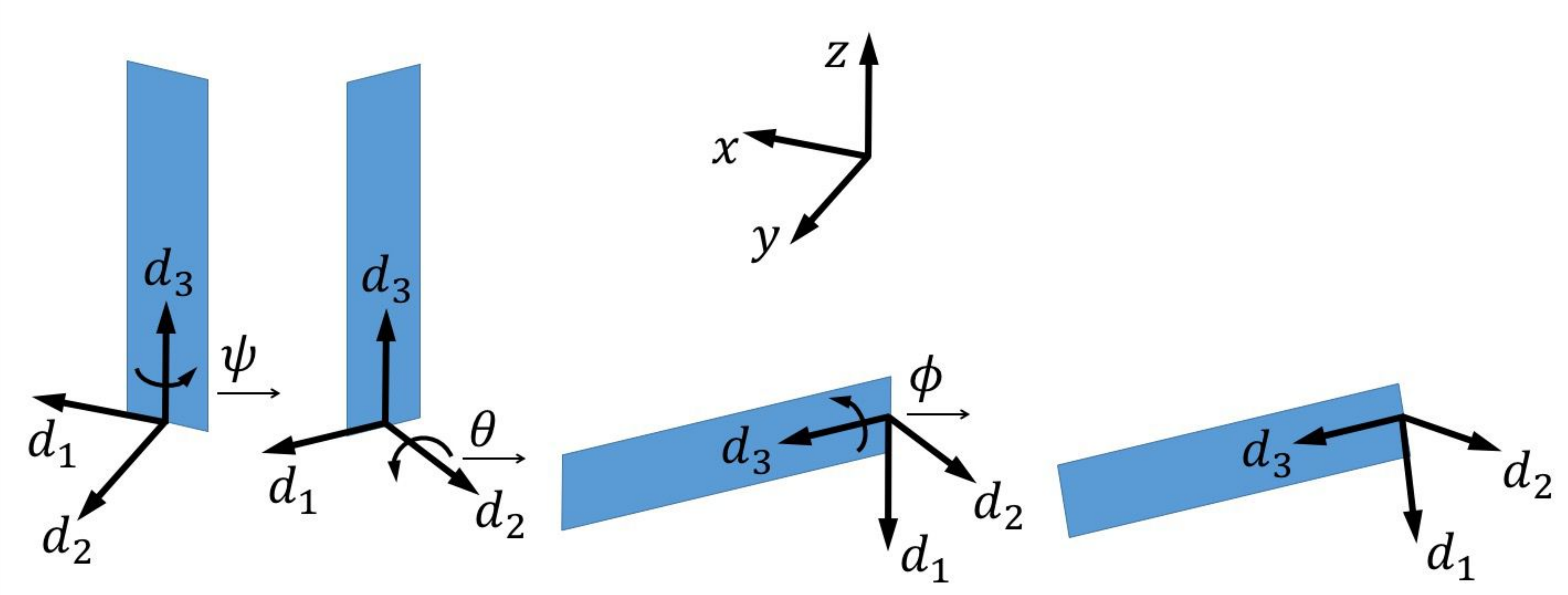}
	\caption{}\label{fig:BCrotation}
\end{subfigure} 
	\caption{Relations between the director frame, Euler angles, and Cartesian coordinates.  (\subref{fig:framecoordinate}) The director frame and Cartesian coordinates superimposed on a thin band experiment.  The right end of the band $\bX(0)$ coincides with the Cartesian origin when zero shear is applied, and is translated in the negative $z$ direction during shearing.   The $z$ axis is aligned with $-\bm{d_1}(0)$ and $-\bm{d_1}(L)$, and the $x$ and $y$ axes are aligned with the sum $\bm{d_3}(0)+\bm{d_3}(L)$ and difference $\bm{d_3}(0)-\bm{d_3}(L)$, respectively.   (\subref{fig:eulerangle}) The Euler angles defined with respect to the director frame and Cartesian coordinates.  After Love  \cite{love1927treatise}.  (\subref{fig:BCrotation}) The sequence of rotations of a strip implied by the Euler angles, using a $z$-$y$-$z$ (3-2-3) rotation convention.  Beginning with a frame ($\bm{d_{1}}$,$\bm{d_{2}}$,$\bm{d_{3}}$) aligned with the ($x$-$y$-$z$) axes, we sequentially rotate around $\bm{d}_3$ by $\psi$, around $\bm{d}_2$ by $\theta$, and around $\bm{d}_3$ by $\phi$.}\label{fig:Eulerdescription}
\end{figure}

To solve the Kirchhoff equations and reconstruct the centerline of the bands, we employ a quaternion representation of the director frame.  Additionally, it is most straightforward to represent some of the boundary conditions using classical Euler angles ($\psi$, $\theta$, $\phi$), which we express with reference to a Cartesian coordinate system ($x$, $y$, $z$), before translating them into the quaternion language.  These coordinates and angles are shown in Figure \ref{fig:Eulerdescription}, along with a sequence of rotations applied to a band.  We place an origin at the starting point of the centerline $\bX(0)$ when the shear $\Delta D = 0$.  We move the point $\bX(0)$ in the negative $z$ direction in the numerics, leaving $\bX(L)$ fixed.  During the experiments, the $z$ axis is aligned with $-\bm{d_1}(0)$ and $-\bm{d_1}(L)$, and the $x$ and $y$ axes are aligned with the sum $\bm{d_3}(0)+\bm{d_3}(L)$ and difference $\bm{d_3}(0)-\bm{d_3}(L)$, respectively.  Thus, the plane of clamping is spanned by $x$ and $z$.  
We follow the $z$-$y$-$z$ (3-2-3) rotation convention, as given in Love \cite{love1927treatise}. 
Our boundary conditions correspond to $\psi(0)=-\psi(L)=\psi_0$, the clamping angle, $\theta(0)=\theta(L)=\pi/2$, and $\phi(0)=\phi(L)=0$.

The relationship between the director frame and the Cartesian frame may be expressed in terms of either the Euler angles or the components $q_i$ of a unit quaternion \cite{love1927treatise, altmann1986rotations}, 
\begin{equation}\label{323rotation} 
\begin{aligned} 
\begin{bmatrix} \bm{d_1} \\ \bm{d_2} \\ \bm{d_3} 
\end{bmatrix}
&=
\begin{bmatrix} cos\phi & \sin\phi & 0 \\ -sin\phi & cos\phi & 0 \\ 0 & 0 & 1 
\end{bmatrix}
\begin{bmatrix} cos\theta & 0 & -sin\theta \\ 0 & 1 & 0 \\ sin\theta & 0 & cos\theta 
\end{bmatrix}
\begin{bmatrix} cos\psi & \sin\psi & 0 \\ -sin\psi & cos\psi & 0 \\ 0 & 0 & 1 
\end{bmatrix}
\begin{bmatrix} \bm{\hat{x}} \\ \bm{\hat{y}} \\ \bm{\hat{z}} 
\end{bmatrix} \\
&=2 \begin{bmatrix} q_0^2+q_1^2-\frac{1}{2} & q_1 q_2 + q_0 q_3 & q_1 q_3 - q_0 q_2\\ q_1 q_2 - q_0 q_3 & q_0^2+q_2^2-\frac{1}{2} &q_2 q_3 +q_0 q_1 \\ q_1 q_3 +q_0 q_2  &  q_2 q_3 - q_0 q_1 & q_0^2+q_3^2-\tfrac{1}{2} 
\end{bmatrix}
\begin{bmatrix} \bm{\hat{x}} \\ \bm{\hat{y}} \\ \bm{\hat{z}} 
\end{bmatrix}
\end{aligned} 
\end{equation}
The Euler angles and quaternion components are related by
\begin{equation}\label{angletoparameter} 
q_0=\cos \tfrac{\theta}{2} \cos \tfrac{\phi + \psi}{2}\,,\; q_1=\sin \tfrac{\theta}{2} \sin \tfrac{\phi - \psi}{2}\,,\; q_2=\sin \tfrac{\theta}{2} \cos \tfrac{\phi - \psi}{2}\,,\; q_3=\cos \tfrac{\theta}{2} \sin \tfrac{\phi + \psi}{2} \, , 
\end{equation}
derivatives of Cartesian components are given by
\begin{equation}\label{inextensibility}
x'=2(q_1 q_3 +q_0 q_2) \,,\; y'=2(q_2 q_3 -q_0 q_1) \,,\; z'=2(q_0^2 +q_3^2 -\tfrac{1}{2}) \, ,
\end{equation}
and derivatives of quaternion components are given by
\begin{equation}\label{Dquaternion} 
\begin{aligned}
q'_{0}&=\tfrac{1}{2}(-q_1 \kappa_1 -q_2 \kappa_2 -q_3 \tau)\,,\; q'_{1}=\tfrac{1}{2}(q_0 \kappa_1 -q_3 \kappa_2 +q_2 \tau)\,,\; \\ 
q'_{2}&=\tfrac{1}{2}(q_3 \kappa_1 + q_0 \kappa_2 - q_1 \tau)\,,\;\;\;\, q'_{3}=\tfrac{1}{2}(-q_2 \kappa_1 +q_1 \kappa_2 +q_0 \tau)\, .
\end{aligned}
\end{equation}

Using nondimensional lengths and curvatures,
\begin{equation}\label{dimensionlessgeometry} 
\begin{aligned}
s\rightarrow s/L \,,\; x\rightarrow x/L\,,\; y\rightarrow y/L\,,\; z \rightarrow z/L\,,\; \kappa_1\rightarrow \kappa_1 L\,,\; \tau\rightarrow \tau L\,,\; 
\end{aligned}
\end{equation}
the boundary conditions for position can be combined with those for clamping angle and translated from the Euler angle to quaternion description, for a total of fourteen boundary conditions,
\begin{equation}\label{boundaryconditions1} 
\begin{aligned}
&q_0(0)=q_0(1)=q_2(0)=q_2(1)=\tfrac{\sqrt{2}}{2} \cos\tfrac{\psi_0}{2}\,,\;\\
  &q_1(0)=-q_1(1)=-q_3(0)=q_3(1)= -\tfrac{\sqrt{2}}{2} \sin\tfrac{\psi_0}{2} \, , \\
&x(0)= y(0)= y(1)=z(1)=0\,,\; 
 x(1)=\Delta L/L\,,\; z(0)=-\Delta D /L\,,\;  
\end{aligned}
\end{equation}
involving three parameters, the compression $\Delta L/L$, shear $\Delta D/L$, and clamping angle $\psi_0$.

For kinematics described by a general Darboux vector $\bm{\omega}=\kappa_1 \bm{d_1} + \kappa_2 \bm{d_2} + \tau \bm{d_3}$, the Kirchhoff equations \eqref{F&Mequilibrium} can be written in components as,
\begin{equation}\label{F&Mequilibrium1} 
\begin{aligned}
N_1'-N_2 \tau+N_3 \kappa_2=0 \, , \\
N_2'+N_1 \tau-N_3 \kappa_1 =0 \, , \\
N_3'+N_2 \kappa_1-N_1 \kappa_2=0 \, , \\
M_1'-M_2 \tau-N_2+M_3 \kappa_2=0 \, , \\
M_2'+M_1 \tau-M_3 \kappa_1+N_1=0 \, , \\
M_3'+M_2 \kappa_1-M_1 \kappa_2=0 \, .
\end{aligned}
\end{equation}
Using nondimensional forces,
\begin{equation}\label{dimensionlessforce} 
\begin{aligned}
N_1\rightarrow N_1 L^2/(GJ),N_2\rightarrow N_2 L^2/(GJ),N_3\rightarrow N_3 L^2/(GJ), 
\end{aligned}
\end{equation}
and linear constitutive laws $M_1=EI_1 \kappa_1$, $M_2=EI_2 \kappa_2$, and $M_3=GJ \tau$, and defining the rigidity ratios $a \equiv EI_1 /(GJ)$ and $b \equiv EI_2/(GJ)$, these equations become
\begin{equation}\label{F&Mequilibrium3} 
\begin{aligned}
N_1' &= N_2 \tau -N_3 \kappa_2 \, ,\\
N_2' &= -N_1 \tau + N_3 \kappa_1 \, , \\
N_3' &=-N_2 \kappa_1 +N_1 \kappa_2 \, , \\
a\kappa_1 ' &=(b-1) \kappa_2 \tau+N_2 \, , \\
b\kappa_2 ' &=(1-a) \kappa_1 \tau -N_1 \, , \\
\tau' &=(a-b) \kappa_1 \kappa_2 \, .\\
\end{aligned}
\end{equation}
However, except in Appendix \ref{anisotropy}, we use a different set of equations that represent a flat band as a perfectly anisotropic rod that cannot bend around $\bm{d_2}$.  Thus, $\kappa_2$ is set to zero, and $M_2\rightarrow M_2 L/(GJ)$ is a Lagrange multiplier.
For a rod with a rectangular cross-section of width $D$ and thickness $t$ composed of elastically isotropic material, the bending and torsional rigidities are \cite{love1927treatise},
\begin{equation}\label{inertiaofmoment} 
\begin{aligned} 
EI_1=\tfrac{1}{12} E D t^3\,,\;  EI_2=\tfrac{1}{12} E D^3 t \,,\; 
GJ=\lambda G D t^3=\lambda D \frac{E}{2(1+\nu)} t^3 \, ,
\end{aligned}
\end{equation}
in which $\lambda$ is a shape factor that is a function of $D/t$.  The ratios of bending to torsional rigidity $a$ and $b$ are thus
\begin{equation}\label{coefficients} 
a=\frac{(1+\nu)}{6\lambda} \,,\;  b=\frac{(1+\nu)}{6\lambda}\left(\frac{D}{t} \right)^2 \, .
\end{equation}
For our narrow bands, $D/t \approx 23.6$, the shape factor is already approaching the asymptotic value $\lambda \rightarrow 1/3$, and the ratio $b/a$ is quite large, justifying our approximation of vanishing $\kappa_2$.
In this limit, the equilibrium equations take the form,
\begin{equation}\label{F&Mequilibrium2} 
\begin{aligned}
N_1'&=N_2 \tau \, , \\
N_2'&=-N_1 \tau+N_3 \kappa_1 \, , \\
N_3'&=-N_2 \kappa_1 \, , \\
M_1'&=M_2 \tau+N_2 \, , \\
M_2'&=-M_1 \tau+M_3 \kappa_1-N_1 \, , \\
M_3'&=-M_2 \kappa_1 \, .
\end{aligned}
\end{equation}
Using the same linear constitutive laws for $M_1$ and $M_3$, and combining with the relationships \eqref{inextensibility} and \eqref{Dquaternion}, we have thirteen equations,
  \begin{equation}\label{Governequation} 
\begin{aligned}
N_1' &= N_2 \tau \, ,\\
N_2' &= -N_1 \tau + N_3 \kappa_1 \, , \\
N_3' &=-N_2 \kappa_1 \, , \\
a\kappa_1 ' &=M_2 \tau+N_2 \, , \\
M_2 ' &=(1-a) \kappa_1 \tau -N_1 \, , \\
\tau' &=-M_2 \kappa_1 \, ,\\
q_{0}' &=\frac{1}{2} (-q_1 \kappa_1 -q_3 \tau) \,,\;
q_{1}' =\frac{1}{2} (q_0 \kappa_1 +q_2 \tau)\,,\; \\
q_{2}' &=\frac{1}{2} (q_3 \kappa_1 -q_1 \tau) \,,\;
q_{3}' =\frac{1}{2} (-q_2 \kappa_1 + q_0 \tau)\,,\; \\
x' &=2(q_1 q_3 + q_0 q_2) \,,\;
y' =2(q_2 q_3 - q_0 q_1) \,,\;
z' =2(q_0^2 + q_3^2-\tfrac{1}{2}) \, .
\end{aligned}
\end{equation}
The system \eqref{Governequation} and \eqref{boundaryconditions1} has one excess boundary condition.  The redundancy comes from the fact that quaternion components satisfy the algebraic constraint $q^2_0+q^2_1+q^2_2+q^2_3=1$.  We need only specify all four components at one end, so we simply do not use the $q_0(1)$ boundary condition.

Adding a nondimensional body force $g$ to the equilibrium equations modifies the first three to
\begin{equation}\label{Gravity} 
\begin{aligned}
N_1' &= N_2 \tau -2 g (q_0 q_2 - q_1 q_3) \, ,\\
N_2' &= -N_1 \tau + N_3 \kappa_1 + 2 g (q_2 q_3 + q_0 q_1) \, ,\\
N_3' &=-N_2 \kappa_1 + 2 g (q^2 _0 +q^2 _3 -\tfrac{1}{2}) \, . \\
\end{aligned}
\end{equation}

A good starting point for continuation is the circular configuration 
\begin{equation}\label{InitialGuess} 
\begin{aligned}
N_{1}=0, N_{2}=0, N_{3}=0\, ,\\
 \kappa_{1}=2\pi, M_{2}=0 , \tau=0\, , \\
q_{0}=\tfrac{\sqrt{2}}{2} \sin(\pi s)\, , \, q_{1}=-\tfrac{\sqrt{2}}{2} \cos(\pi s)\, ,\\
q_{2}=\tfrac{\sqrt{2}}{2} \sin(\pi s)\, , \, q_{3}= \tfrac{\sqrt{2}}{2} \cos(\pi s)\, ,\\
x=-\tfrac{\sin 2 \pi s}{2 \pi}\, , \, y=\tfrac{1-\cos 2 \pi s}{2 \pi},z=0\, ,\\
\end{aligned}
\end{equation}
from which the ends can be opened and rotated to achieve various buckled states. Note that we cannot apply compression directly to a straight configuration, because the initial guess for Auto 07P cannot be a branch point \cite{doedel2007auto}.   

The rod frame renderings are achieved by sweeping the director $\bm{d}_1(s)$ along the center line $\bX(s)$.  The resulting surface can be parameterized as
\begin{equation}\label{reconstruction}
\begin{aligned} 
\bm{R}(s,v)&=\bX(s)+v \bm{d}_1(s) \, ,\\
&=\left[x+2v\left(q_0^2+q_1^2-\tfrac{1}{2}\right)\right]\bm{\hat{x}} + \left[y+2v\left(q_1 q_2 + q_0 q_3\right)\right]\bm{\hat{y}} + \left[z+2v\left(q_1 q_3 -q_0 q_2\right)\right]\bm{\hat{z}} \, ,
\end{aligned}
\end{equation}
with $-\frac{D}{2L} \le v \le \frac{D}{2L}$.  
Note that a finite-width band will deform into a different surface.  The above surface corresponding to the rod frame is thus only a rough approximation to the actual shape of a narrow band.  In a developable model, the generators do not coincide with the material directors $\bm{d}_1$.

\section{Poisson's ratio}\label{poisson}  

\begin{figure}[h!]
	\centering
	\includegraphics[width=0.85\textwidth]{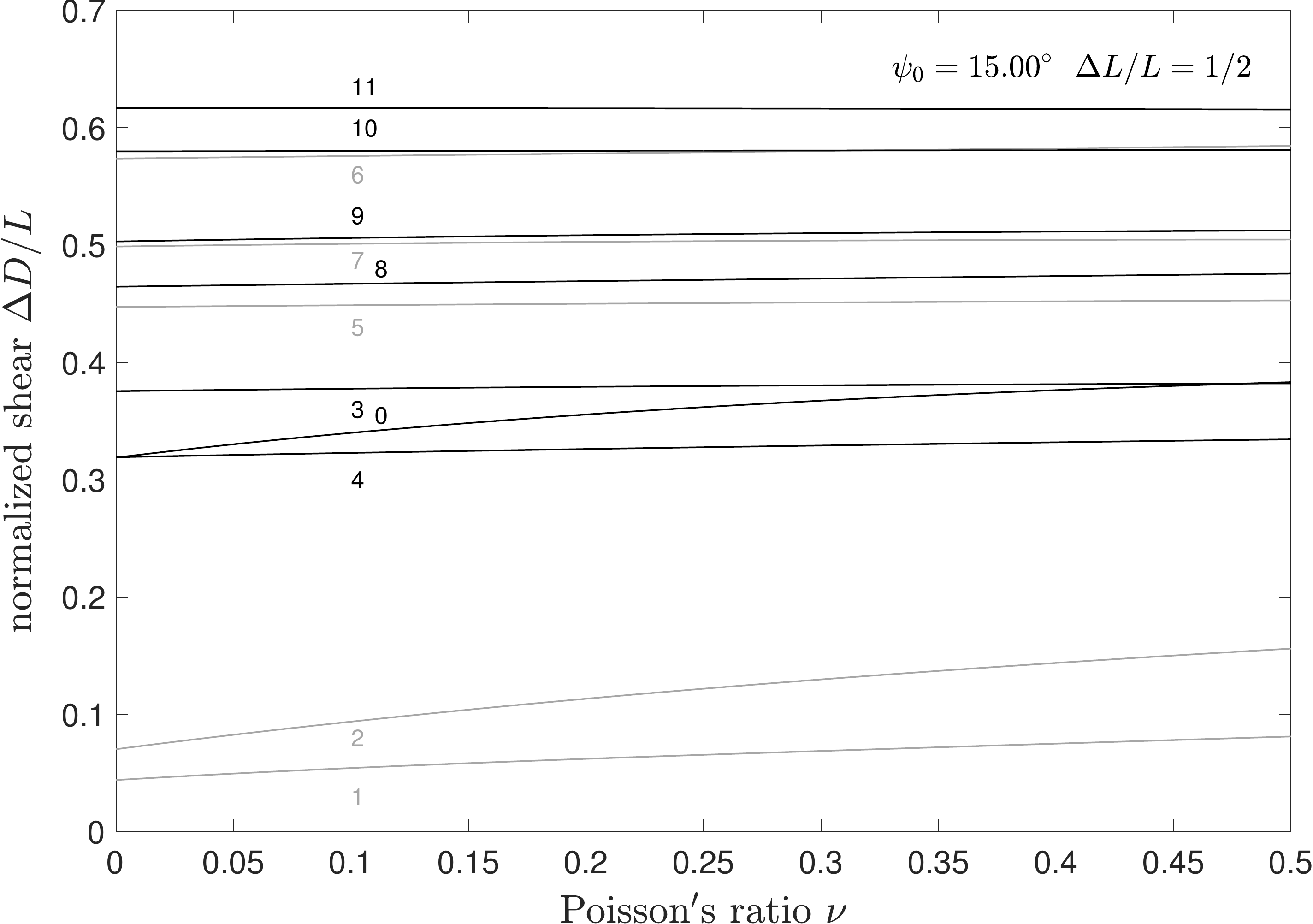}
	\caption{Effect of Poisson's ratio $\nu$ on the shear value of bifurcations with clamping angle $\psi_0=15^{\circ}$.  Points $0$, $1$, and $2$ are associated with highly twisted states, and are affected the most.  A $\nu$ of $0.25$ was used in this work.}
	\label{fig:poissonratio}
\end{figure}

Using two-parameter continuation in shear-Poisson's ratio space, we have confirmed that the effects of Poisson's ratio $\nu$ on the results for perfectly anisotropic rods are minimal, with the exception of bifurcations such as $0$, $1$, and $2$, which are associated with highly twisted states.  This is consistent with Goss's results for isotropic rods \cite{Goss2003snap}.
Figure \ref{fig:poissonratio} shows a representative example for the clamping angle $\psi_0=15^{\circ}$, showing the shear value of the bifurcation for admissible positive values of Poisson's ratio.
Poisson's ratio affects the ratios of bending to torsional rigidity \eqref{coefficients}.  A larger $\nu$ means a relatively lower torsional rigidity, and thus a wider range of stability for highly twisted states.

\section{Anisotropy }\label{anisotropy}

\begin{figure}[h!]
	\centering
	\includegraphics[width=0.85\textwidth]{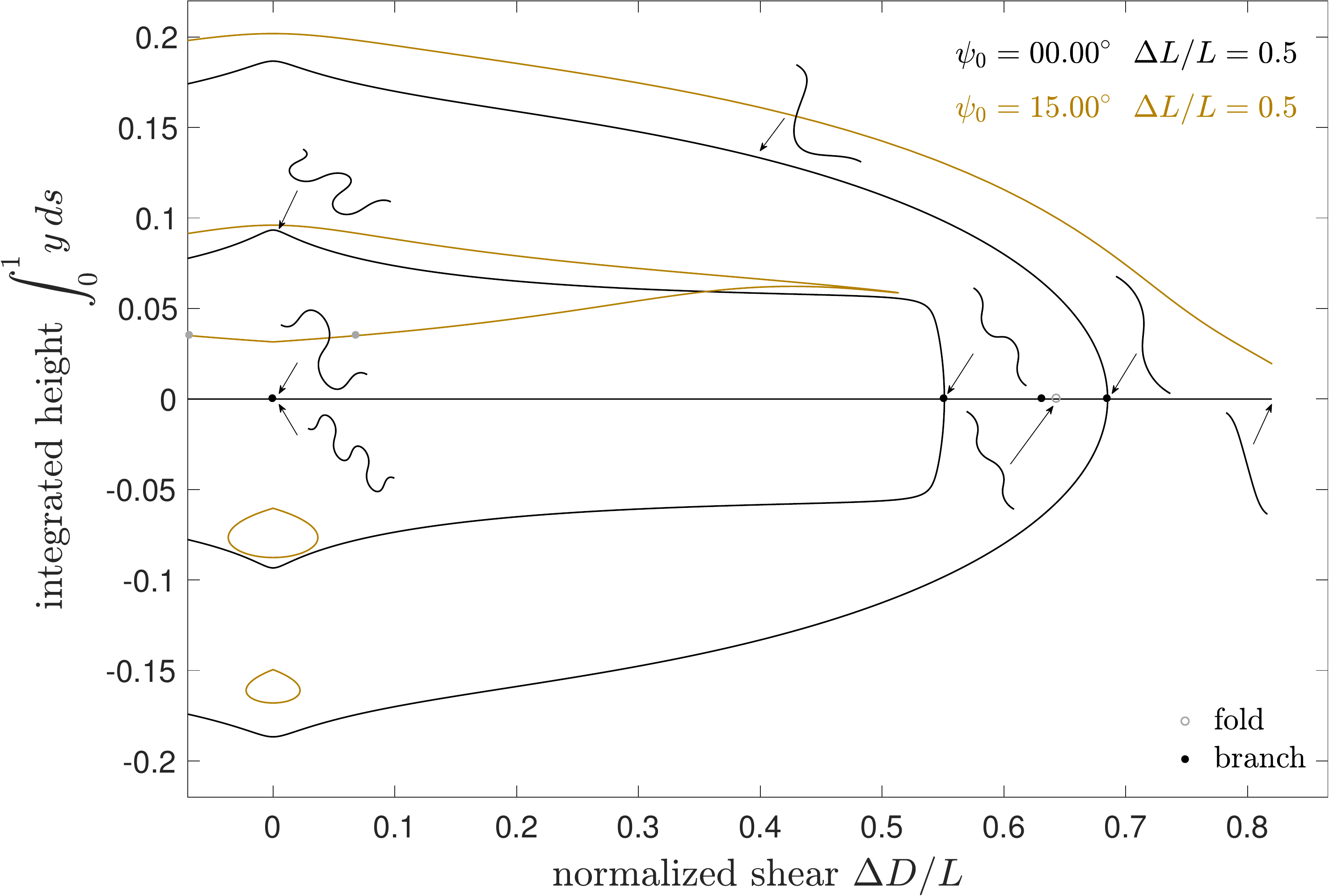}
	\caption{Solution curves and bifurcations (open and closed disks) for square cross section Kirchhoff rods for clamping angles $\psi_0=0^{\circ}$ and $\psi_0=15^{\circ}$, along with drawn curves from the arrowed locations for $\psi_0=0^{\circ}$. Compare with Figures \ref{fig:Config0} and \ref{fig:Solution15}. }
	\label{fig:isotropic}
\end{figure}

Using the full rod equations \eqref{F&Mequilibrium3}, we confirm that much of the complexity of our solution manifolds is due to the anisotropy of the rod.  For a rod with square cross section, $\lambda \approx 0.141$ \cite{love1927treatise}, $a=b\approx1.478$ for $\nu=0.25$, and $\tau$ is a constant.  Analytical solutions exist for such systems, but for brevity we present numerical continuation results in Figure \ref{fig:isotropic} for clamping angles of $\psi_0=0^{\circ}$ (black curves) and $\psi_0=15^{\circ}$ (brown curves).  This can be compared with Figures \ref{fig:Config0}
and \ref{fig:Solution15} in Section \ref{results}.  Aside from the absence of twisted states, the connectivity of the solution manifolds is quite different, and fewer states exist at high shear.

For zero clamping angle $\psi_0=0^{\circ}$, shearing the first mode of planar \emph{elastica} deforms the rod into a spatial configuration with two nonzero curvatures and a constant twist.    At a supercritical pitchfork bifurcation at $\Delta D/L \approx 0.685$, the shape eventually changes to the second mode of planar \emph{elastica} in a plane perpendicular to that of the initial configuration.  This state has only one curvature and zero twist.  We thus realize that for an isotropic rod, the shear is simply equivalent to a meaningless spatial rotation plus changes in compression and clamping angle, the latter serving to exchange the stability of first and second mode \emph{elastica}.  This behavior can be imperfectly observed in a smartphone charger cord, despite the inevitable presence of rest curvature in such objects.  For nonzero clamping angle, this pitchfork is broken and the configurations remain non-planar at high shears.

Further tinkering with the numerics shows that at zero clamping angle, the bifurcations on the first mode ($U$) branch to the $US\pm$ and $uUu$ branches appear when the anisotropy $D/t$ reaches moderate values.
Buzano \cite{buzano1986secondary} studied the appearance of secondary bifurcations of compressed, clamped rectangular cross section rods when the cross section became infinitesimally anisotropic.   van der Heijden and Thompson \cite{van1998lock} discuss a transition in the behavior of elliptical cross section rods when the anisotropy reaches a moderate value.

\section{Comparison with a strip model}\label{stripmodel}

Here we use one of our experimental deformation paths to compare the anisotropic Kirchhoff rod model with a strip model implemented in a manner following van der Heijden and Starostin \cite{starostin2015equilibrium}.
In strip models, the description of inflection points is singular, and in this implementation, such points are inserted by hand.  
We display two consequences of this situation.
 One is that numerical continuation will not lead to the creation of new inflection points.  We can allow pre-existing inflection points to move as internal boundaries within the strip by using the approach of Ascher and Russell \cite{ascher1981reformulation}, but they cannot spontaneously appear or disappear. 
It seems that this issue did not arise in the work of Dias and Audoly \cite{dias2014non, dias2015wunderlich} or Audoly and Seffen \cite{audoly2015buckling}, where continuation was performed along branches that did not involve creation of new inflection points.
  Second, at the pre-set inflection points, the curvature and torsion behave strangely.  As the limit of a narrow strip is approached, these quantities develop sharp gradients and do not approach those of a rod, but instead a singular limit involving a jump in curvature \cite{hinz2015translation}, which would indicate a singular source of moment in a rod.
While the first issue may arise from our choice of implementation, the second is a characteristic of inextensible strip models themselves, and can be observed in the prior results of several groups \cite{starostin2015equilibrium, moore2015computation}, including two-dimensional simulations of strips in the inextensible limit \cite{kleiman2016influence}.
Both issues are such that strip models cannot capture some of the behavior in our experiments, which involve the creation and destruction of inflection points, and in which the curvature and torsion appear to be well-behaved.
Regularization of the strip energy may resolve the first issue, but does not resolve the second \cite{moore2015computation}. 
Others have suggested the use of a piecewise-continuous energy density, with modifications in regions of small curvature \cite{freddi2016corrected}.
We conclude that there is a need for new models of strips that interpolate between rod models and wide, inextensible strip models.
A good testing ground for such a model might be the simple case of a body subject to simple tensile end constraints and twist.  A rod model admits twisted solutions with a straight centerline \cite{goriely2001dynamics}, which do not correspond to developable surfaces.

\begin{figure}[h!]
	\centering
	\includegraphics[width=0.95\textwidth]{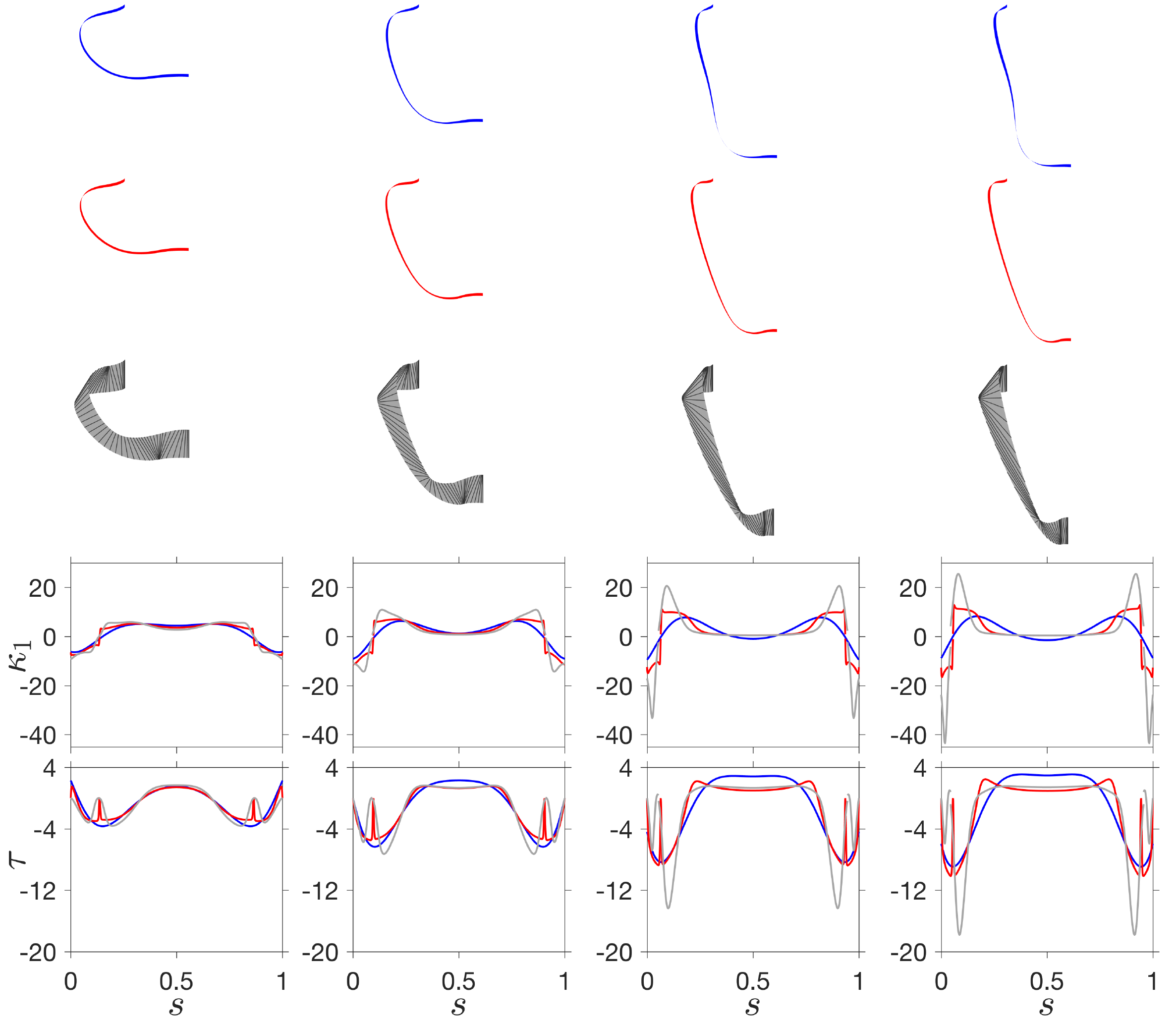}
	\caption{Numerically determined configurations, curvature, and torsion (twist) corresponding to a path along the $U$-$w$-$uUui$ branch shown in the supplementary video {\texttt{widtheffect45.mp4}}, with compression $\Delta L/L=0.5$, clamping angle $\psi_0=45^{\circ}$, and shear $\Delta D/L = 0.2, 0.4, 0.56, 0.6$ increasing from left to right.  Shown are renderings of the rod frame based on the perfectly anisotropic Kirchhoff rod equations (blue) and renderings of the rectifying developable corresponding to the Wunderlich strip equations for narrow (red, $D/L=1/80$) and wide (grey, $D/L=1/8$) strips, with generators drawn on the wide strip.
As explained in the text, continuation of the Wunderlich model following the approach of \cite{starostin2015equilibrium} does not capture either the appearance of inflection points in narrow bands or the occurrence of a snap-through in wide bands.  All solutions begin with two inflection points near the ends, and the rod solution gains two additional inflection points near the middle.  Note the singular behavior of the narrow band curvature and torsion near the two inflection points.  Jumps in curvature and torsion are introduced to handle the inflection points in the strip model \cite{starostin2015equilibrium}.}
	\label{fig:comparison}
\end{figure}

Figure \ref{fig:comparison} compares results from the rod model (blue) and a Wunderlich strip model \cite{starostin2015equilibrium} applied to narrow (red, $D/L=1/80$) and wide (grey, $D/L=1/8$) strips, for values of compression $\Delta L/L=0.5$, clamping angle $\psi_0=45^{\circ}$, and a sequence of shears $\Delta D/L = 0.2, 0.4, 0.56, 0.6$.
This corresponds to a path along the $U$-$w$-$uUui$ branch shown in the supplementary video {\texttt{widtheffect45.mp4}} \cite{videos}.  
The configurations begin with two inflection points near the ends.
In experiments, narrow bands smoothly gain two additional inflection points near the middle, and wide bands experience a snap-through instability which also creates two inflection points.  The numerical rod solutions behave like the experimental narrow bands.  The numerical strip solutions do not pick up new inflection points or detect a snap-through bifurcation.  
Recall that the single curvature and twist of a perfectly anisotropic rod can be identified with the curvature and torsion for a developable strip.  The singular behavior of the curvature and torsion are most apparent for narrow strips.

The rod model and the rendering of the rod frame have already been discussed in the main text and Appendix \ref{bvp}.  
The strip model employs the Wunderlich functional proportional to $\kappa_1^2 (1+\eta ^2)^2 \tfrac{1}{D \eta'} \ln (\tfrac{1+D \eta' /2}{1-D \eta' /2})$, where $\eta \equiv \tfrac{\tau}{\kappa_1}$.  The limit of this model as the strip width $D$ vanishes is known as the Sadowsky functional, which lacks the logarithmic term and its prefactor.  The Wunderlich functional leads to a set of Euler-Lagrange equations, equivalent up to a sign to equations (29) in \cite{starostin2015equilibrium}.  The strips are rendered as rectifying developables
\begin{equation}\label{reconstructionstrip}
\bm{R}(s,v)=\bX(s)+v \left[ \bm{d}_1(s) + \eta(s)\bm{d}_3 \right] \, ,
\end{equation}
with $-\frac{D}{2L} \le v \le \frac{D}{2L}$.  Generators are drawn on the wide strips, showing the focusing of bending energy akin to that shown in Figure \ref{fig:WidExpConfiguration} and in the video.

The solution of the strip model involves several steps, which we merely sketch here, referring the reader to the necessary references for extensive details.  First, a starting solution is obtained from the Euler \emph{elastica} or rod model configuration at zero shear.  
In the Wunderlich model, but not the Sadowsky model, the description of the energy density is singular, and a Taylor expansion in $\eta'$ is applied to find a suitable starting solution for continuation, until the shear is a small nonzero value.  The solution is partitioned into pieces by its inflection points.  The resulting multi-point variable-arc-length boundary value problem is reformulated as a standard two-point boundary value problem by the introduction of scalar variables \cite{ascher1981reformulation} representing the dynamically updating lengths of segments between the inflection points, and solved using the continuation package AUTO 07P \cite{doedel2007auto}.
At each inflection point, a small jump in curvature and torsion is introduced \cite{starostin2015equilibrium} such that the ratio $\eta$ is continuous; this jump is decreased as far as numerical stiffness will allow.
Near these jumps, ``ears'' develop in the curvature profile and a spurious extra oscillation in the torsion arises due to the requirement that the torsion vanish with the curvature.  These features, which are not seen in experiments, become more extreme as the strip width shrinks, as can be clearly seen in Figure \ref{fig:comparison} and in 
numerical results on one- and two-dimensional inextensible models in the literature, such as Figure 7 of \cite{starostin2015equilibrium}, Figure 6 of \cite{moore2015computation}, and Figure 4 of \cite{kleiman2016influence}.
The Sadowsky limit of the ``ears'' is a jump in curvature \cite{hinz2015translation}, something that cannot exist in a Kirchhoff rod model without an applied moment.  Note that this jump is orders of magnitude larger than the artificial jump introduced in order to handle the numerics near the inflection point.

\clearpage

\bibliographystyle{unsrt}
\bibliography{ThinbandsReferences} 

\end{document}